\documentclass[journal]{IEEEtran}
\usepackage{cite}
\usepackage{amsmath}
\usepackage{amsfonts}
\usepackage{color}
\usepackage{amssymb}
\usepackage{array}
\usepackage{url}
\usepackage{graphicx}
\usepackage{subfigure}
\usepackage{longtable}
\usepackage{multirow}
\usepackage{marvosym}
\usepackage{textcomp}
\usepackage{threeparttable, tablefootnote}
\usepackage{enumerate}
\usepackage[font=small,labelsep=period]{caption}
\usepackage{dsfont}
\usepackage{soul}
\soulregister\cite7
\soulregister\ref7

\allowdisplaybreaks[4]

\usepackage{xcolor}
\usepackage{soul}
\soulregister\cite7
\soulregister\ref7
\usepackage{algorithm,algpseudocode}
\makeatletter
\newcommand{\algmargin}{\the\ALG@thistlm}
\makeatother
\newlength{\whilewidth}
\algdef{SE}[parWHILE]{parWhile}{EndparWhile}[1]
  {\parbox[t]{\dimexpr\linewidth-\algmargin}{%
     \hangindent\whilewidth\strut\algorithmicwhile\ #1\ \algorithmicdo\strut}}{\algorithmicend\ \algorithmicwhile}%
\algnewcommand{\parState}[1]{\State%
  \parbox[t]{\dimexpr\linewidth-\algmargin}{\strut #1\strut}}

\algdef{SE}[parIF]{parIf}{EndparIf}[1]
  {\parbox[t]{\dimexpr\linewidth-\algmargin}{%
     \hangindent\ifwidth\strut\algorithmicif\ #1\ \algorithmicdo\strut}}{\algorithmicend\ \algorithmicif}%

\allowdisplaybreaks[4]
\renewcommand{\arraystretch}{1.2}
\newtheorem{lemma}{Lemma}

\newtheorem{theorem}{Theorem}

\ifCLASSINFOpdf
\else
\fi
\begin{document}
\title{History-Aware Online Cache Placement in Fog-Assisted IoT Systems: An Integration of Learning and Control}

%\author{Michael~Shell,~\IEEEmembership{Member,~IEEE,}
%        John~Doe,~\IEEEmembership{Fellow,~OSA,}
%        and~Jane~Doe,~\IEEEmembership{Life~Fellow,~IEEE}% <-this % stops a space
%\thanks{M. Shell was with the Department
%of Electrical and Computer Engineering, Georgia Institute of Technology, Atlanta,

\author{Xin~Gao,~\IEEEmembership{Student Member,~IEEE,}
		Xi~Huang,~\IEEEmembership{Member,~IEEE,}
		Yinxu~Tang,
        Ziyu~Shao$^{*}$,~\IEEEmembership{Senior Member,~IEEE,}
        Yang~Yang,~\IEEEmembership{Fellow,~IEEE}                
\thanks{
%This work was partially supported by Nature Science Foundation of Shanghai under Grant 19ZR1433900. (*Corresponding author: Ziyu Shao)

%Partial results in this work have been published in IEEE International Conference on Communications (ICC), 2020\cite{gao2020proactive}.

X. Gao is with the School of Information Science and Technology, ShanghaiTech University, Shanghai 201210, China, 
also with the Shanghai Institute of Microsystem and Information Technology, Chinese Academy of Sciences, Shanghai 200050, China, 
and with the University of Chinese Academy of Sciences, Beijing 100049, China. (E-mail: gaoxin@shanghaitech.edu.cn)

X. Huang, Y. Tang and Z. Shao are with the School of Information Science and Technology, ShanghaiTech University, Shanghai 201210, China. (E-mail: \{huangxi, tangyx, shaozy\}@shanghaitech.edu.cn)

Y. Yang is with Shanghai Institute of Fog Computing Technology (SHIFT), ShanghaiTech University, Shanghai 201210, China, and the Research Center for Network Communication, Peng Cheng Laboratory, Shenzhen 518000, China. (E-mail: yangyang@shanghaitech.edu.cn)
}
}

\maketitle

\begin{abstract}
In Fog-assisted IoT systems, it is a common practice to cache popular content at the network edge to achieve high quality of service. Due to uncertainties in practice such as unknown file popularities, cache placement scheme design is still an open problem with unresolved challenges: 1) how to maintain time-averaged storage costs under budgets, 2) how to incorporate online learning to aid cache placement to minimize performance loss (\textit{a.k.a.} regret), and 3) how to exploit offline historical information to further reduce regret. In this paper, we formulate the cache placement problem with unknown file popularities as a constrained combinatorial multi-armed bandit (CMAB) problem. 
To solve the problem, we employ virtual queue techniques to manage time-averaged storage cost constraints, and adopt history-aware bandit learning methods to integrate offline historical information into the online learning procedure to handle the exploration-exploitation tradeoff. 
With an effective combination of online control and history-aware online learning, we devise a  Cache Placement scheme with History-aware Bandit Learning called \textit{CPHBL}.
Our theoretical analysis and simulations show that CPHBL achieves a sublinear time-averaged regret bound.
Moreover, the simulation results verify CPHBL's advantage over the deep reinforcement learning based approach.
%To our best knowledge, our work provides the first systematic study on the synergy of online control, online learning, and offline historical information. 
\end{abstract}


% \fi
\IEEEpeerreviewmaketitle

\section{Introduction}

\begin{figure}[!t]
 \center
 \includegraphics[scale=0.36]{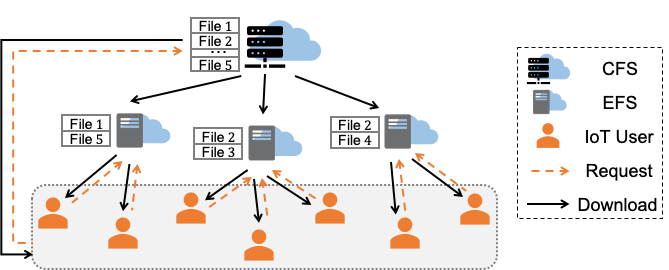}
 \caption{An illustration of caching-enabled Fog-assisted IoT systems.}
 \label{fig: model}
\end{figure}

During recent years, the proliferation of \textit{Internet of Things} (IoT) devices such as smart phones and the emerging of IoT applications such as video streaming have led to an unprecedented growth of data traffic\cite{bastug2014living}.
To meet the explosively growing traffic demands at the network edge and facilitate IoT applications with high \textit{quality of service} (QoS), caching popular contents at fog servers has emerged as a promising solution\cite{zhao2017online, jiang2017novel, zhao2018femos,gao2020pora}.
Figure \ref{fig: model} shows an example of wireless caching in a multi-tier Fog-assisted IoT system. As shown in the figure, by utilizing the storage resources on fog servers that are close to IoT devices, popular contents (\textit{e.g.}, files) can be cached to achieve timely content delivery.
Due to resource limit, each edge fog server (EFS) can cache only a subset of files to serve its associated IoT users. 
If a user's requested file is found on the corresponding EFS (\textit{a.k.a.} a \textit{hit}), then it can be downloaded directly;
otherwise, the file needs to be fetched from the central fog server (CFS) in the upper fog tier with extra bandwidth consumption and latency.  
Therefore, the key to maximize the benefits of caching in Fog-assisted IoT systems lies in the selection of a proper set of cached files (\textit{a.k.a.} \textit{cache placement}) on each EFS.  

However, 
the effective design for cache placement remains as a challenging problem 
due to the uncertainty of file popularities in such systems. 
Specifically, as an important ingredient for cache placement optimization, file popularities are usually unknown in practice\cite{bharath2016learning}. 
Such information can only be inferred implicitly from feedback information such as cache hit signals for user requests. 
Meanwhile, in practice, it is common for Fog-assisted IoT systems to retain offline historical observations (in terms of file request logs) on each EFS. 
Such offline information can also be exploited to estimate the file popularities in the procedure of cache placement.
%be explored to achieve a more accurate inference for the unknown file popularities.
Nonetheless, it remains non-trivial about how to integrate both online feedback and offline historical information to reduce uncertainties in decision making and minimize the resulting performance loss (\textit{a.k.a.} regret).
%For example, the historical observations about the requests of IoT users are provided prior to the start of the cache placement scheme.
If such an integration can be achieved, then each EFS can proactively update cache placement based on its learned popularity statistics to improve system performances.

Towards such a joint design, three challenges must be addressed.
The first is concerning the tradeoff between conflicting performance metrics.
%When caching a file, the EFS will receive a reward for each request for the the file, but pay a \textit{storage cost} (\textit{e.g.}, storage footprint) for caching it. 
On one hand, caching more popular files on each EFS conduces to higher \textit{cache hit rewards} (\textit{e.g.}, the total size of files served by wireless caching).
On the other hand, the number of cached files should be limited to avoid excessive \textit{storage costs} (\textit{e.g.}, memory footprint)\cite{pang2016joint}.
Such a tradeoff between cache hit rewards and storage costs should be carefully considered for cache placement.
The second is regarding the \textit{exploration-exploitation} dilemma encountered in the online learning procedure; \textit{i.e.}, for each EFS, should it cache the files with empirically high estimated popularities (exploitation) or those files with inadequate feedback but potentially high popularities (exploration)?
The third is {about how to} leverage offline historical information to further improve learning efficiency, which serves as a new degree of freedom in the design space of cache placement. Faced with such challenges, the interplays among online control, online learning, and offline historical information deserve a systematic investigation.

In this paper, we focus on the problem of proactive cache placement in caching-enabled Fog-assisted IoT systems with offline historical information and unknown file popularities under constraints on time-averaged storage costs of EFSs.
We summarize our contributions and key results as follows.

\begin{itemize}
	\item[$\diamond$] \textbf{Problem Formulation:}
We formulate the problem as a stochastic optimization problem under uncertainties, 
with the aim to maximize the total cache hit reward in terms of the total size of files directly fetched from EFSs to IoT users over a finite time horizon.
Meanwhile, we also consider the time-averaged storage cost constraint on each EFS. 
By exploiting the problem structure, we extend the settings of the recently developed bandit model \cite{li2019combinatorial} and reformulate the problem as a constrained combinatorial multi-armed bandit (CMAB) problem.

	\item[$\diamond$] \textbf{Algorithm Design:}
 	To solve the formulated problem, we propose \textit{CPHBL} (Cache Placement with History-aware Bandit Learning), a learning-aided cache placement scheme that conducts proactive and effective cache placement under time-averaged storage cost constraints. 
 	In general, CPHBL consists of two interacting procedures: the online learning procedure and the cache update procedure. 
 	Particularly, in the online learning procedure, we adopt the HUCB1 (UCB1 with Historic Data) method \cite{shivaswamy2012multi} to leverage both offline historical information and online feedback to learn the unknown file popularities with a decent exploration-exploitation tradeoff.
	In the cache update procedure, 
	we leverage Lyapunov optimization method \cite{neely2010stochastic} to update cached files on EFSs in an adaptive manner, so that cache hit rewards can be maximized subject to the storage cost constraints.

	\item[$\diamond$] \textbf{Theoretical Analysis:}
	To the best of our knowledge, our work conducts the first systematic study on the integration of online control, online learning, and offline historical information.
%	Besides, we provide theoretical analysis to capture the interplay between online control and history-aware online learning for cache placement in Fog-assisted IoT systems. We also characterize how online feedback and offline historical information jointly affect the performance of such systems.
	In particular, our theoretical analysis shows that our devised scheme achieves a near-optimal total cache hit reward under time-averaged storage cost constraints with a time-averaged regret bound of order $O(1/V+1/T+\sqrt{(\log T)/(T+H_{\min})})$. Note that $V$ is a positive tunable parameter, $T$ is the length of time horizon, and $H_{\min}$ is the minimum number of offline historical observations among different EFSs.

	\item[$\diamond$] \textbf{Numerical Evaluation:} 
	We conduct extensive simulations to investigate the performances of CPHBL and its variants. Moreover, we devise a novel deep reinforcement learning (DRL) based scheme as one of the baselines to be compared with CPHBL.
	Our simulation results not only verify our theoretical analysis, but also show the advantage of CPHBL over the baseline schemes.

	\item[$\diamond$] \textbf{New Degree of Freedom in the Design Space of Fog-Assisted IoT Systems}: We systematically investigate the fundamental benefits of offline historical information in Fog-assisted IoT systems. We provide both theoretical analysis and numerical simulations to evaluate such benefits. Our results reveal novel insights to system designers to improve their systems.
\end{itemize}

The rest of this paper is organized as follows. 
Section \ref{sec: related work} discusses the related works.
Section \ref{sec: model} illustrates our system model and problem formulation. Section \ref{sec: algorithm} shows our algorithm design, followed by the performance analysis in Section \ref{sec: analysis}. 
Section \ref{sec: drl} proposes a novel DRL based scheme as a baseline for evaluation and then Section \ref{sec: simulation} discusses our simulation results.
Finally, Section \ref{sec: conclusion} concludes this paper. 
%Proofs and more results are delegated to our technical report \cite{gao2020proactivetech}.

\begin{table*}[!h]
\centering
\caption{Comparison between our work and related works}
\label{table: related works}
%\begin{tabular}{|p{0.56cm}<{\centering}|p{0.56cm}<{\centering}|p{0.7cm}<{\centering}|p{0.9cm}<{\centering}|p{0.9cm}<{\centering}|p{1.75cm}<{\centering}|p{1.0cm}<{\centering}|}
\renewcommand\arraystretch{1}
\setlength{\tabcolsep}{1.7mm}{	
\begin{threeparttable}	
\begin{tabular}{|c|c|c|c|c|c|c|}
\hline
& \multirow{2}{*}{Optimization Metrics} & \multicolumn{2}{c|}{Resource Constraints} & Online & Online & Offline History \\ \cline{3-4}
& & Per-time-slot Constraints & Long-term Constraints & Control & Learning & Information \\
\hline
\cite{pang2016joint} & Revenue and cost of caching \& delivery cost & $\bullet$ & $\bullet$ & $\bullet$ & & \\
\hline
\cite{kwak2018hybrid} & Service rates for file requests & $\bullet$ & $\bullet$ & $\bullet$ & & \\
\hline
\cite{wang2018distributed} & Queueing delay \& energy consumption & $\bullet$ & $\bullet$ & $\bullet$ & & \\
\hline
\cite{xu2018joint} & Task delay \& energy consumption & $\bullet$ & $\bullet$ & $\bullet$ & & \\
\hline
\cite{blasco2014learning} & Cache hit reward & $\bullet$ & &  & $\bullet$ & \\
\hline
\cite{blasco2014multi} & Cache hit reward \& file downloading cost & $\bullet$ &  &  & $\bullet$ & \\
\hline
\cite{muller2016smart} & Number of cache hits & $\bullet$ & &  & $\bullet$ & \\
\hline
\cite{zhang2019learning} & Weighted network utility & $\bullet$ & &  & $\bullet$ & \\
\hline
\cite{song2017learning} & Revenue of caching \& content sharing cost & $\bullet$ & &  & $\bullet$ & \\ 
\hline
\cite{xu2020collaborative} & Network transmission delay & $\bullet$ & &  & $\bullet$ & \\
\hline
Our Work & Cache hit reward \& storage cost & $\bullet$ & $\bullet$ & $\bullet$ & $\bullet$ & $\bullet$ \\
\hline
\end{tabular}
%\begin{tablenotes}
%	\footnotesize
%	 \item[1] The reward is defined in terms of whether the task latency is below a pre-specified threshold.
%\end{tablenotes}
\end{threeparttable}}
\end{table*}

% Related work
\section{Related Work}\label{sec: related work}

In the past decades, cache placement has been widely studied to improve the performance of {wireless networks such as IoT networks\cite{ajmal2019survey} and cellular networks\cite{li2018survey}}.
{Among existing works, those that are most relevant to our work are generally carried out from two perspectives: the \textit{online control} perspective and the \textit{online learning} perspective.}
%Solutions of this category leverage the content request patterns to make the optimal placement decisions. By prefetching popular files ahead of time based on the file popularities, proactive schemes can achieve more efficient content deliveries. 

\textbf{Online Control based Cache Placement:}
Most works that take the online control perspective formulated cache placement problems as stochastic network optimization problems with respect to different metrics. 
For example, in \cite{pang2016joint}, Pang \textit{et al.} jointly studied the cache placement and data sponsoring problems in mobile video content delivery networks. 
Their solution aimed to maximize the overall content delivery payoff with budget constraints on caching and delivery costs. 
Kwak \textit{et al.}\cite{kwak2018hybrid} devised a dynamic cache placement scheme to optimize service rates for user requests in a hierarchical wireless caching network.
Wang \textit{et al.}\cite{wang2018distributed} developed a joint traffic forwarding and cache placement scheme to optimize the queueing delay and energy consumption of caching-enabled networks.
In \cite{xu2018joint}, Xu \textit{et al.} proposed an online algorithm to jointly optimize wireless caching and task offloading with the goal of ultra-low task computation delays under a long-term energy constraint.
In general, such works adopted Lyapunov optimization method\cite{neely2010stochastic} to solve their formulated problems through a series of per-time-slot adaptive control.
Although the effectiveness of their solutions has been well justified, they generally assumed that file popularities or file requests are readily given prior to the cache placement procedure. Such assumptions are usually not the case in practice\cite{bharath2016learning}.

\textbf{Online Learning based Cache Placement:}
Faced with constantly arriving file requests and unknown file popularities, a number of works adopted various learning techniques such as deep learning\cite{pang2018toward, chen2019echo, ndikumana2020deep, liu2020distributed}, transfer learning\cite{bharath2016learning}\cite{bacstuug2015transfer}, and reinforcement learning\cite{blasco2014learning, blasco2014multi, muller2016smart, xu2020collaborative, song2017learning, zhang2019learning, sengupta2014learning, sadeghi2019reinforcement} to improve the performance of wireless caching networks. 
%Works \cite{blasco2014learning, sengupta2014learning, blasco2014multi, song2017learning} applied the multi-armed bandit (MAB) method, a classic reinforcement learning method that is usually used to address the exploration-exploitation dilemma, to learn the file popularities.
However, existing solutions in such works cannot handle time-averaged constraints. Besides, they mainly resorted to time-consuming offline pre-training and heuristic hyper-parameter tuning to produce their solutions. Moreover, they generally provided no theoretical guarantee but limited insights for the resulting performance.

Bandit learning is another method that is widely adopted to promote the performance of such systems.
So far, it has been applied to solve scheduling problems such as
task offloading\cite{zhu2018learn}, 
task allocation\cite{yao2019energy}, 
and path selection\cite{mukherjee2018resource}.
%, and cache placement\cite{bharath2016learning}\cite{lei2017deep, bacstuug2015transfer, blasco2014learning, blasco2014multi, sengupta2014learning, song2017learning, sadeghi2019reinforcement}.
The most relevant to our work are those which consider optimizing proactive cache placement in terms of different performance metrics.
For example, Blasco \textit{et al.}\cite{blasco2014learning}\cite{blasco2014multi} studied the cache placement problem for a single caching unit with multiple users. 
By considering the problem as a CMAB problem, in \cite{blasco2014learning} they aimed to maximize the amount of served traffic through wireless caching, while in \cite{blasco2014multi} they further took file downloading costs into the account for optimization.
%In \cite{blasco2014multi}, they extended their results in \cite{blasco2014learning} by consider the cost of downloading new files to the cache storage. 
%Following the assumption in \cite{blasco2014learning}, they modeled the file popularity using a Zipf distribution. 
In \cite{muller2016smart}, M{\"u}ller \textit{et al.} proposed a cache placement scheme based on contextual bandits, which learns the context-dependent content popularity to maximize the number of cache hits.
%In \cite{sengupta2014learning}, the authors considered the case when there are multiple caching units and reduced the cache placement problem to a linear program by adopting coded caching schemes. 
Zhang \textit{et al.}\cite{zhang2019learning} studied the network utility maximization problem in the context of cache placement with a non-fixed content library over time. 
Song \textit{et al.}\cite{song2017learning} proposed a joint cache placement and content sharing scheme among cooperative caching units to maximize the content caching revenue and minimize the content sharing expense.
In \cite{xu2020collaborative}, Xu \textit{et al.} modeled the procedure of cache placement with multiple caching units from the perspective of multi-agent multi-armed bandit (MAMAB) and devised an online scheme to minimize the accumulated transmission delay over time.
Such works generally do not consider the storage costs on EFSs in terms of memory footprint. In practice, without such a consideration, caching files with excessively high storage costs may offset the benefits of wireless caching.
Moreover, none of such works exploits offline historical information in their learning procedures.

\textbf{Novelty of Our Work:}
Different from existing works, to our best knowledge, our work presents the first systematic study on the synergy of online control, online learning, and offline historical information. 
%To address the cache placement problem in Fog-assisted IoT systems, our devised scheme proactively leverages both offline historical information and collected online feedback to carry out online cache update in an adaptive manner.  
In particular, we conduct theoretical analysis to characterize the joint impacts of online control, online learning, and offline information on the performances of cache placement. Our results also provide novel insights to the designers of Fog-assisted IoT systems.  
The comparison between our work and existing works is presented in Table \ref{table: related works}.

% System model
\section{System Model and Problem Formulation}\label{sec: model}
%\vspace{-0.2em}

In this section, we describe our system model in detail. Then we present our problem formulations. Key notations in this paper are summarized in Table \ref{table: key notations}.

\begin{table}[!t]
%%% increase table row spacing, adjust to taste
\renewcommand{\arraystretch}{1.3}
%% if using array.sty, it might be a good idea to tweak the value of
%% \extrarowheight as needed to properly center the text within the cells
\caption{Key notations}
\label{table: key notations}
\centering
\begin{tabular}{p{0.89cm} l}
    \hline\hline
    Notation & Description  \\ \hline
    $T$ & Length of time horizon \\ \hline
    $\mathcal{N}$ & Set of EFSs with $|\mathcal{N}|\triangleq N$ \\ \hline
    $\mathcal{K}$ & Set of IoT users with $|\mathcal{K}|\triangleq K$  \\ \hline
    $\mathcal{K}_{n}$ & Set of IoT users served by EFS $n$  \\ \hline    
    $\mathcal{F}$ & Set of files with $|\mathcal{F}|\triangleq F$ \\ \hline
    $L_{f}$ & Size of file $f$ \\ \hline
    $M_{n}$ & {Storage capacity} of EFS $n$ \\ \hline
    \multirow{2}*{$\theta_{k,f}(t)$} & Indicator of whether file $f$ is requested by IoT user $k$ in  \\ 
    & time slot $t$ \\ \hline
    \multirow{2}*{$D_{n,f}(t)$} & Total number of IoT users in set $\mathcal{K}_{n}$ who request for file \\ 
    & $f$ in time slot $t$ \\ \hline
    $d_{n,f}$ & Popularity of file $f$ on EFS $n$, $d_{n,f}\triangleq \mathbb{E}[D_{n,f}(t)]$ \\ \hline
    \multirow{2}*{$H_{n,f}$} & Number of offline historical observations with respect to \\
    & the popularity of file $f$ on EFS $n$ \\ \hline
    \multirow{2}*{$D^{h}_{n,f}(s)$} & Total number of IoT users in set $\mathcal{K}_{n}$ who request for file \\ 
    & $f$ according to the $s$-th offline historical observation \\ \hline
    $\tilde{d}_{n,f}(t)$ & Estimated popularity of file $f$ on EFS $n$ in time slot $t$ \\ \hline
    \multirow{2}*{$X_{n,f}(t)$} & Cache placement decision for caching file $f$ on EFS $n$ in \\ 
    & time slot $t$ \\ \hline
    $C_{n}(t)$ & Storage cost of EFS $n$ in time slot $t$ \\ \hline
    \multirow{2}*{$R_{n,f}(t)$} & Cache hit reward of EFS $n$ with respect to file $f$ in time \\ 
    & slot $t$ \\ \hline
    $R_{n}(t)$ & Total cache hit reward of EFS $n$ in time slot $t$ \\ \hline
    $b_{n}$ & Storage cost budget for EFS $n$ \\ \hline
	\hline
\end{tabular}
\end{table}

% Basic model
\subsection{Basic Model}\label{subsec: basic model}
%\vspace{-0.2em}
%Table \ref{table: key notations} summarizes the key notations in this paper.
We consider a caching-enabled Fog-assisted IoT system that operates over a finite time horizon of $T$ time slots. 
In the system, there is a central fog server (CFS) that manages $N$ edge fog servers (EFSs) to serve $K$ IoT users. 
The fog servers and IoT users communicate with each other through wireless connections. We assume that OFDM (orthogonal frequency division multiplexing)\cite{lopez2009ofdma} is employed as the underlying wireless transmission mechanism. Under such a mechanism, the co-channel interference among fog servers and IoT users can be eliminated by the orthogonal subcarrier allocation. 
Based on such an assumption, we abstract the physical-layer wireless links as bit pipes and focus on the network-layer data communications between servers and IoT users.
We denote the sets of EFSs and users by $\mathcal{N} \triangleq \{1,2,\cdots,N\}$ and $\mathcal{K} \triangleq \{1,2,\cdots,K\}$, respectively.
For each EFS $n$, we define $\mathcal{K}_{n}$ ($\mathcal{K}_{n} \subseteq \mathcal{K}$, $|\mathcal{K}_{n}|=K_{n}$) as the set of IoT users within its service range. Note that each IoT user is served by one and only one EFS and thus the sets $\{\mathcal{K}_{n}\}_{n}$ are disjoint. 

Particularly, we focus on the scenario in which IoT users request to download files from EFSs. We assume that the CFS has stored all of $F$ files (denoted by set $\mathcal{F}\triangleq\{1,2,\cdots,F\}$) that could be requested within the time horizon. Each file $f$ has a fixed size of $L_{f}$ storage units. 
Due to caching capacity limit, each EFS $n$ only has $M_{n}$ units of storage to cache a portion of the files and $M_{n} < \sum_{f\in\mathcal{F}} L_{f}$.  
Accordingly, if a user cannot find its requested file on its associated EFS, it will request to download the file directly from the CFS. 
We assume that the CFS can provide simultaneous and independent file deliveries to all EFSs and IoT users. 
An example which illustrates our system model is shown in Figure \ref{fig: model}.
%In this work, we consider the cache placement on each EFS while taking its cache hit reward and storage cost into account.

%\vspace{-0.2em}
% File popularity
\subsection{File Popularity}
%\vspace{-0.2em}

On each EFS $n$, we consider the popularity of each file $f$ as the expected number of users to request file $f$ per time slot, whose ground-truth value is denoted by $d_{n, f}$.
We assume that each file's popularity remains constant within the time horizon. In practice, such file popularities are usually unknown \textit{a priori} and can only be inferred based on online feedback information collected after user requests have been served. 

Next, we introduce some variables to characterize user dynamics with respect to file popularity. 
We define binary variable $\theta_{k,f}(t) \in \{0, 1\}$ such that $\theta_{k,f}(t)=1$ if IoT user $k$ requests for file $f$ in time slot $t$ and $\theta_{k,f}(t)=0$ otherwise.
Then we denote the file requests of IoT user $k$ during time slot $t$ by vector $\boldsymbol{\theta}_{k}(t) \triangleq (\theta_{k,1}(t),\theta_{k,2}(t),\cdots,\theta_{k,F}(t))$.
Meanwhile, we use $D_{n,f}(t) \triangleq \sum_{k\in\mathcal{K}_{n}}\theta_{k,f}(t)$ to denote the total number of IoT users in set $\mathcal{K}_{n}$ who request for file $f$ on EFS $n$ in time slot $t$.
Note that $D_{n,f}(t)$ is a discrete random variable over a support set $\{0,1,\cdots,K_{n}\}$ and assumed to be \textit{i.i.d.} across time slots with a mean of $d_{n,f}$.

Besides, we assume that initially (\textit{i.e.}, $t=0$), each EFS is provided with a fixed set of offline historical observations with respect to the number of requests for each file. Specifically, the offline historical observations for file $f$ on EFS $n$ are denoted by $\{D_{n,f}^{h}(0),D_{n,f}^{h}(1),\cdots,D_{n,f}^{h}(H_{n,f}-1)\}$, where we define $H_{n,f}\geq 0$ as the number of offline historical observations about file $f$ on EFS $n$. When $H_{n,f}=0$, there is no offline historical information. 
Let $D_{n,f}^{h}(s)$ denote the $s$-th offline historical observation. Here we use superscript $h$ to indicate that $D_{n,f}^{h}(s)$ belongs to offline historical information. 
Note that such observations are given as prior information when $t=0$. Their values are assumed to follow the same distribution as the file popularities over the time horizon.
%We further assume that the distribution of $D_{n,f}(t)$ and the mean $d_{n,f}$ are unknown. 

% Cache placement state
%\subsection{Cache State}
%We define $S_{n,f}(t)$ as the variable to indicate whether file $f$ has been cached by EFS $n$ at the beginning of  time slot $t$ ($S_{n,f}(t)=1$) or not ($S_{n,f}(t)=0$). For notational simplicity, we define $\boldsymbol{S}_{n}(t) \triangleq (S_{n,1}(t),S_{n,2}(t),\cdots,S_{n,F}(t))$. Note that $S_{n,f}(0)=0$ for each $n\in\mathcal{N}$ and $f\in\mathcal{F}$.

%\vspace{-0.2em}
% Workflow
\subsection{System Workflow}
%\vspace{-0.2em}
During each time slot $t$, the system operates across two phases: the caching phase and the service phase.
\begin{itemize}
\item [$\diamond$] \textit{Caching phase:} 
At the beginning of time slot $t$, each EFS $n$ updates its cached files and consumes a storage cost for each cached file. Then each EFS $n$ broadcasts its cache placement to all IoT users in set $\mathcal{K}_{n}$.
\item [$\diamond$] \textit{Service phase:} 
Each IoT user generates file requests. For each request, if it is not cached on the EFS, then the user will fetch the file from the CFS. Otherwise, the user directly downloads the file from the EFS and the EFS will receive a corresponding cache hit reward.
\end{itemize}

In the next few subsections, we present the definitions of cache placement decisions, storage costs, and cache hit rewards, respectively.

%\vspace{-0.2em}
% Cache placement decision
\subsection{Cache Placement Decision} \label{subsec: decision var}
%\vspace{-0.2em}

For each EFS $n$, we denote its cache placement decision made during each time slot $t$ by a binary vector $\boldsymbol{X}_{n}(t)\triangleq (X_{n,1}(t),X_{n,2}(t),\cdots,X_{n,F}(t))$.
Each entry $X_{n,f}(t)=1$ if EFS $n$ decides to cache file $f$ during time slot $t$ and $X_{n,f}(t)=0$ otherwise. Note that the total size of cached files on EFS $n$ does not exceed its storage capacity, \textit{i.e.},
\begin{equation}\label{constraint: storage}
 \sum_{f\in\mathcal{F}}L_{f}X_{n,f}\left(t\right)\leq M_{n},\ \forall n\in\mathcal{N},t.
\end{equation}

%\vspace{-0.2em}
% storage cost
\subsection{Storage Cost} \label{subsec: storage cost}
%\vspace{-0.2em}

For each EFS $n$, caching file $f$ during a time slot $t$ will incur a storage cost of $\alpha L_{f}$, where $\alpha>0$ is the unit storage cost. The storage cost can be viewed as the memory footprint for maintaining the file which is proportional to the size of file $f$.
Accordingly, given decision $\boldsymbol{X}_{n}(t)$, we define the total storage cost on EFS $n$ during time slot $t$ as
\begin{equation}\label{eq: EFS cost}
 C_{n}\left(t\right)\triangleq \sum_{f\in\mathcal{F}}\alpha L_{f}X_{n,f}\left(t\right).
\end{equation} 

%\vspace{-0.2em}
% Reward
\subsection{Cache Hit Reward} \label{subsec: reward}
%\vspace{-0.2em}

Recall that during each time slot $t$, for \textit{each} requested file $f$, if $X_{n, f}(t)=1$, 
then EFS $n$ will receive a reward $L_{f}$ for the corresponding cache hit \cite{blasco2014learning} (in terms of amounts of traffic to fetch file $f$ from EFS $n$).
Then given the cache placement $X_{n,f}(t)$ and user demand $D_{n, f}(t)$ during time slot $t$, we define the cache hit reward of EFS $n$  with respect to file $f$ as
\begin{equation}
\begin{split}
 R_{n,f}\left(t\right) \triangleq L_{f}D_{n,f}\left(t\right)X_{n,f}\left(t\right).
\end{split}
\end{equation}
Note that the cache hit reward $R_{n,f}(t)=0$ if file $f$ is not cached on EFS $n$ during time slot $t$ (\textit{i.e.}, $X_{n,f}(t)=0$).
Accordingly, we define the total cache hit reward of EFS $n$ during time slot $t$ as
%\vspace{-0.3em}
\begin{equation}\label{eq: EFS reward}
 R_{n}(t) = \hat{R}_{n}(\boldsymbol{X}_n(t))\triangleq\sum_{f\in\mathcal{F}}L_{f}D_{n,f}(t)X_{n,f}(t).
\end{equation}
%The total cache hit reward $R_{n}(t)$ is indeed the total traffic volume served by EFS $n$ during time slot $t$.

%\vspace{-0.2em}
% Problem formulation
\subsection{Problem Formulation}\label{sec: problem}
%\vspace{-0.2em}

To achieve effective cache placement with a high QoS, two goals are considered in our work. 
One is to maximize the total size of transmitted files from all EFSs so that requests from IoT users can obtain timely services. 
In our model, this is equivalent to maximizing the time-averaged cache hit reward of all EFSs over a time horizon of $T$ time slots.
The other is to guarantee a budgeted usage of storage costs over time. 
To this end, for each EFS $n$, we first define $b_{n}$ as the storage cost budget for caching files.
Then we impose the following constraint to ensure the time-averaged storage costs under the budget in the long run:
\begin{equation}\label{constraint: cost}
 \limsup_{t\rightarrow\infty}\frac{1}{t}\sum_{\tau=0}^{t-1}\mathbb{E}\left[C_{n}\left(\tau\right)\right]\leq b_{n},\ \forall n\in\mathcal{N}.
\end{equation}
Based on the above system model and constraints, our problem formulation is given by
\begin{subequations}\label{p: goal}
\begin{align}
 \underset{\{\boldsymbol{X}(t)\}_{t}}{\text{maximize}}~~
 &\displaystyle \frac{1}{T}\sum_{t=0}^{T-1}\sum_{n\in\mathcal{N}}\mathbb{E}\left[R_{n}\left(t\right)\right] \label{objective: maximize reward} \\
 \text{subject to}~~
 &\displaystyle X_{n,f}(t)\in\{0,1\}, \forall n\in\mathcal{N}, f\in\mathcal{F}, t,\label{constraint: 0 or 1}\\
 &\displaystyle (\ref{constraint: storage}),(\ref{constraint: cost}).\nonumber
\end{align}
\end{subequations}
In the above formulation, the objective (\ref{objective: maximize reward}) is to maximize the time-averaged expectation of total cache hit reward of all EFSs. Constraint (\ref{constraint: 0 or 1}) states that each cache placement decision $X_{n,f}(t)$ should be a binary variable. 
The constraint in (\ref{constraint: storage}) guarantees that the total size of cached files on each EFS should not exceed the storage capacity. The constraint in (\ref{constraint: cost}) ensures the budget constraint on the storage cost of each EFS.

% Algorithm design
%\vspace{-0.8em}
\section{Algorithm Design}\label{sec: algorithm}
%\vspace{-0.2em}

\begin{figure}[!t]
	\centering
	\includegraphics[scale=0.34]{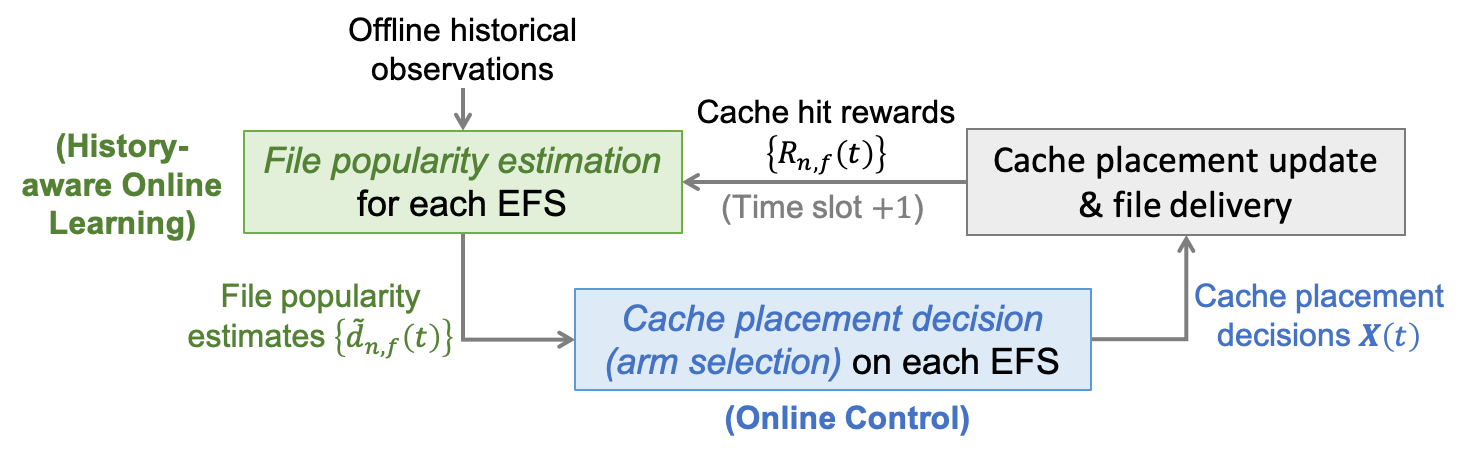}
	\caption{An illustration of our algorithm design.}
	\label{fig: algorithm}
\end{figure}

For problem (\ref{p: goal}), given the full knowledge of user demands $\{D_{n,f}(t)\}_{n,f}$, it can be solved asymptotically optimally by Lyapunov optimization methods \cite{neely2010stochastic}. 
However, file popularities are usually not given as prior information in practice.
Faced with such uncertainties, online learning needs to be incorporated to guide the decision-making process by estimating the statistics of file popularities 
from both online feedback and offline historical information.
To this end, we need to deal with the well-known \textit{exploration-exploitation} dilemma, \textit{i.e.}, how to balance the decisions made to acquire new knowledge about file popularity to improve learning accuracy (\textit{exploration}) and the decisions made to leverage current knowledge to select the empirically most popular files (\textit{exploitation}).
For such a decision-making problem under uncertainty, we consider it through the lens of combinatorial multi-armed bandit (CMAB) with extended settings.
With an effective integration of online bandit learning, online control, and offline historical information,
%Following the idea of integrating bandit learning with virtual queue techniques in \cite{li2019combinatorial} and the idea of exploiting offline historical information in the learning procedure in \cite{shivaswamy2012multi}, 
we devise a history-aware learning-aided cache placement scheme called CPHBL (Cache Placement with History-aware Bandit Learning) to solve problem (\ref{p: goal}). 
Figure \ref{fig: algorithm} depicts the design of CPHBL. 
During each time slot, under CPHBL, each EFS first estimates the popularity of different files based on both offline historical information and collected online feedback. 
Based on such estimates, the EFS determines and updates its cache placement in the current time slot. After the update, each EFS delivers requested cached files to IoT users. For each cache hit, a reward will be credited to the EFS. 

In the following subsections, we extend the settings of the existing CMAB model and demonstrate the reformulation of problem (\ref{p: goal}) under such settings. Then we articulate our algorithm design with respect to online learning and online control procedures, respectively. Finally, we discuss the computational complexity of our devised algorithm. 

%\vspace{-0.3em}
\subsection{Problem Reformulation}
%\vspace{-0.3em}

The basic settings of CMAB \cite{chen2013combinatorial} consider a sequential interaction between a player and its environment with multiple actions (\textit{a.k.a.} arms) over a finite number of rounds.
During each round, the player selects a subset of available arms to play.
For each selected arm, the agent will receive a reward that is sampled from an unknown distribution.
The overall goal of the player is to find an effective arm-selection scheme to maximize its expected cumulative reward.

Based on the CMAB model, Li \textit{et al.}\cite{li2019combinatorial} extended the settings of classical CMAB by allowing the temporary unavailability of arms while considering the fairness of arm selection.
Inspired by their work, we reformulate problem (\ref{p: goal}) as a constrained CMAB problem in the following way.
We view each EFS as a distinct player and each file as an arm. 
During each time slot $t$, each player $n\in\mathcal{N}$ selects a subset of arms to play.
If player $n$ chooses to play arm $f\in\mathcal{F}$ in time slot $t$, then file $f$ will be cached on EFS $n$ and a reward $R_{n,f}(t)=L_{f}D_{n,f}(t)$ will be received by the player.
Recall that the file demand $D_{n,f}(t)$ during each time slot $t$ is a random variable with an unknown mean $d_{n,f}$ and is \textit{i.i.d.} across time slots. 
Accordingly, reward $R_{n,f}(t)$ is also an \textit{i.i.d.} random variable with an unknown mean $r_{n,f}=\mathbb{E}[R_{n,f}(t)]=L_{f}d_{n,f}$.
Meanwhile, the cache placement decision $\boldsymbol{X}_{n}(t)=(X_{n,1}(t),X_{n,2}(t),\cdots,X_{n,F}(t))$ of EFS $n$ corresponds to the arm selection of player $n$ in time slot $t$. Specifically, $X_{n,f}(t)=1$ if arm $f$ is chosen and $X_{n,f}(t)=0$ otherwise.
Our goal is to devise an arm selection scheme for the players to maximize their expected cumulative rewards subject to the constraints in (\ref{constraint: storage}) and (\ref{constraint: cost}).

\textit{Remark:} Our model extends the settings of the bandit model proposed by \cite{li2019combinatorial} in the following four aspects.
First, we consider multiple players instead of one player.
Second, the storage cost constraints in our problem are more challenging to handle than the arm fairness constraints in \cite{li2019combinatorial}. Specifically, under our settings, the selection of each arm for a player is coupled together under storage cost constraints, whereas in \cite{li2019combinatorial} there is no such coupling among arm selections.
Third, we consider the storage capacity constraint for each player during each time slot, which is ignored in \cite{li2019combinatorial}.
Last but not least, we consider a more general reward function with respect to file uncertainties. 
The above extensions make our reformulated problem more challenging than the problem in \cite{li2019combinatorial}.
 
To characterize the performance loss (\textit{a.k.a.} regret) due to decision making under such uncertainties, we define the regret with respect to a given scheme (denoted by decision sequence $\{\boldsymbol{X}(t)\}_{t}$) as
\begin{equation}\label{def: regret}
 Reg\left(T\right)\triangleq R^{*}-\frac{1}{T}\sum_{t=0}^{T-1}\sum_{n\in\mathcal{N}}\mathbb{E}\left[\hat{R}_{n}\left(\boldsymbol{X}_{n}\left(t\right)\right)\right],
\end{equation}
where constant $R^{*}$ is defined as the optimal time-averaged total expected reward for all players.
In fact, maximizing the time-averaged expected reward is equivalent to minimizing the regret. 
Therefore, we can rewrite problem (\ref{p: goal}) as follows:
\begin{subequations}\label{p: minimize regret}
\begin{align}
 \underset{\{\boldsymbol{X}(t)\}_{t}}{\text{minimize}}~~
 &\displaystyle Reg\left(T\right)\\
 \text{subject to}~~
 &(\ref{constraint: storage})(\ref{constraint: cost})(\ref{constraint: 0 or 1}).
\end{align}
\end{subequations}
 
To solve problem (\ref{p: minimize regret}), we integrate history-aware bandit learning methods and virtual queue techniques to handle the exploration-exploitation tradeoff and the time-averaged storage cost constraints, respectively. In the following subsections, we demonstrate our algorithm design in detail.

%\vspace{-0.4em}
\subsection{Online Bandit Learning with Offline Historical Information} \label{sec: online learning}
%\vspace{-0.4em}

%We employ the classical UCB1\cite{auer2002finite} algorithm to learn the unknown file popularities and address the exploration-exploitation tradeoff.
By (\ref{eq: EFS reward}), the regret defined in (\ref{def: regret}) can be rewritten as
\begin{align}\label{eq: regret reform}
 Reg\left(T\right) \nonumber = &R^{*}-\frac{1}{T}\sum_{t=0}^{T-1}\sum_{n\in\mathcal{N}}\sum_{f\in\mathcal{F}}L_{f}\mathbb{E}\left[D_{n,f}\left(t\right)X_{n,f}\left(t\right)\right] \nonumber \\
 =&R^{*}-\frac{1}{T}\sum_{t=0}^{T-1}\sum_{n\in\mathcal{N}}\sum_{f\in\mathcal{F}}L_{f}d_{n,f}\mathbb{E}\left[X_{n,f}\left(t\right)\right], 
\end{align}
where the last equality holds due to the independence between user demand $D_{n,f}(t)$ and cache placement $X_{n,f}(t)$, and the fact that $\mathbb{E}[D_{n,f}(t)]=d_{n,f}$.
By (\ref{eq: regret reform}) and our previous analysis, to solve problem (\ref{p: minimize regret}), each EFS $n$ should learn the unknown file popularity $d_{n,f}$ with respect to each file $f$.

During each time slot $t$, after updating cached files according to decision $\boldsymbol{X}_{n}(t)$, each EFS $n$ observes the current demand $D_{n,f}(t)$ for each cached file $f$. 
Then EFS $n$ transmits requested files to IoT users and acquires cache hit rewards. 
Based on the pre-given offline historical information and cache hit feedback from IoT users, we have the following estimate for each file popularity $d_{n,f}$:
\begin{equation}\label{eq: ucb estimate}
 \tilde{d}_{n,f}(t)\!=\!\min\!\left\{\! \bar{d}_{n,f}(t)\!+\!\!K_{n}\sqrt{\frac{3\log t}{2(h_{n,f}(t)\!+\!H_{n,f})}},\ K_{n}\!\right\}.
\end{equation}
In (\ref{eq: ucb estimate}), $\bar{d}_{n,f}(t)$ is the empirical mean of the number of requests for file $f$ that involves both offline historical observations and collected online feedbacks; 
$h_{n,f}(t)$ counts the number of time slots (within the first $t$ time slots) during which file $f$ is chosen to be cached on EFS $n$; and $K_{n}$ denotes the number of users served by EFS $n$.
Specifically, the number of observations $h_{n,f}(t)$ and the empirical mean of file popularity $\bar{d}_{n,f}(t)$ by time slot $t$ are defined as follows, respectively:
\begin{align}
&h_{n,f}(t)\! \triangleq \!\sum_{\tau=0}^{t-1}\!X_{n,f}(\tau),\\
&\bar{d}_{n,f}(t) \!\triangleq\! \frac{\sum_{\tau=0}^{t-1}\!D_{n,f}(\tau)X_{n,f}(\tau)\!+\!\sum_{s=0}^{H_{n,f}-1}\!D_{n,f}^{h}(s)}{h_{n,f}(t)\!+\!H_{n,f}}.
\end{align}

\textit{Remark:} 
In (\ref{eq: ucb estimate}), the term $K_{n}\sqrt{\frac{3\log t}{2(h_{n,f}(t)+H_{n,f})}}$ denotes the \textit{confidence radius}\cite{slivkins2019introduction} that represents the degree of uncertainty with respect to the empirical estimate $\bar{d}_{n,f}(t)$. 
The larger the confidence radius, the greater the value of the estimate (\ref{eq: ucb estimate}) and thus the greater the chance for file $f$ to be cached on EFS $n$.
In the confidence radius, the term $h_{n,f}(t)+H_{n,f}$ is the total number of observations (including both  online observations and offline historical observations) for the popularity of file $f$ on EFS $n$. Given a small number of observations (\textit{i.e.}, $h_{n,f}(t)+H_{n,f}\ll t$), the confidence radius for the empirical estimate $\bar{d}_{n,f}(t)$ will be large, which implies that the file is rarely cached and hence a great uncertainty about the estimate.
In this case, the confidence radius plays a dominant role in the estimate $\tilde{d}_{n,f}(t)$. As a result, file $f$ will be more likely to be cached on EFS $n$. 
In contrast, if a file has been cached for an adequate number of times, its popularity estimate (\ref{eq: ucb estimate}) will be close to its empirical mean and the role of confidence radius will be marginalized.
Besides, the estimate (\ref{eq: ucb estimate}) also characterizes the effects of offline historical information and online feedback information. 
Particularly, in the early stage (when $t$ is small), suppose that the number of online observations is much smaller than the number of offline historical observations, \textit{i.e.}, $h_{n,f}(t)\ll H_{n,f}$. In this case, the estimate (\ref{eq: ucb estimate}) mainly depends on offline historical information. 
However, as more and more online feedbacks are collected, the impact of online information becomes more dominant.

%\vspace{-0.2em}
\subsection{Storage Cost Budgets with Virtual Queue Technique}\label{sec: virtual queue}
%\vspace{-0.2em}

\begin{figure}[!t]
	\centering
	\includegraphics[scale=0.4]{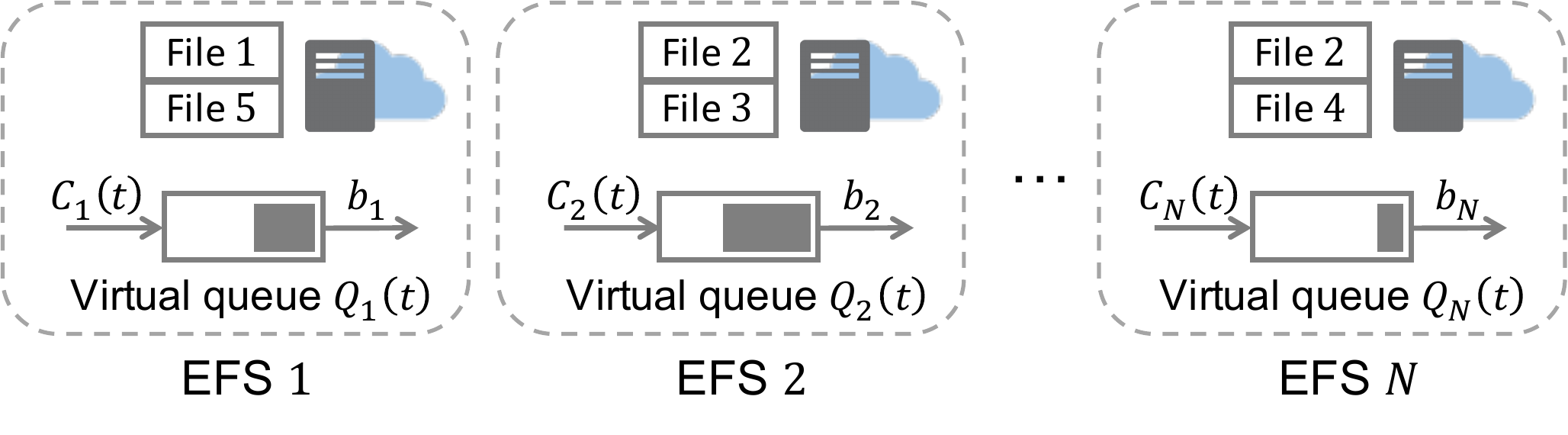}
	\caption{An illustration of virtual queues for storage cost on each EFS. Each EFS $n\in\mathcal{N}$ maintains a virtual queue $Q_{n}(t)$ with an input of $C_{n}(t)$ and an output of $b_{n}$ during each time slot $t$. If the queueing process $\{Q_{n}(t)\}_{t}$ is strongly stable, then the time-averaged storage cost constraint (\ref{constraint: cost}) on EFS $n$ can be satisfied.}
	\label{fig: virtual queue}
\end{figure} 

By leveraging Lyapunov optimization techniques\cite{neely2010stochastic}, we transform the time-averaged storage cost constraints into queue stability constraints.
Specifically, we introduce a virtual queue $Q_{n}(t)$ for each EFS $n\in\mathcal{N}$ with $Q_{n}(0)=0$ to handle the time-averaged constraints (\ref{constraint: cost}) on storage costs.
As illustrated in Figure \ref{fig: virtual queue}, each virtual queue $Q_{n}(t)$ is updated during each time slot $t$ as follows:
\begin{equation}\label{eq: queue update}
 Q_{n}\left(t+1\right)=\left[Q_{n}\left(t\right)-b_{n}\right]^{+}+C_{n}\left(t\right),
\end{equation}
in which we define $[\cdot]^{+}\triangleq \max\{\cdot, 0\}$.
Note that the constraints in (\ref{constraint: cost}) are satisfied only when the queueing process $\{Q_{n}(t)\}_{t}$ for each EFS $n$ is strongly stable \cite{neely2010stochastic}.
%Formally, we write
%\begin{equation}\label{eq: queue stability}
%	 \limsup_{t\rightarrow\infty}\frac{1}{t}\sum_{\tau=0}^{t-1}\mathbb{E}\left[Q_{n}\left(\tau\right)\right]< \infty.
%\end{equation}
Intuitively, the mean queue inputs (\textit{i.e.}, storage costs) should not be greater than the mean queue outputs (\textit{i.e.}, cost budgets). Otherwise, virtual queues will be overloaded, thereby violating the constraints in (\ref{constraint: cost}).
To maintain the stability of virtual queues and minimize the regret, we transform problem (\ref{p: minimize regret}) into a series of per-time-slot subproblems. We show the detailed derivation in Appendix \ref{appendix: alg develop}.
Specifically, during each time slot $t$, we aim to solve the following problem for each EFS $n\in\mathcal{N}$:
\begin{subequations}\label{p: one-slot subproblem}
\begin{align}
 \underset{\boldsymbol{X}_{n}(t)}{\text{maximize}}~~
 &\displaystyle \sum_{f\in\mathcal{F}}\tilde{w}_{n,f}\left(t\right)X_{n,f}\left(t\right)\\
 \text{subject to}~~
 &\displaystyle \sum_{f\in\mathcal{F}}L_{f}X_{n,f}\left(t\right)\leq M_{n}, \label{constraint: storage capacity} \\
 &\displaystyle X_{n,f}(t)\in\{0,1\}, \forall f\in\mathcal{F},
\end{align}
\end{subequations}
where $\tilde{w}_{n,f}(t)$ is defined as 
\begin{equation}\label{def: value}
 \tilde{w}_{n,f}(t) \triangleq L_{f}(V\tilde{d}_{n,f}(t)-\alpha Q_{n}(t) ).
\end{equation}
In (\ref{def: value}), parameter $V$ is a tunable positive constant; weight $\tilde{w}_{n,f}(t)$ can be viewed as the gain of caching file $f$ on EFS $n$ during time slot $t$;
and the objective of problem (\ref{p: one-slot subproblem}) is to maximize the total gain of caching files on EFS $n$ under the storage capacity constraint in (\ref{constraint: storage capacity}).

During each time slot $t$, we solve problem (\ref{p: one-slot subproblem}) for each EFS $n$ to determine its cache placement $\boldsymbol{X}_{n}(t)$. 
We split set $\mathcal{F}$ into two disjoint sets $\mathcal{F}_{n,1}(t)=\{f\in\mathcal{F}:\tilde{w}_{n,f}(t)\geq 0\}$ and $\mathcal{F}_{n,2}(t)=\{f\in\mathcal{F}:\tilde{w}_{n,f}(t)< 0\}$ for each EFS $n$.
Specifically, for each file $f \in \mathcal{F}$,
\begin{enumerate}
 \item if $\tilde{d}_{n,f}(t)\!\geq \!\alpha Q_{n}(t)/V$, then $\tilde{w}_{n,f}(t)\!\geq\! 0$ and $f\in\mathcal{F}_{n,1}(t)$;
 \item if $\tilde{d}_{n,f}(t)\!<\! \alpha Q_{n}(t)/V$, then $\tilde{w}_{n,f}(t)\!<\! 0$ and $f\in\mathcal{F}_{n,2}(t)$.
\end{enumerate}
For each file $f\in\mathcal{F}_{n,2}(t)$, the corresponding optimal placement decision is $X_{n,f}(t)=0$ since caching file $f$ on EFS $n$ will incur a negative gain, \textit{i.e.}, $\tilde{w}_{n,f}(t)< 0$.
By setting $X_{n,f}(t)=0$ for each file $f\in\mathcal{F}_{n,2}(t)$, we can regard problem (\ref{p: one-slot subproblem}) as a classical Knapsack problem\cite{martello1999dynamic}
\begin{equation}\label{p: knapsack problem}
\begin{array}{cl}
 \underset{\left\{ X_{n,f}(t)\right\}_{f\in\mathcal{F}_{n,1}(t)}}{\text{maximize}}
 &\displaystyle \sum_{f\in\mathcal{F}_{n,1}(t)}\tilde{w}_{n,f}\left(t\right)X_{n,f}\left(t\right)\\
 \text{subject to}
 &\displaystyle \sum_{f\in\mathcal{F}_{n,1}(t)}L_{f}X_{n,f}\left(t\right)\leq M_{n},\\
 &\displaystyle X_{n,f}(t)\in\{0,1\}, \forall f\in\mathcal{F}_{n,1}(t).
\end{array}
\end{equation}

Intuitively, from the lens of Knapsack problem, we have a number of items (files) in set $\mathcal{F}_{n,1}(t)$ and a knapsack (EFS $n$'s cache) with a capacity of $M_{n}$. 
The weight of each item $f\in\mathcal{F}_{n,1}(t)$ is $L_{f}$, while the value of putting item $f$ in the knapsack is $\tilde{w}_{n,f}(t)$. 
Given the weights and values of all items, our goal is to select and put a subset of the items from $\mathcal{F}_{n,1}(t)$ into the knapsack with the maximum total value. 
Such a problem can be solved optimally by applying dynamic programming (DP) algorithm \cite{martello2000new}. 

\subsection{Integrated Algorithm Design}
Based on the design presented in the previous two subsections, we propose a novel learning-aided proactive cache placement scheme called CPHBL (Cache Placement with History-aware Bandit Learning). The pseudocode of CPHBL is presented in Algorithm \ref{alg: learning-aided CP}.
In particular, we denote the file indices in set $\mathcal{F}_{n,1}(t)$ by $\phi_{n,1}(t),\phi_{n,2}(t),\cdots,\phi_{n,|\mathcal{F}_{n,1}(t)|}(t)$, respectively. 
We use $v(i,m)$ to denote the optimal value of problem (\ref{p: knapsack problem}) when only the first $i$ files (\textit{i.e.}, files indexed by $\phi_{n,1}(t), \cdots, \phi_{n,i}(t)$) in $\mathcal{F}_{n,1}(t)$ can be selected to store in the remaining memory capacity of $m$ storage units.
Regarding CPHBL, we have the following remarks.

\textit{Remark 1:}
In (\ref{def: value}), the value of parameter $V$ in weight $\tilde{w}_{n,f}(t)$ measures the relative importance of achieving high cache hit rewards to ensuring storage cost constraints.
Note that the value of $\tilde{w}_{n,f}(t)$ is positively proportional to the value of parameter $V$.
Therefore, for each file $f\in\mathcal{F}$, the gain $\tilde{w}_{n,f}(t)$ of caching file $f$ on EFS $n$ during time slot $t$ will increase as the value of $V$ increases. 
Under CPHBL, EFS $n$ will cache more files to achieve not only a higher gain but also a larger storage cost. 
Moreover, files with high estimated mean cache hit rewards would be the first to be cached.

\textit{Remark 2:}
To ensure the storage cost constraints in (\ref{constraint: cost}), 
CPHBL would restrict each EFS to cache limited files as its virtual queue backlog size becomes large.
Intuitively, for each EFS $n$, if its time-averaged storage cost tends to exceed the cost budget $b_{n}$, its corresponding virtual queue backlog size $Q_{n}(t)$ will be large. 
By the definition of weight $\tilde{w}_{n,f}(t)$ in (\ref{def: value}), the value of $\tilde{w}_{n,f}(t)$ is negatively proportional to the virtual queue backlog size $Q_{n}(t)$.
Therefore, when the value of $Q_{n}(t)$ increases, the weight $\tilde{w}_{n,f}(t)$ of caching file $f$ on EFS $n$ tends to be negative. 
Under CPHBL, files with negative weights will not be cached, which conduces to a low time-averaged storage cost.

%\vspace{-0.2em}
\subsection{Computational Complexity of CPHBL}
%\vspace{-0.2em}

The computational complexity of CPHBL mainly lies in the decision making for cache placement on each EFS $n\in\mathcal{N}$ (line \ref{line: set cache placement} in Algorithm \ref{alg: learning-aided CP}). In this process, DP is adopted to solve problem (\ref{p: one-slot subproblem}) with a computational complexity of {$O(FM_{n})$\cite{martello2000new}.} Note that $F$ denotes the total number of files on the CFS and $M_{n}$ denotes the storage capacity of EFS $n$. In practice, the cache placement procedure can be implemented in a distributed fashion over EFNs; accordingly, the total computational complexity of CPHBL is $O(F\max_{n\in\mathcal{N}}M_{n})$.

\begin{algorithm}[!t]
\caption{Cache Placement with History-aware Bandit Learning (CPHBL)}
\label{alg: learning-aided CP}
{\small{
\begin{algorithmic}[1]
  \State Initialize $h_{n,f}(0)=0$, $\bar{d}_{n,f}(0)=\frac{1}{H_{n,f}}\sum_{s=0}^{H_{n,f}-1}D_{n,f}^{h}(s)$ and $\tilde{d}_{n,f}(0)=K_{n}$ for each EFS $n\in\mathcal{N}$ and each file $f\in\mathcal{F}$. In each time slot $t\in\{0,1,\cdots\}$:
  \Statex \textit{\%History-aware Online Learning}
  \For {each EFS $n\in\mathcal{N}$ and each file $f\in\mathcal{F}$}
      \If {$h_{n,f}(t)+H_{n,f}>0$ and $t>0$}
        \State $\tilde{d}_{n,f}(t)\!\leftarrow\!\min\big\{\bar{d}_{n,f}(t)+K_{n}\sqrt{\frac{3\log t}{2(h_{n,f}(t)+H_{n,f})}},K_{n}\big\}$.\label{algline: ucb1}
      \EndIf
  \EndFor
  \Statex \textit{\%Cache Placement}
  \For {each EFS $n\in\mathcal{N}$}\label{line: cache placement start}
    \State {\sc SetCachePlacement}($t$, $n$, $\{\tilde{d}_{n,f}(t)\}_{f}$). \label{line: set cache placement}\label{algline: cache placement}
  \EndFor \label{line: cache placement end}
  
  \Statex \textit{\%Update of Statistics and Virtual Queues}
  \State Update cached files according to $\boldsymbol{X}(t)$ and virtual queues $\boldsymbol{Q}(t)$ according to (\ref{eq: queue update}).

  \For {each EFS $n\in\mathcal{N}$ and each file $f\in\mathcal{F}$}
      \State $h_{n,f}(t+1)\leftarrow h_{n,f}\left(t\right)+X_{n,f}\left(t\right)$.
      \State $\bar{d}_{n,f}(t+1)\leftarrow\frac{h_{n,f}(t)+H_{n,f}}{h_{n,f}(t+1)+H_{n,f}}\bar{d}_{n,f}(t)+\frac{D_{n,f}(t)X_{n,f}(t)}{h_{n,f}(t+1)+H_{n,f}}$.
  \EndFor
\end{algorithmic}
}}
\end{algorithm} 

\begin{algorithm}[!h]
{\small{
\begin{algorithmic}[1]
\Function{\sc SetCachePlacement}{$t$, $n$, $\{\tilde{d}_{n,f}(t)\}_{f}$}
\label{func: set}
\parState{\textbf{Inputs:} At the beginning of time slot $t$, for EFS $n$, given file demand estimate $\{\tilde{d}_{n,f}(t)\}_{f}$. }
  \State Set $\mathcal{F}_{n,1}(t)\leftarrow\emptyset$.
    \For {each file $f\in\mathcal{F}$}
      \State Set $\tilde{w}_{n,f}(t)\!\leftarrow\! L_{f}\big(V\tilde{d}_{n,f}(t)-Q_{n}(t) \big)$.
      \If {$\tilde{w}_{n,f}(t)< 0$}\label{line: neg_weight begin}
        \State Set $X_{n,f}(t)\leftarrow 0$. \label{line: neg_weight end}
      \Else
        \State Set $\mathcal{F}_{n,1}(t)\leftarrow \mathcal{F}_{n,1}(t)\cup \{f\}$.
      \EndIf
    \EndFor
    
    \parState{Initialize $v_{n}(i,m)=0$ for $i\in\{0,1,2,\cdots,|\mathcal{F}_{n,1}(t)|\}$ and $m\in\{0,1,\cdots,M_{n}\}$.} \label{line: dp start}
    \For {each $i\in\{1,2,\cdots,|\mathcal{F}_{n,1}(t)|\}$}
      \For {each $m\in\{1,\cdots,M_{n}\}$}
        \If {$L_{\phi_{n,i}(t)}>m$}
          \State Set $v_{n}(i, m)\leftarrow v_{n}(i-1,m)$.
        \Else
          \parState{Set $v_{n}(i,m)\leftarrow \max\big\{v_{n}(i-1,m), v_{n}(i-1,m-L_{\phi_{n,i}(t)})+\tilde{w}_{n,\phi_{n,i}(t)}(t)\big\}$.}
        \EndIf
      \EndFor 
    \EndFor
  
  \State {\sc SetOptPlacement}($n$, $|\mathcal{F}_{n,1}(t)|$, $M_{n}$).\label{line: dp end}
\EndFunction
\end{algorithmic}
}}
\end{algorithm}

\begin{algorithm}[!h]
{\small{
\begin{algorithmic}[1]
\Function{\sc SetOptPlacement}{$n$, $i$, $m$}
\label{func: find}
\parState{\textbf{Inputs:} For EFS $n$, given the number of files $i$ and the remaining storage size $m$. }
\If {$i\geq 1$}
  \If {$v_{n}(i,m)=v_{n}(i-1,m-L_{\phi_{n,i}(t)})+\tilde{w}_{n,\phi_{n,i}(t)}(t)$ and $~~~~~~~~~~m-L_{\phi_{n,i}(t)}\geq 0$}
      \State Set $X_{n,\phi_{n,i}(t)}(t)\leftarrow 1$.
      \State {\sc SetOptPlacement}($n$, $i-1$, $m-L_{\phi_{n,i}(t)}$).
  \ElsIf {$v_{n}(i,m)=v_{n}(i-1,m)$}
      \State Set $X_{n,\phi_{n,i}(t)}(t)\leftarrow 0$.
      \State {\sc SetOptPlacement}($n$, $i-1$, $m$).  
  \EndIf
\EndIf
\EndFunction
\end{algorithmic}
}}
\end{algorithm}

% Theoretical analysis
%\vspace{-0.2em}
\section{Performance Analysis}\label{sec: analysis}
%\vspace{-0.2em}

For each EFS $n$, given the number $K_{n}$ of its served users and its storage capacity $M_{n}$, 
as well as the size $L_{f}$ of each file $f\in\mathcal{F}$, we establish the following two theorems to characterize the performance of CPHBL.

%\vspace{-0.2em}
\subsection{Storage Cost Constraints}
%\vspace{-0.2em}

A budget vector $\boldsymbol{b}=(b_{1},b_{2},\dots,b_{N})$ of storage costs is said to be \textit{feasible} if there exists a feasible cache placement scheme under which all storage cost constraints in (\ref{constraint: cost}) can be satisfied. 
We define the set of all feasible budget vectors as the \textit{maximal feasibility region} of the system, which is denoted by the set $\mathcal{B}$.
The following theorem shows that all virtual queues are strongly stable under CPHBL when $\boldsymbol{b}$ is an interior point of $\mathcal{B}$.

%\vspace{-1em}

\begin{theorem}\label{theorem: feasibility}
	Suppose that the budget vector $\boldsymbol{b}$ lies in the interior of $\mathcal{B}$, then the time-averaged storage cost constraints in (\ref{constraint: cost}) are satisfied under CPHBL. Moreover, the virtual queues defined in (\ref{eq: queue update}) are strongly stable and there exists some constant $\epsilon >0$ such that 
	\begin{equation}\label{ineq: queue upper bound}
		\limsup_{t\rightarrow\infty}\frac{1}{t}\!\sum_{\tau=0}^{t-1}\sum_{n\in\mathcal{N}}\mathbb{E}[Q_{n}(\tau)]\!\leq\!\frac{B\!+\!V\!\sum_{n\in\mathcal{N}}2K_{n}M_{n}}{\epsilon},
	\end{equation}
	where $B \triangleq \sum_{n\in\mathcal{N}}(b_{n}^{2}+\alpha^{2}M_{n}^{2})/2$.
\end{theorem}

The proof of Theorem \ref{theorem: feasibility} is given in Appendix \ref{proof: feasibility}.

\textit{Remark 1:}
Theorem \ref{theorem: feasibility} shows that CPHBL ensures the stability of virtual queue backlogs $\{ Q_{n}(t) \}_{n}$. Moreover, the time-averaged total backlog size of such virtual queues is linearly proportional to the value of parameter $V$.
In other words, given that vector $\boldsymbol{b}$ is interior to the maximal feasibility region, 
under CPHBL, the time-averaged total storage cost is tunable and guaranteed to be under the given budget. 

%\textit{Proof Sketch:} We adopt the standard Lyapunov drift analysis in \cite{neely2010stochastic} to conduct the proof of Theorem \ref{theorem: feasibility}. Here is the sketch of our proof: First, we design a quadratic Lyapunov function of virtual queue backlogs. Then we compute the change in the Lyapunov function from one time slot to the next, which is called the Lyapunov drift. Next, by adding the per-time-slot regret to Lyapunov drift, we obtain the drift-plus-regret. 
%%By maintaining the dominant cross terms and bounding the rest terms of drift-plus-regret, we develop an upper bound for it. 
%With the aid of an auxiliary scheme, we bound the drift-plus-regret with a linear function of the sizes of virtual queue backlogs. 
%Finally, we apply iterated expectations and telescoping sums to complete the proof.

%\vspace{-0.2em}
\subsection{Regret Bound}
%\vspace{-0.2em}

Our second theorem provides an upper bound for the regret incurred by CPHBL over time.
%The proof of the second theorem is shown in Appendix \ref{proof: regret bound}.
\begin{theorem}\label{theorem: regret bound}
	Under CPHBL, the regret (\ref{def: regret}) over time horizon $T$ is upper bounded as follows:
	\begin{equation}\label{ineq: regret upper bound}
	Reg\left(T\right)\leq \frac{B}{V}+\frac{4\sum_{n\in\mathcal{N}}K_{n}M_{n}}{T}
	+\varGamma \sqrt{\frac{\log T}{T+H_{\min}}},
\end{equation}
	in which we define the constants $B\triangleq \sum_{n\in\mathcal{N}}(b_{n}^{2}+\alpha^{2}M_{n}^{2})/2$ and $\varGamma\triangleq 2\sum_{n\in\mathcal{N}}K_{n}\sqrt{6M_{n}\sum_{f\in\mathcal{F}}L_{f}}$. Here $T$ is the time horizon length and $H_{\min}\triangleq \min_{n,f}H_{n,f}$ is the the minimal number of offline historical observation among all EFSs and files.
\end{theorem}

The proof of Theorem \ref{theorem: regret bound} is given in Appendix \ref{proof: regret bound}.

\textit{Remark 2-1:}
In (\ref{ineq: regret upper bound}), the term $B/V$ is mainly incurred by balancing the cache hit reward and the storage cost constraints. 
	Intuitively, the larger the value of $V$, the more focus CPHBL puts on maximizing cache hit rewards, and hence a smaller regret.	
	Nonetheless, this also comes with an increase in the total size of virtual queue backlogs, which is unfavorable for keeping storage costs under the budget.
	In contrast, the smaller the value of $V$, the more sensitive CPHBL would be to the increase in the storage costs. As a result, each EFS would constantly update its cached file set with files of different storage costs, leading to inferior cache hit rewards.  
	In practice, the selection of the value of $V$ depends on the design tradeoff of real systems.

\textit{Remark 2-2:}
	The last two terms of the regret bound are in the order of $O(1/T+\sqrt{(\log T)/(T+H_{\min})})$. These two terms are mainly incurred by the online learning procedure with offline historical information and collected online feedback.  
	In the following, we first consider the impact of $H_{\min}$ on the regret bound under a fixed value of time horizon length $T$.
	Note that when $H_{\min}=0$, our problem degenerates to the special case without offline historical information, as considered in our previous work\cite{gao2020proactive}. In this case, the whole regret bound is in the order of $O(1/V+\sqrt{(\log T)/T})$.
	When the offline historical information is available (\textit{i.e.}, $H_{\min}>0$), the regret bound would be even lower. Specifically, we consider the following four cases under a fixed value of $T$.\footnote{The notations $O$, $\Theta$, and $\Omega$ are all asymptotic notations introduced in \cite{cormen2009introduction}.}
\begin{enumerate}
	\item The first case is when $H_{\min}=O(1)$, \textit{i.e.}, a constant value unrelated to $T$. Compared to the scenario without offline historical information, though the value of regret bound reduces in this case, its order remains to be $O(1/V+\sqrt{(\log T)/T})$.
	\item The second case is when $H_{\min}=\Theta (T)$, \textit{i.e.}, the number of offline historical observations is comparable to the length of time horizon. In this case, the regret bound is still in the order of $O(1/V+\sqrt{(\log T)/T})$.
	\item The third case is when $H_{\min}=\Theta (T\log T)$. In this case, under a sufficiently great length of time horizon $T$, the regret bound approaches $O(1/V+\sqrt{1/T})$. 
	\item The fourth case is when $H_{\min}=\Omega (T^{2}\log T)$, \textit{i.e.}, there is adequate offline historical information. 
	In this case, each EFS proactively leverages offline historical information to acquire highly accurate estimations on file popularities. As a result, the last term in the regret bound becomes even smaller, and the second term becomes dominant. Therefore, the order of the regret decreases to $O(1/V+1/T)$.
\end{enumerate}
When it comes to the impact of time horizon length $T$, the regret bound decreases and approaches $B/V$ as the value of $T$ increases.
In summary, given a longer time horizon length and more historical information (\textit{i.e.}, larger values of $T$ and $H_{\min}$), CPHBL achieves a better regret performance.
Such results are also verified by numerical simulation in Section \ref{subsec: different T and H}.

%\textit{Proof Sketch:} 
%The proof of Theorem \ref{theorem: regret bound} follows the idea of regret analysis in \cite{li2019combinatorial}.
%First, we introduce the notion of Lyapunov drift to characterize the change of virtual queue backlog sizes (in terms of the Lyapunov function) between consecutive time slots. 
%Next, by taking the per-time-slot regret of CPHBL into the Lyapunov drift, 
%we upper bound such a drift-plus-regret term by a linear function of the difference between the performances incurred by CPHBL and the optimal scheme. 
%With the aid of an auxiliary scheme, we further transform the previous upper bound into a linear function of the difference between the performances of CPHBL and the auxiliary scheme. 
%Finally, for the terms in the resulting bound, we leverage Chernoff-Hoeffding techniques and Jensen's inequality to bound each of them to complete the proof. 
%More details of the proof are given in Appendix \ref{proof: regret bound}.

% DRL design
\section{DRL Based Benchmark Design}\label{sec: drl}

\begin{figure}[!t]
%	\captionsetup{labelfont={color=blue}}
	\centering
	\includegraphics[scale=0.46]{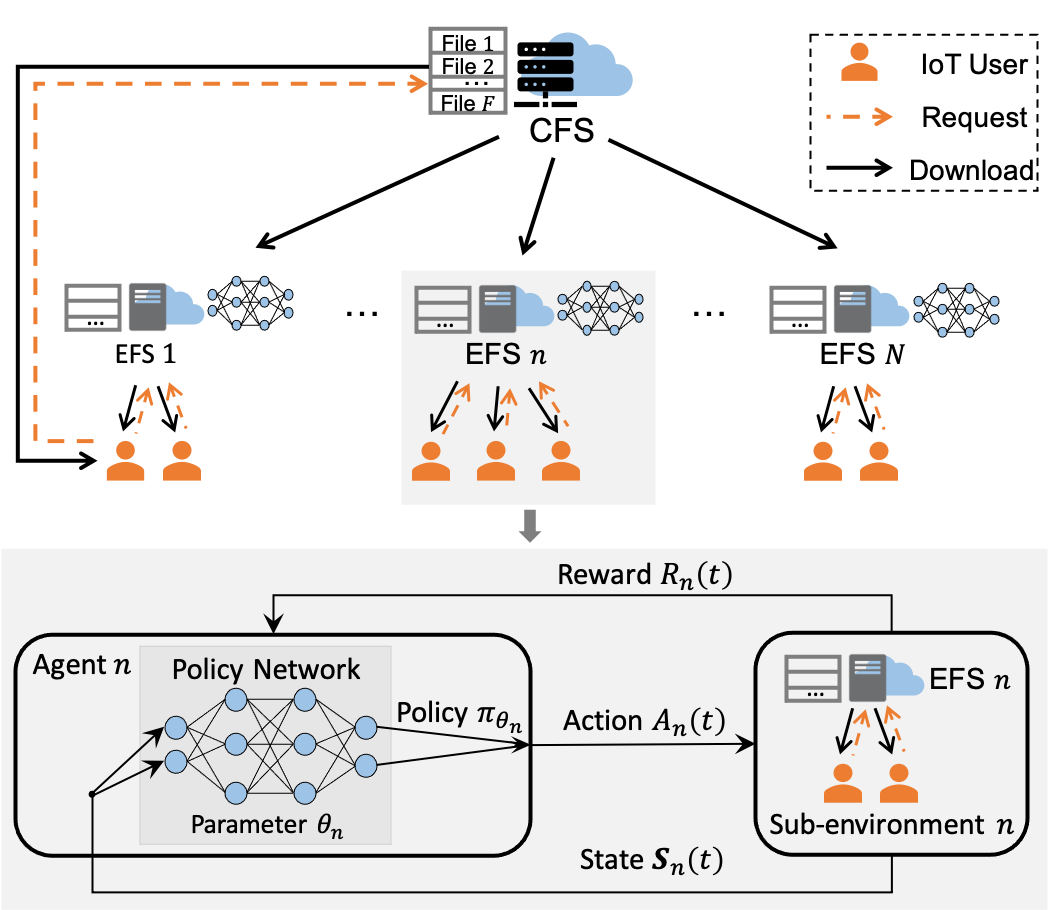}
	\caption{Overview of CPDRL design. The environment is partitioned into $N$ independent sub-environments, each for an agent (EFS). Note that we do not show the CFS in the sub-environment block. However, in each time slot, each EFS may interact with the CFS for file downloading. In our model, the CFS is assumed to provide simultaneous and independent file deliveries to all EFSs.}
	\label{fig: drl}
\end{figure}

\begin{figure}[!t]
%	\captionsetup{labelfont={color=blue}}
	\centering
	\includegraphics[scale=0.5]{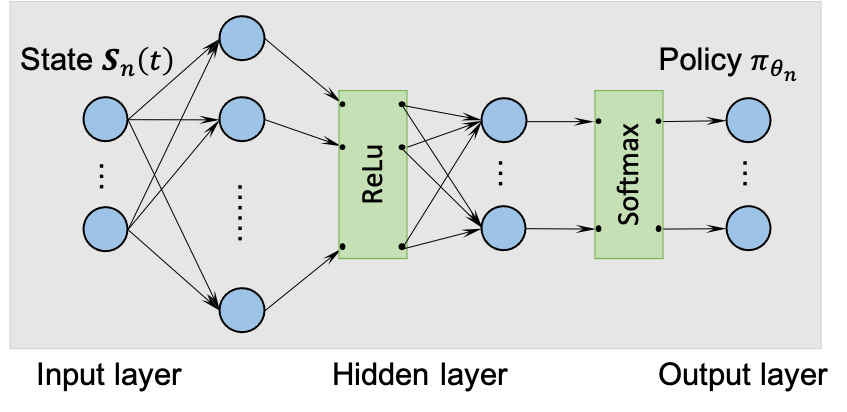}
	\caption{Design of the policy network for agent $n$. The cache placement scheme of agent $n$ is designed as a feedforward neural network (FNN) with one hidden layer of dimension $512$, followed by a ReLU activation function.}
	\label{fig: neural net}
\end{figure}

In recent years, deep reinforcement learning (DRL) has been widely adopted in various fields to conduct goal-directed learning and sequential decision making\cite{bian2019neural}\cite{pei2019optimal}. 
It deals with agents that learn to make better sequential decisions by interacting with the environment without complicated modeling and too much domain knowledge requirement.
In this section, to compare our scheme CPHBL with DRL based approaches, we propose a novel Cache Placement scheme with DRL called \textit{CPDRL} as a baseline for evaluation.

\subsection{Overall Design of CPDRL}

Under CPDRL, we view each EFS $n\in\mathcal{N}$ as a DRL agent $n$ which interacts with the environment over time slots. As a result, the original problem turns into a multi-agent DRL problem with $N$ agents. 
Note that under our settings, such an $N$-agent DRL problem can be decomposed into $N$ single-agent DRL subproblems since there is no coupling among the agents' decision makings. The reasons are shown as follows.
First, the CFS provide simultaneous and independent file deliveries to all EFSs. 
Second, recall that each IoT user is served by one and only one EFS and thus the subsets of IoT users that are associated with EFSs are disjoint.
Based on the above two properties, the decision making on each EFS has no impact on the decisions on other EFSs. 
Therefore, the environment can be partitioned into $N$ independent sub-environments and each agent $n$ only interacts with its related sub-environment $n$. As a result, under CPDRL, each agent (EFS) solves for a single-agent DRL subproblem independently. Next, we introduce the basic settings of the single-agent DRL system.

In a classical single-agent DRL system, there is an agent which interacts with its environment over iterations. At the beginning of each iteration $t$, the agent observes some representation of the environment's state $S(t)$. In response, the agent takes an action $A(t)$ based on its maintained policy $\pi_{\theta}$.
The policy $\pi_{\theta}$ is parameterized by a deep neural network (DNN) with parameter $\theta$. After the agent performs the action $A(t)$, it observes a new state $S(t+1)$ and receives a reward $R(t)$. Based on the gained information, the agent improves its policy $\pi_{\theta}$ to maximize the time-averaged expected reward it receives, \textit{i.e.}, $\mathbb{E}[\frac{1}{T}\sum_{t=0}^{T-1}\gamma^{t}R(t)]$. Here $\gamma\in [0,1]$ is called the discount rate and it determines the present value of future rewards.

\subsection{Detailed Design of CPDRL}

Considering the limitation of existing DRL techniques, when solving for the cache placement problem (\ref{p: goal}), we ignore the storage cost constraints (\ref{constraint: cost}) in the design of CPDRL. In this subsection, we show our detailed design of CPDRL in terms of the state representation, agent action, and reward signal for a particular agent $n$.\footnote{In this work, for simplicity, we assume that all of the $N$ agents share the same DRL design, including the same policy network structure and training parameters. In practice, one can employ heterogeneous DRL designs for different agents to adapt to more general scenarios.}

\subsubsection{State Representation}
We define the observed environment state by agent $n$ in time slot $t$ as $\boldsymbol{S}_{n}(t)\triangleq \boldsymbol{X}_{n}(t-1)$, \textit{i.e.}, the cache placement on the EFS $n$ in the previous time slot $(t-1)$.

\subsubsection{Agent Action}
We define the action of agent $n$ in time slot $t$ as a tuple $A_{n}(t)\in\mathcal{A}\triangleq\{(f,x)\vert f\in\mathcal{F},x\in\{0,1\}\}$. Action $A_{n}(t)=(f,x)$ means that the agent $n$ updates the cache placement decision for file $f$ on EFS $n$ in time slot $t$ to $x$. When $x=1$, file $f$ will be cached on EFS $n$; otherwise, file $f$ will not be cached on EFS $n$.

\subsubsection{Reward Design}
The reward received by agent $n$ in time slot $t$ is set as the cache hit reward $R_{n}(t)$ defined in (\ref{eq: EFS reward}) of Section \ref{subsec: reward}.

\subsubsection{Policy Network}
We design each agent $n$'s cache placement scheme as a feedforward neural network (FNN)\cite{irie1988capabilities} with one hidden layer of dimension $512$, followed by a ReLU activation function. We show such a network design in Figure \ref{fig: neural net}. As shown in the figure, the policy network takes the observed environment state as input. When given input $\boldsymbol{S}_{n}(t)$, a probability distribution $\pi_{\theta_{n}}(\cdot\vert\boldsymbol{S}_{n}(t))$ over the action space $\mathcal{A}$ will be output from the network. Note that such a policy network design requires the number of files $F$ to be fixed. The change in the value of $F$ would require the reconstruction and retraining of the policy network. In each time slot, a candidate action will be sampled from set $\mathcal{A}$ according to the distribution $\pi_{\theta_{n}}(\cdot\vert\boldsymbol{S}_{n}(t))$. The cache placement will be updated accordingly if the sampled action satisfies the storage capacity constraint in (\ref{constraint: storage}); otherwise, the cache placement on EFS $n$ will remain unchanged.

\subsection{CPDRL Workflow}
We show the pseudocode of CPDRL in Algorithm \ref{alg: cpdrl}. The operation of CPDRL is composed of two procedures: the cache placement procedure and the policy update procedure.
In the \textit{cache placement} procedure, under CPDRL, each EFS makes cache placement decisions based on its current policy network.
In the \textit{policy update} procedure, each EFS adopts the policy gradient\cite{sutton2000policy} method to train its policy network with the collected online feedback. Note that each of the networks is trained during the first $T_{0}$ time slots on a batch basis, and the length of each batch is set uniformly as $l$.

\subsection{Comparison with CPHBL}
In comparison with CPHBL, CPDRL has the following limitations. First, it requires the heuristic techniques of network-training or hyper-parameter tuning. Second, its effectiveness can only be justified by experimental simulations without theoretical performance guarantee. Third, it cannot deal with the stochastic time-averaged storage cost constraints. Lastly, it provides few insightful explanations for the resulting decision makings and system performances. In comparison, by employing MAB methods, CPHBL enjoys the advantages of a more lightweight implementation, theoretical tractability, and the applicability to time-averaged constraints. Besides, the design of CPHBL also leads to insightful explanations for the online decision making in previous sections (see remarks in Sections \ref{sec: algorithm}--\ref{sec: analysis}).
We further compare the performance of CPHBL and CPDRL with numerical simulations in Section \ref{subsec: simulation under fixed history}.

%\floatname{algorithm}{\color{blue}Algorithm}
\begin{algorithm}[!t]
%\captionsetup{labelfont={color=blue}}
\caption{Cache Placement with Deep Reinforcement Learning (CPDRL)}
\label{alg: cpdrl}
{\small{
\begin{algorithmic}[1]
  \State Initialize $\boldsymbol{X}_{n}(-1)=\boldsymbol{0}$ and the policy network $\pi_{\theta_{n}}$ for each EFS $n\in\mathcal{N}$. 
  \For {each time slot $t\in\{0,1,\cdots,T-1\}$}
      \For {each EFS $n\in\mathcal{N}$}
          \Statex \textit{~~~~~~~~\%Cache Placement}
          \State Observe state $\boldsymbol{S}_{n}(t)\leftarrow\boldsymbol{X}_{n}(t-1)$.
          \parState{Sample a candidate action $(f, X'_{n,f}(t))$ from $\mathcal{A}$ according to $\pi_{\theta_{n}}(\cdot\vert\boldsymbol{S}_{n}(t))$.}
          \State Set $\boldsymbol{x}'_{n}\leftarrow(X_{n,1}(t-1),\dots, X'_{n,f}(t),\dots,X_{n,F}(t-1))$.
          \If {$\boldsymbol{x}'_{n}$ satisfies the constraint (\ref{constraint: storage})}
              \State Set $A_{n}(t)\leftarrow(f, X_{n,f}(t-1))$.
          \Else
              \State Set $A_{n}(t)\leftarrow(f, X'_{n,f}(t))$.
          \EndIf
          \parState{Perform action $A_{n}(t)$ and then receive a reward of $R_{n}(t)$.}
          \Statex \textit{~~~~~~~~\%Policy Update}
          \If {$1\leq t\leq T_{0}$ and $t\ \%\ l=0$}
              \parState{Train policy network $\pi_{\theta_{n}}$ using the collected information from time slots $(t-l+1)$ to $t$.}
          \EndIf
      \EndFor
  \EndFor
\end{algorithmic}
}}
\end{algorithm}

% Numerical results
%\vspace{-0.2em}
\section{Numerical Results}\label{sec: simulation}
%\vspace{-0.2em}

\subsection{Simulation Settings}
%\vspace{-0.2em}

We consider a Fog-assisted IoT system with {$1$ CFS, $4$ EFSs ($N=4$) and $20$ IoT users} ($K=20$).
Each user is uniformly randomly assigned to one of the EFSs.
The file set $\mathcal{F}$ on the CFS consists of 20 files ($F=20$) with different file sizes $L_{f}\in\{1,2,4,8\}$. 
The storage capacity of each EFS is $M_{n}=16$ units. 
We set the unit storage cost as $\alpha=1$.
We assume that each user $k$'s requests are generated from a Zipf distribution with a skewness parameter $\gamma\in [0.56,1.2]$. Note that such skewness parameters are fixed but \textit{unknown} to the EFSs. 
We set the storage cost budget $b_n$ to be $8$ units for each EFS $n\in\mathcal{N}$.

\begin{figure}[!t]
\centering 
  \subfigure[Time-averaged storage cost on {EFS $1$}.]{
    \includegraphics[scale=0.24]{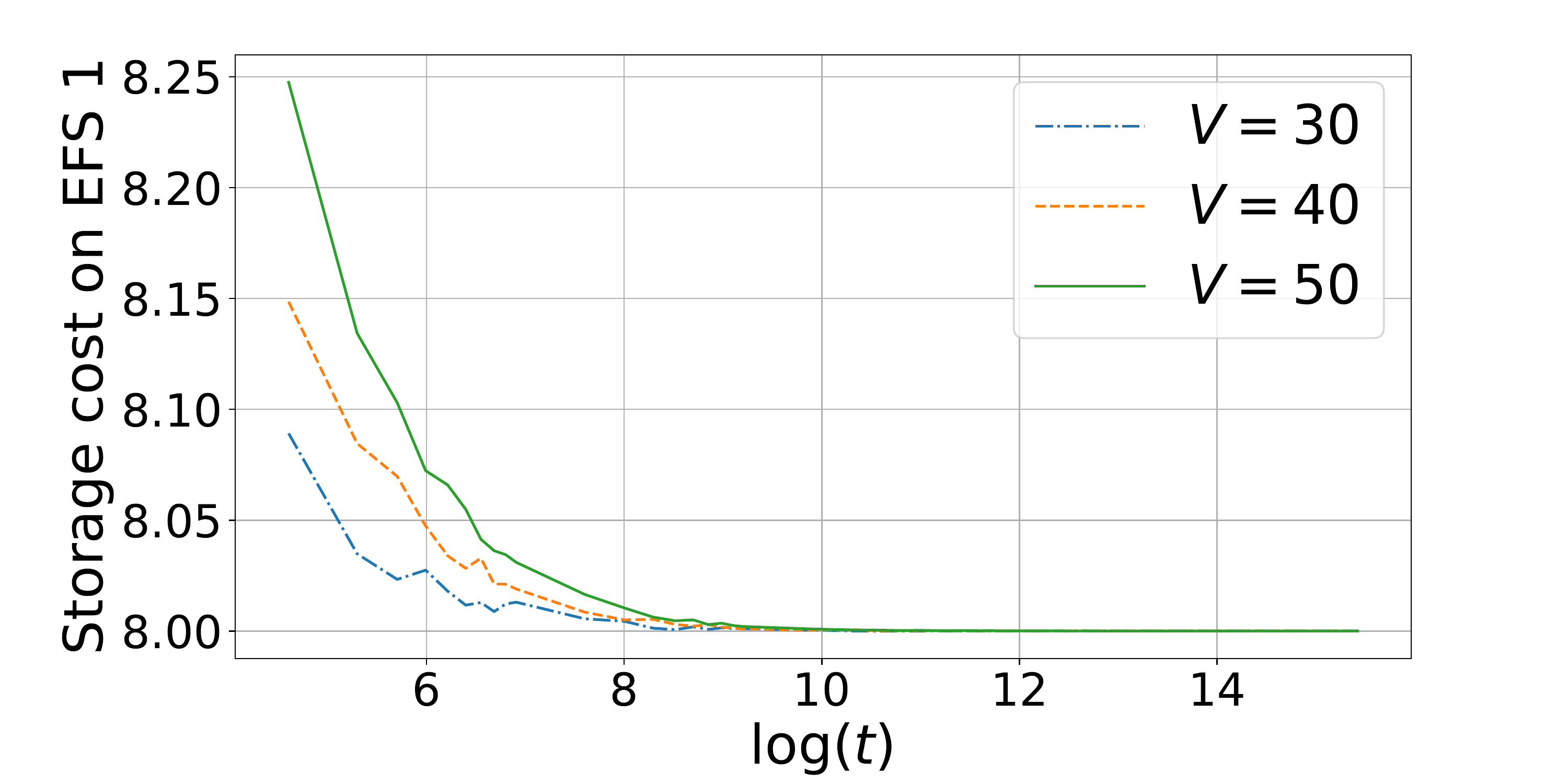} 
    \label{fig: CPHBL convergence rate}
  }
  \subfigure[Storage costs on each EFS.]{ 
	\includegraphics[scale=0.24]{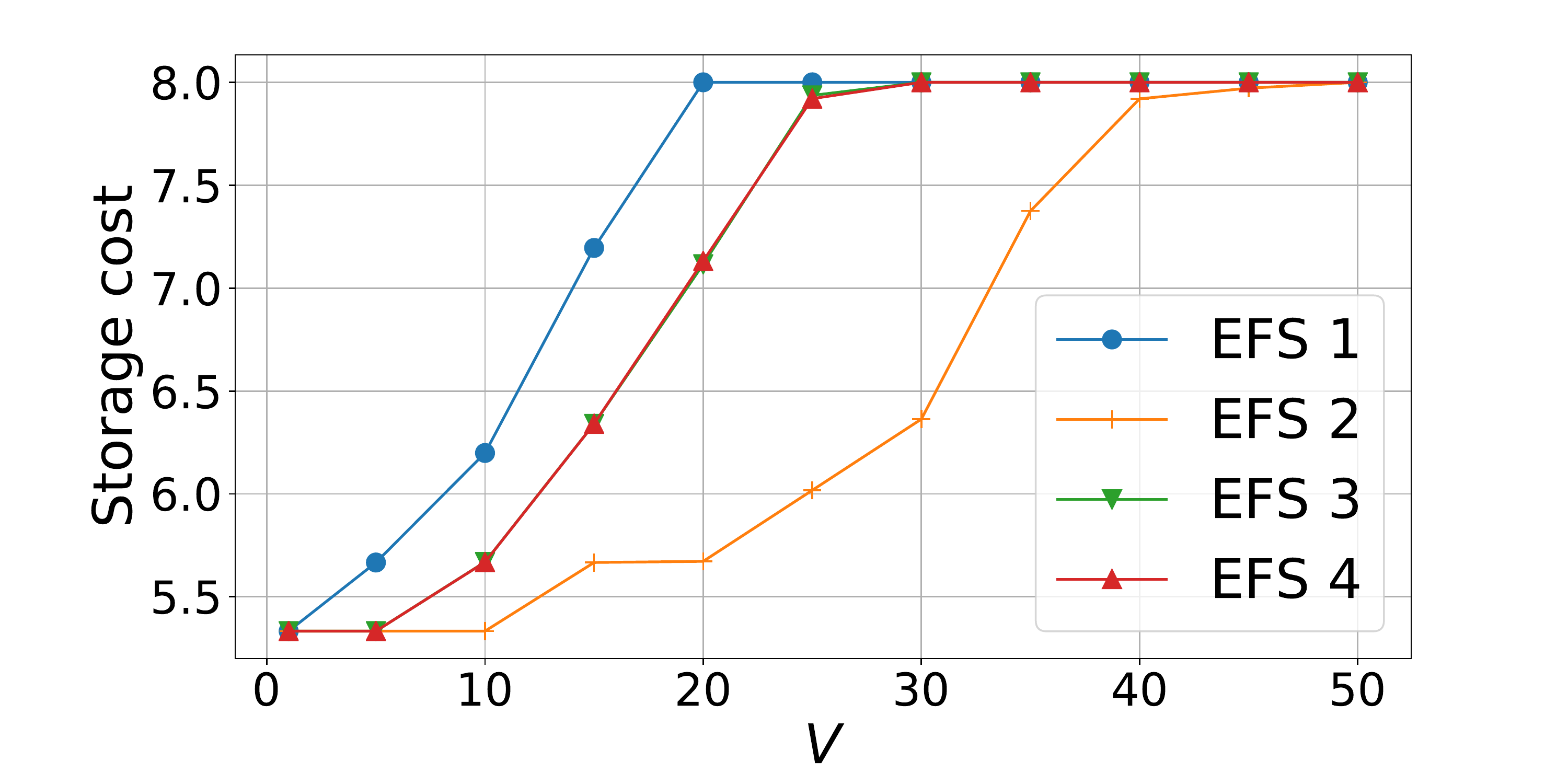}
	\label{fig: CPHBL cost}
  } 
  \subfigure[Regret \& total storage costs.]{
    \includegraphics[scale=0.24]{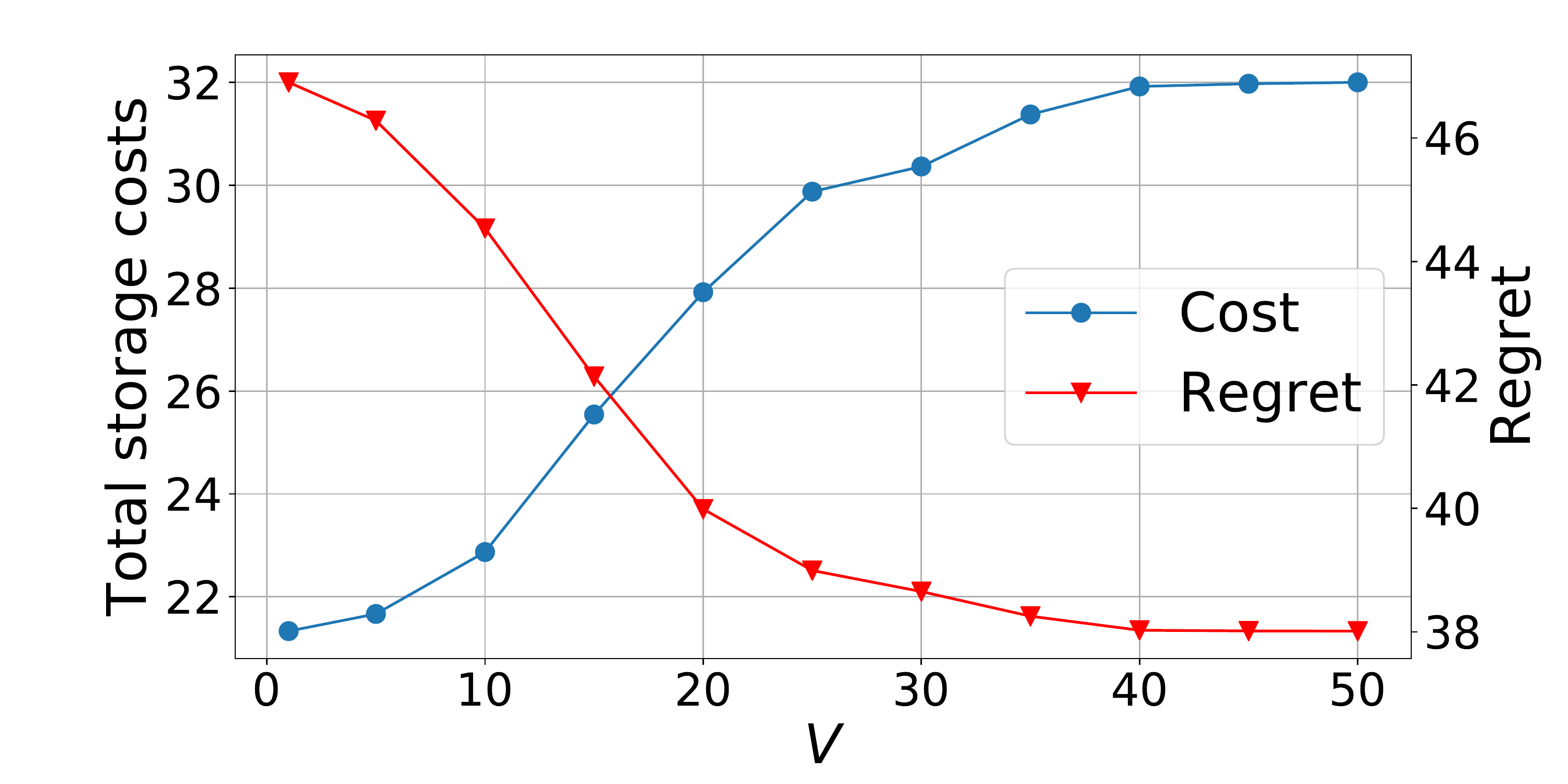} 
    \label{fig: CPHBL cost & reward}
  } 
  \caption{Performance of CPHBL with different values of $V$.}
  \label{fig: CPHBL performance}
\end{figure}

\subsection{Performance of CPHBL with A Fixed Time Horizon Length and A Fixed Number of Offline Historical Observations}\label{subsec: simulation under fixed history}

In this subsection, we investigate the performance of CPHBL by fixing the time horizon length $T$ as $5\times 10^{6}$ time slots and the number of offline historical observations $H_{n,f}$ as $1000$ for all $n\in\mathcal{N}$, $f\in\mathcal{F}$ (\textit{i.e.}, $H_{\min}=1000$).

\textbf{Performance of CPHBL under Different Values of $\boldsymbol{V}$:}
In Figure \ref{fig: CPHBL convergence rate}, we take the first EFS (EFS $1$) as an example to illustrate how the time-averaged storage cost on each EFS changes over time under different values of $V$.
Particularly, on EFS $1$, the time-averaged storage cost approaches the cost budget $b_{1}=8$ units as time goes by. 
Moreover, the greater the value of $V$, the longer the convergence time. 
For example, the convergence time extends from $4000$ time slots to about $10000$ time slots as the value of $V$ increases from $30$ to $50$.
This shows that the larger values of $V$ lead to a longer time for convergence.
Figure \ref{fig: CPHBL cost} evaluates the time-averaged storage cost on each EFS incurred by CPHBL under different values of $V$.
As the value of parameter $V$ increases, the storage cost on each EFS keeps increasing until it reaches the budget $b_{n}=8$ units. 
Such results show that the time-averaged storage cost constraints in (\ref{constraint: cost}) are strictly satisfied under CPHBL.

Next, we switch to the evaluations of regrets and total storage costs incurred by CPHBL with different values of $V$. 
As shown in Figure \ref{fig: CPHBL cost & reward}, there is a notable reduction in the regret as the value of $V$ increases. 
Such results imply that {CPHBL can achieve a lower regret with a larger value of $V$}. Moreover, when the value of $V$ is sufficiently large ($V\geq 40$), 
the regret value stabilizes at around $38.01$. 
This verifies our previous analysis in Theorem \ref{theorem: regret bound} about the term $B/V$ in the regret bound (\ref{ineq: regret upper bound}).
Besides, as the value of $V$ increases, we also see an increase in the total storage costs which eventually reach the budget when $V \ge 40$.
Overall, the results in Figures \ref{fig: CPHBL cost} and \ref{fig: CPHBL cost & reward} verify the tunable tradeoff between the regret value and total storage costs.

\begin{figure}[!t]
\centering 
  \subfigure[Time-averaged total storage costs.]{ 
	\includegraphics[scale=0.24]{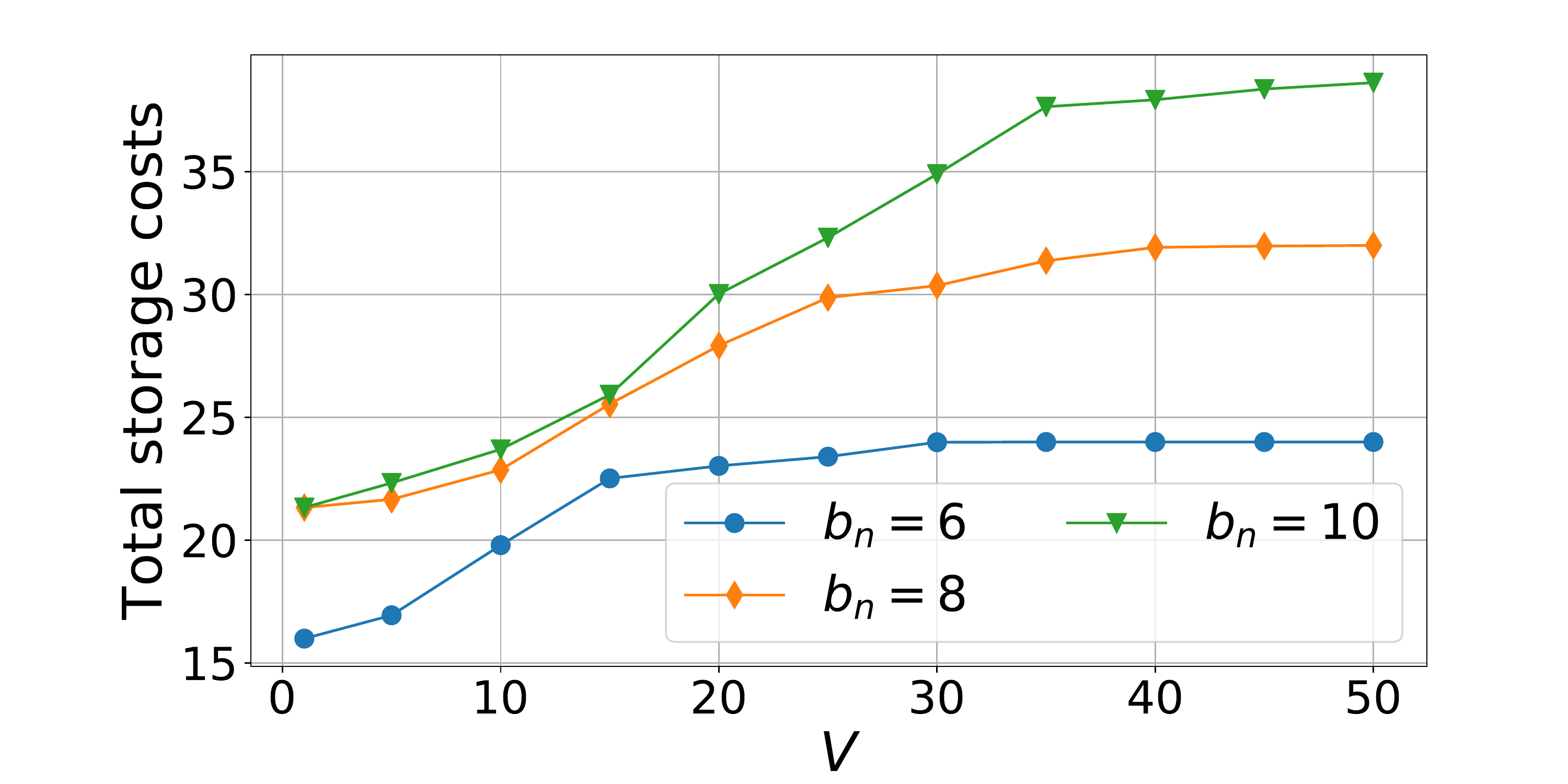}
	\label{fig: cost vs b}
  } 
  \subfigure[Time-averaged total cache hit reward.]{
    \includegraphics[scale=0.24]{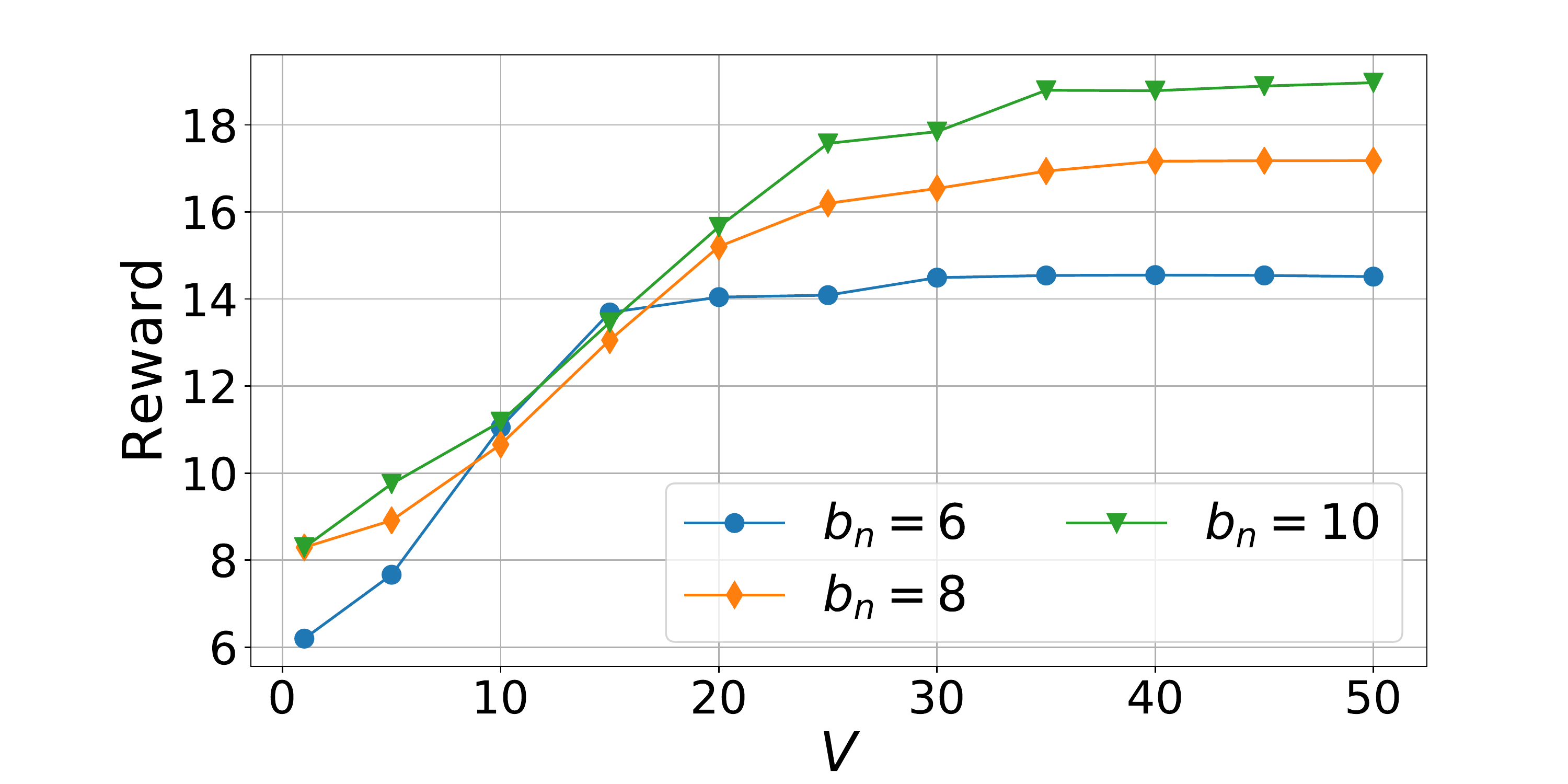} 
    \label{fig: reward vs b}
  } 
  \subfigure[Regret.]{
    \includegraphics[scale=0.24]{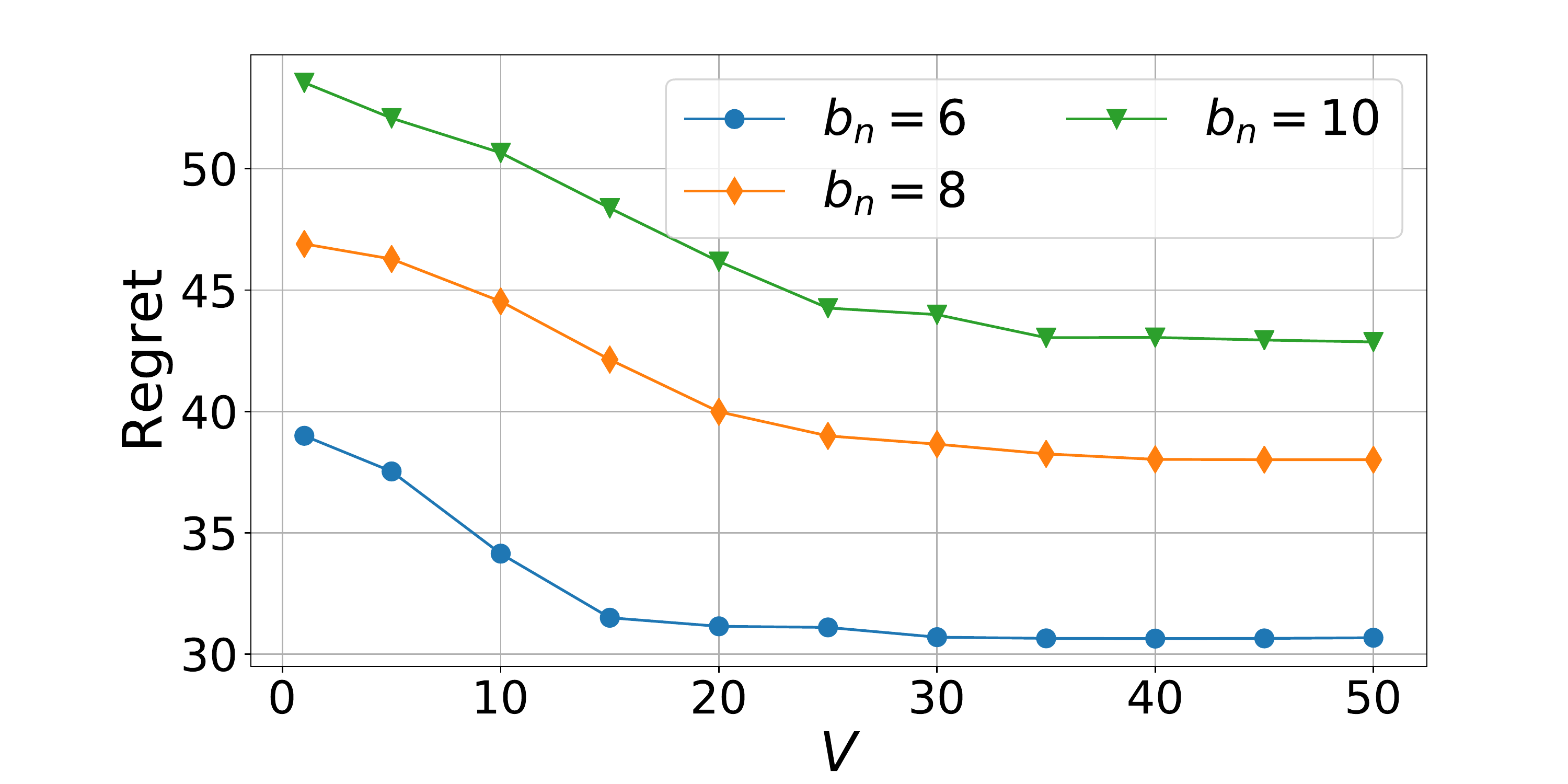} 
    \label{fig: regret vs b}
  } 
  \caption{{Performance of CPHBL given different storage budgets ($b_{n}$).}}
  \label{fig: CPHBL performance vs b}
\end{figure}

\textbf{Performance of CPHBL under Different Settings of Storage Cost Budget $\boldsymbol{b_{n}}$:}
Next, we select different values for the storage cost budget $b_{n}$ of each EFS $n$ to investigate their impacts on system performances. 
Figure \ref{fig: CPHBL performance vs b} shows our simulation results. 
From Figure \ref{fig: cost vs b}, we see that given $V=50$, 
the time-averaged total storage costs increase by $60.93\%$ as the value of $b_{n}$ increases from $6$ to $10$. 
Under the same settings, Figure \ref{fig: reward vs b} shows that the time-averaged total cache hit reward increases by $30.71\%$. 
The reason is that with more budget, each EFS would store more files to further maximize the cache hit rewards. 
Figure \ref{fig: regret vs b} illustrates the regret under different storage cost budgets. 
The results verify our theoretical analysis in (\ref{ineq: queue upper bound}) about the proportional growth of regret with respect to the storage cost budget.
The reason is that under CPHBL, each EFS would explore more files when given more budget, thereby resulting in {a higher regret}.

\textbf{CPHBL vs. Its Variants:}
In Section \ref{sec: online learning}, the confidence radius in (\ref{eq: ucb estimate}) measures the uncertainty about the empirical reward estimate. The larger the confidence radius, the greater the necessity of exploration for the corresponding file. 
Accordingly, each EFS is more prone to caching under-explored files. 
To investigate how the regret changes under different exploration strategies,
we propose two types of variants for CPHBL: one leveraging $\varepsilon$-greedy method and the other employing UCB-like methods. More detail is specified as follows.
\begin{itemize}
	\item[$\diamond$] \textit{CPHBL-greedy}: CPHBL-greedy differs from CPHBL in the cache placement phase (lines \ref{line: cache placement start}-\ref{line: cache placement end} in Algorithm \ref{alg: learning-aided CP}). Specifically, it replaces the HUCB1 estimates $\{\tilde{d}_{n,f}(t)\}_{n,f}$ with empirical means $\{\bar{d}_{n,f}(t)\}_{n,f}$ in function {\sc SetCachePlacement}. 
	Recall that $\bar{d}_{n,f}(t)$ denotes the empirical mean that involves both offline historical observations and online feedbacks.
	Then it adopts $\varepsilon$-greedy method within the cache placement phase.
	With probability $\varepsilon$, each EFS $n$ selects files uniformly randomly from subset $\mathcal{F}_{n,1}(t)$ to cache.
	With probability $1 - \varepsilon$, files with the empirically highest reward estimates are chosen to be cached.  
	Intuitively, CPHBL-greedy spends about a proportion $\varepsilon$ of time for uniform exploration and the rest $(1 - \varepsilon)$ proportion of time for exploitation. 
%	with probability $1-\epsilon$, the cache placement is decided by function {\sc SetCachePlacement}($t, n, \{\bar{d}_{n,f}(t)\}_{n,f}$).
	\item[$\diamond$] \textit{CPHBL-UCBT}: CPHBL-UCBT replaces the HUCB1 estimate (line \ref{algline: ucb1} in Algorithm \ref{alg: learning-aided CP}) with \textit{UCB1-tuned (UCBT)} estimate\cite{auer2002finite} while the rest remains the same as CPHBL.
\end{itemize}

\begin{figure}[!t]
 \setlength{\abovecaptionskip}{-0.1cm}
 \centering
 \includegraphics[scale=0.22]{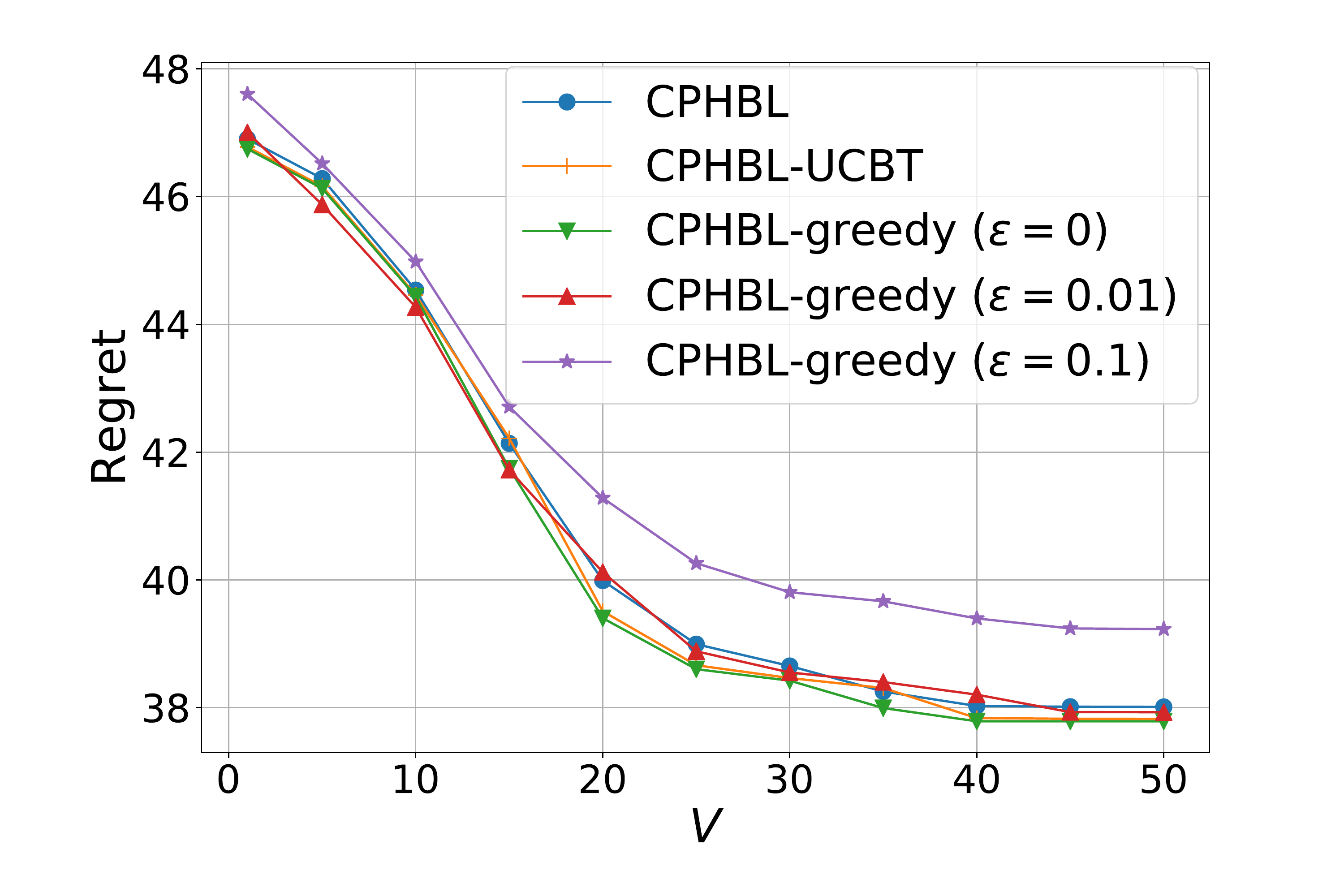}
 \caption{Regret of CPHBL and its variants.}
% \vspace{-1em}
 \label{fig: regret vs v}
 \setlength{\belowcaptionskip}{-1cm}
\end{figure}

We compare the regret value of CPHBL against CPHBL-greedy ($\varepsilon\in\{0, 0.01, 0.1\}$) and CPHBL-UCBT in Figure \ref{fig: regret vs v} under different values of $V$.
Regarding the variants of CPHBL, 
interestingly, although CPHBL-greedy with $\varepsilon=0$ intuitively discards the chance of uniform exploration in the online learning phase, 
it still achieves a regret performance that is close to CPHBL, CPHBL-UCBT, and CPHBL-greedy with $\varepsilon=0.01$.
The reason is that CPHBL-greedy with $\varepsilon=0$ can resort to the storage cost constraint guarantee in the online control phase to conduct enforced exploration.
%As a result, CPHBL-greedy with $\epsilon=0$ incurs the maximum regret among such schemes due to insufficient exploration. 
%In such a case, as the value of $V$ increases, there is even less exploration, 
%since each EFS would put more focus on caching the files with the empirically highest rewards. Accordingly, more regret would be incurred.
%Besides, the results also imply that, with the confidence radius in (\ref{eq: ucb estimate}), CPHBL takes advantage of adaptive exploration and thus achieves a better regret performance.
%It is interesting that though these  schemes leverage different exploration strategies, they achieve comparable regret values.}
%: CPHBL and CPHBL-UCBT utilize different confidence radius, 
%while CPHBL-greedy explores uniformly with a probability of $\epsilon$ during each time slot.
In comparison, the regret of CPHBL-greedy with $\varepsilon=0.1$ still performs inferior to other schemes due to its over-exploration.

\begin{figure}[!t] 
%\captionsetup{labelfont={color=blue}}
%\setlength{\abovecaptionskip}{-0.1cm}
\centering 
  \subfigure[Time-averaged total cache hit reward.]{
    \includegraphics[scale=0.24]{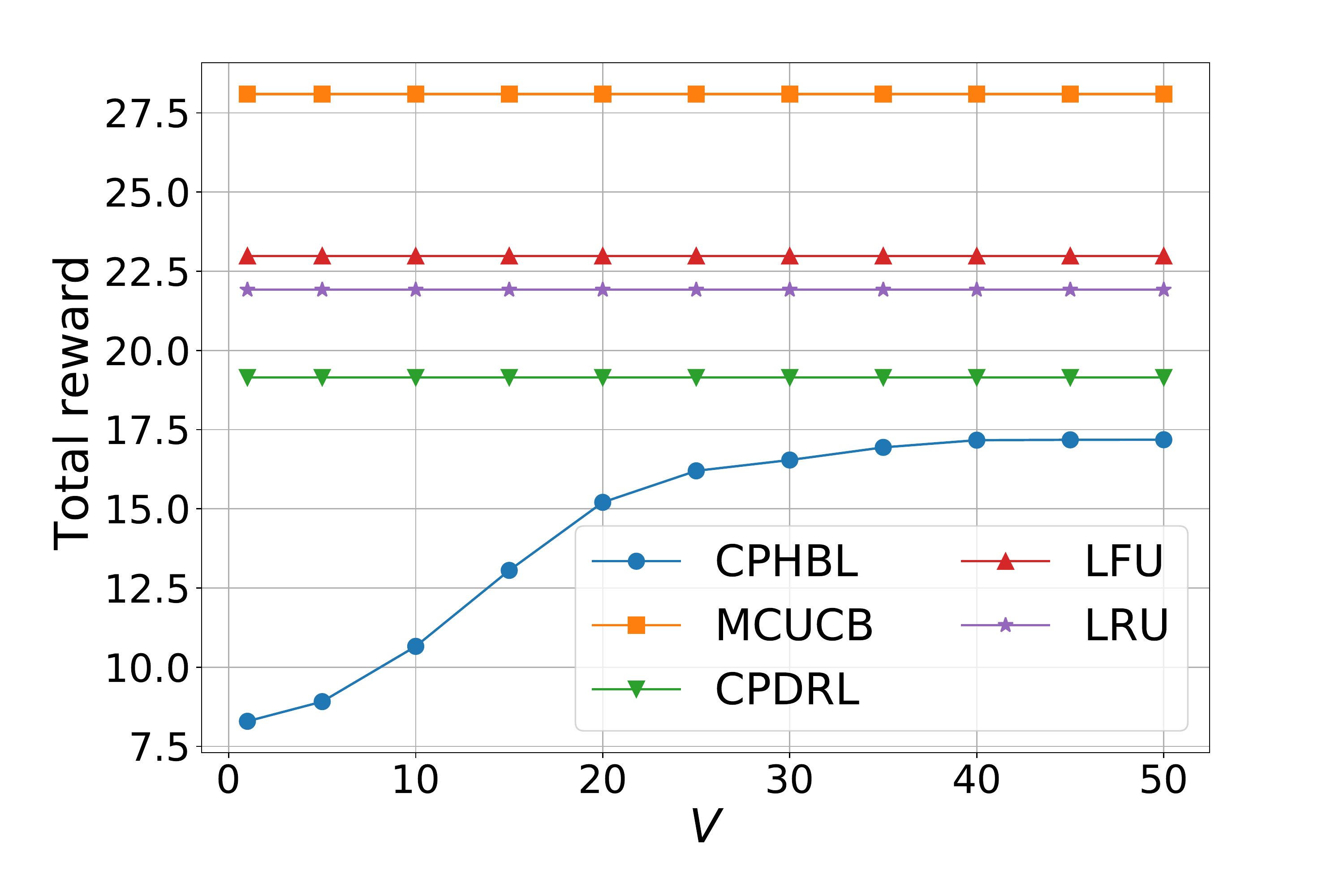} 
    \label{fig: benchmark reward}
  } 
  \subfigure[Time-averaged total storage costs.]{
    \includegraphics[scale=0.24]{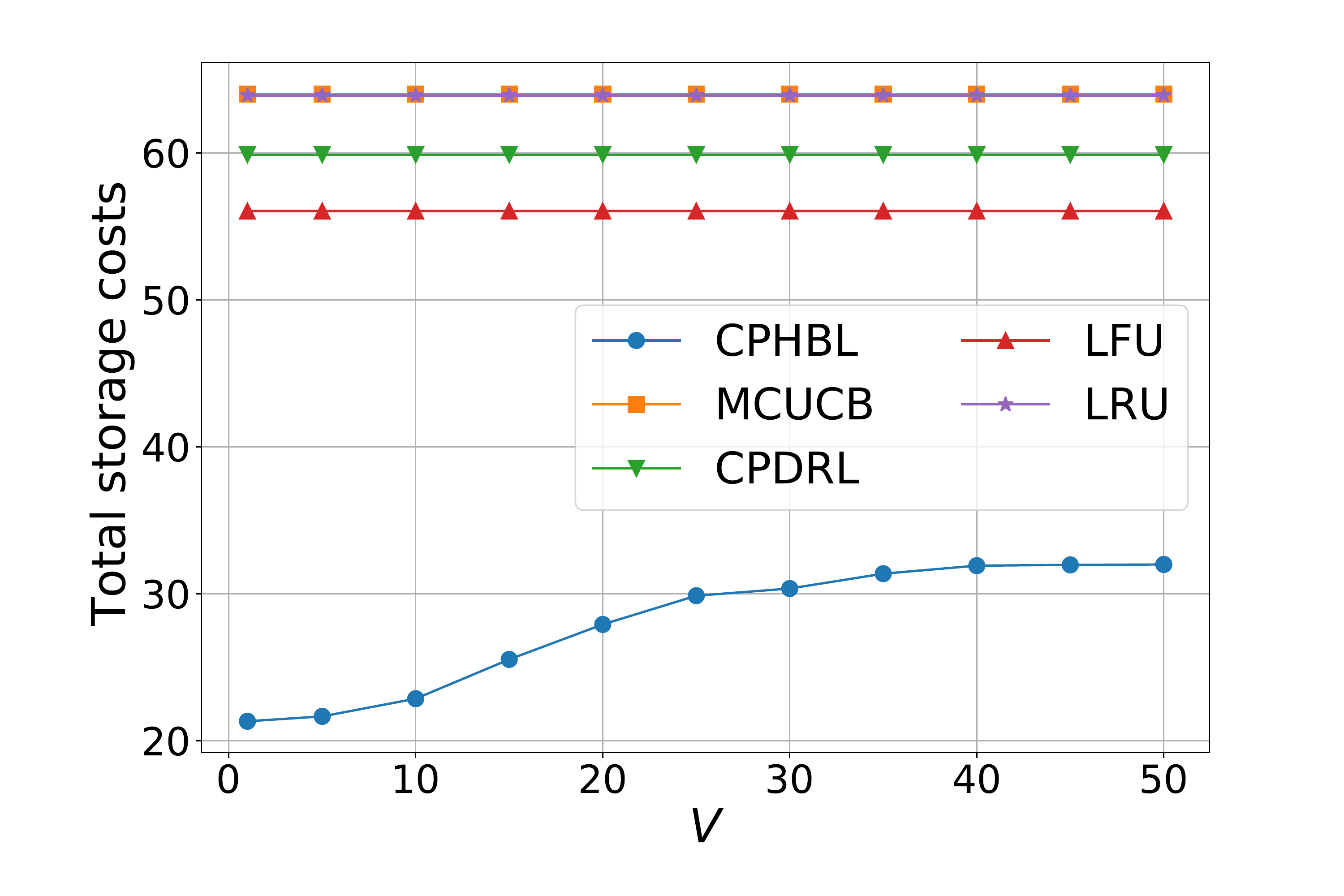} 
    \label{fig: benchmark cost}
  }
  \caption{Comparison of CPHBL and baseline schemes.}
  \label{fig: benchmarks}
\end{figure}

\textbf{CPHBL vs. Other Baseline Schemes:}
	We also compare the performances of CPHBL with four baseline schemes: MCUCB\cite{chen2013combinatorial}, CPDRL (\textit{Cache Placement with Deep Reinforcement Learning}), LFU (\textit{Least Frequently Used})\cite{lee2001lrfu}, and LRU (\textit{Least Recently Used})\cite{lee2001lrfu}. Below we demonstrate how each of them proceeds in detail, respectively.
\begin{itemize}
	\item[$\diamond$] \textit{MCUCB}: Under MCUCB\cite{blasco2014learning}, a modified combinatorial UCB scheme is used to estimate file popularities and decide cache placement during each time slot.
	\item[$\diamond$] \textit{CPDRL}: The detailed design of CPDRL is presented in Section \ref{sec: drl}. In the simulation, we set the network training parameter as $T_{0}=10^{6}$ and $l=10$. The policy network parameters are  updated using the RMSprop\cite{hinton2012overview} algorithm with a learning rate of $10^{-5}$.
	\item[$\diamond$] \textit{LFU}: Under LFU, each EFS maintains a counter for each of its cached files. 
	Each counter records the number of times that its corresponding file has been requested on the EFS. 
	If a requested file is not in the cache, the requested file would be downloaded from the CFS and  cached on the EFS by replacing the least frequently used files therein.
	\item[$\diamond$] \textit{LRU}: Under LRU, each EFS records the most recently requested time slot for each of its cached files. If a requested file is not in the cache, the requested file would be downloaded from the CFS and cached on the EFS by replacing the least recently used files.
\end{itemize}
%Schemes MFU and MRU are motivated by the LFU (\textit{Most Frequently Used}) and the LRU (\textit{Least Rencently Used}) schemes proposed by Lee \textit{et al.} in \cite{lee2001lrfu}

We show the simulation results in Figure \ref{fig: benchmarks}. The cache hit rewards and total storage costs of the four baseline schemes (MCUCB, CPDRL, LFU, and LRU) remain constant given different values of $V$. This is because their decision making does not involve parameter $V$. 
From Figure \ref{fig: benchmarks}, we see that CPHBL achieves the lowest cache hit reward while MCUCB achieves the highest cache hit reward. 
Particularly, given $V = 50$, compared to MCUCB, CPHBL achieves a $38.85\%$ lower total cache hit reward.
In comparison with the other three baseline schemes, the DRL based scheme CPDRL achieves the worst performance in terms of the cache hit reward. The reason is that it can not learn efficiently from limited online feedback information.

However, except CPHBL, the other four schemes fail to ensure the storage cost constraints in (\ref{constraint: cost}).\footnote{Recall that the storage cost budget on each EFS is set as $b_{n}=8$ units in our simulations. Accordingly, the total time-averaged storage costs of the four EFSs should not exceed $32$ units. However, the total time-averaged storage costs all exceed $55$ units under the four baseline schemes.}
More specifically, given $V=50$, when compared to the four baseline schemes (MCUCB, CPDRL, LFU, and LRU), CPHBL achieves $50.00\%$, $46.56\%$, $42.90\%$ and $49.93\%$ reductions in the total storage costs, respectively.
%With the increase of the value of $V$, the total storage costs under CPHBL reaches the maximal value of $32$ units. In the meanwhile, the total reward also increases and reaches almost the maximal value $111.38$ that can be achieved under CPHBL. 
%The result implies that to satisfy the storage cost constraints, total reward should be sacrificed.
Note that such results verify the advantage of our scheme   over DRL based approaches.

\subsection{Performance of CPHBL with Different Values of Time Horizon Length and Numbers of Offline Historical Observations}\label{subsec: different T and H}

\begin{figure}[!t]
\centering 
  \subfigure[Regret with fixed values of $H_{\min}$ when $V=50$.]{
    \includegraphics[scale=0.24]{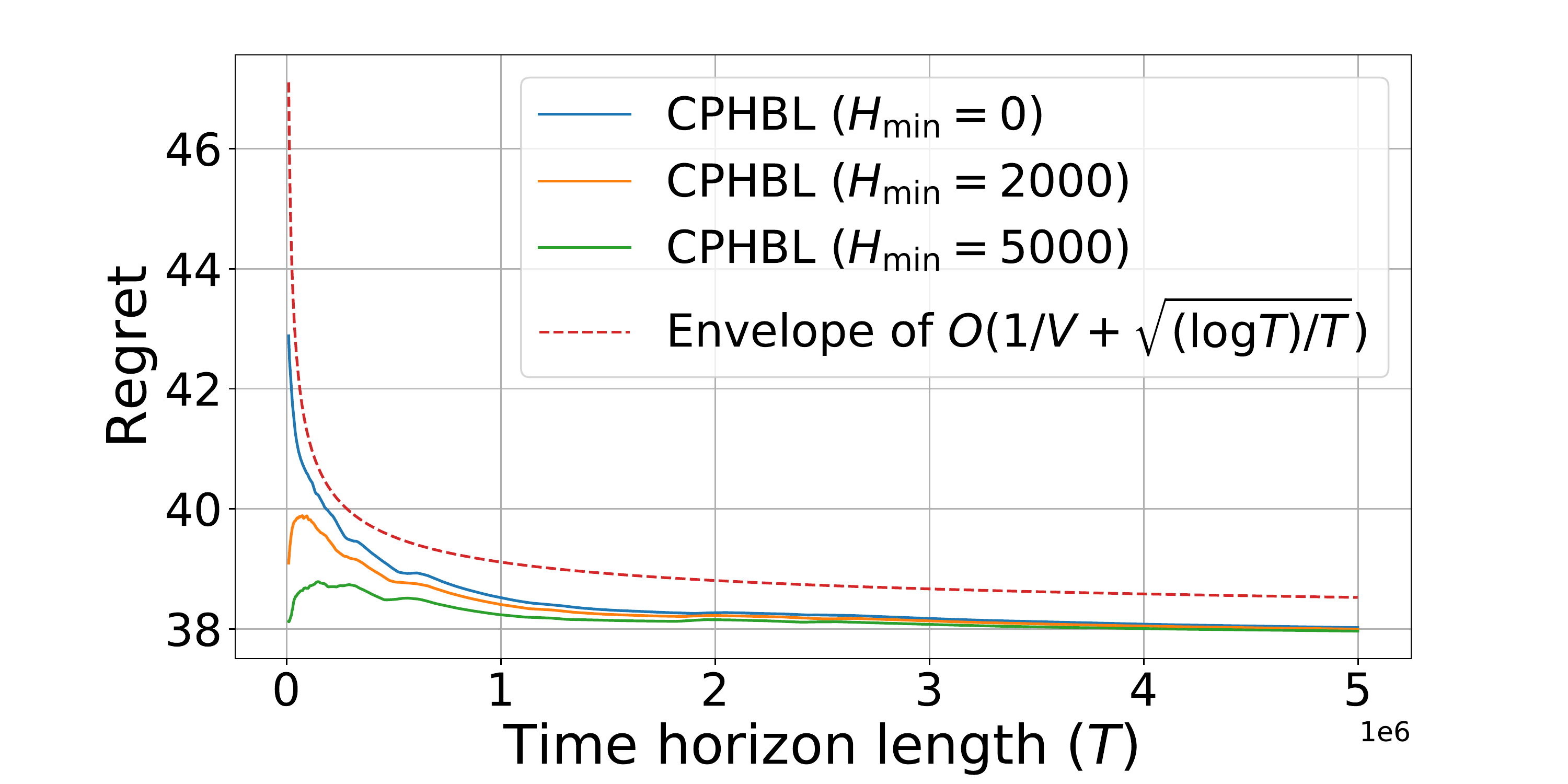} 
    \label{fig: CPHBL regret}
  }
  \subfigure[Regret with different values of $H_{\min}$ when $V=50$.]{
    \includegraphics[scale=0.24]{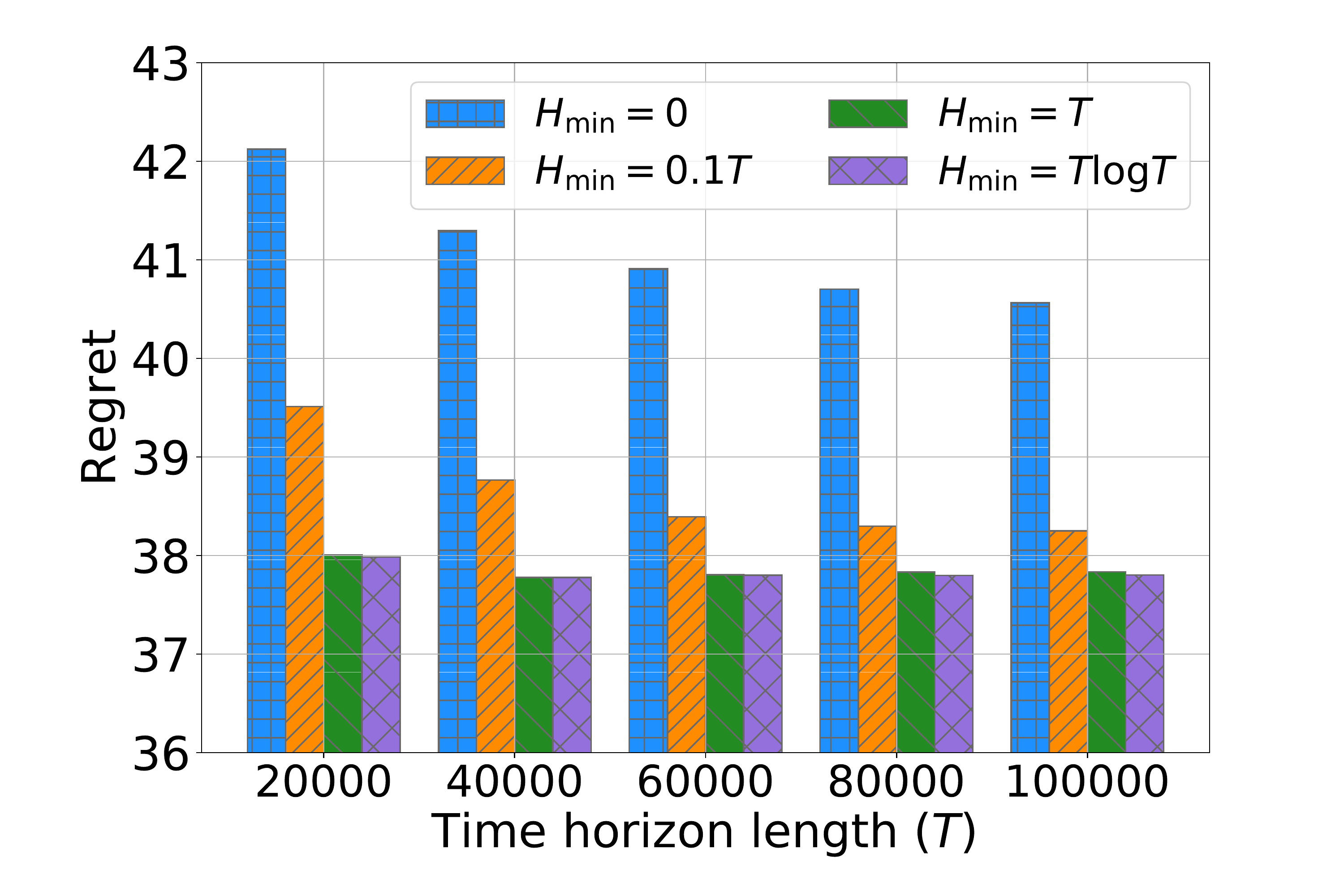} 
    \label{fig: regret_historic_time}
  }
  \caption{Regret of CPHBL.}
  \label{fig: historic}
\end{figure}

In this subsection, we investigate the impacts of time horizon length $T$ and the number $H_{\min}$ of offline historical observations\footnote{In our simulations, the number $H_{n,f}$ of offline historical observations on EFS $n$ for file $f$ is set to be identical for all $n\in\mathcal{N}$ and $f\in\mathcal{F}$. Therefore, by the definition that $H_{\min}\triangleq \min_{n,f}H_{n,f}$, we have $H_{n,f}=H_{\min}$ for all $n\in\mathcal{N}$ and $f\in\mathcal{F}$.} on the regret of CPHBL. We take the case when $V=50$ as an example for illustration. The results are shown in Figure \ref{fig: historic}. 

In Figure \ref{fig: CPHBL regret}, we present the regret performances over a constant time horizon length $T$ under fixed values of $H_{\min}$. Specifically, each curve corresponds to the result under a constant value of $H_{\min}\in\{0, 2000, 5000\}$ (independent of $T$).
Note that when $H_{\min}=0$, there is no offline historical information.
On one hand, given a fixed number $H_{\min}$ of offline historical observations, the results show that the regret value is reduced by an order of $O(1/V+\sqrt{(\log T)/T})$.\footnote{In Figure \ref{fig: CPHBL regret}, we provide a curve of $38+300\sqrt{(\log T)/T}$ as an envelope of $O(1/V+\sqrt{(\log T)/T})$ for illustration. Note that since $V$ is fixed, $1/V$ can be viewed as a constant term.}
On the other hand, given a fixed value of $T$, CPHBL achieves a lower regret with more offline historical observations. However, as the value of $T$ becomes sufficiently large, the regret reduction turns negligible.
For example, as the value of $H_{\min}$ increases from $0$ to $5000$, the regret reduces by $0.74\%$ when $T=10^{6}$, but only by $0.15\%$ when $T=5\times 10^{6}$.
%We also note that when $H_{n,f}\in\{2000, 5000\}$, the regret increases first and then decreases as $T$ grows.

In Figure \ref{fig: regret_historic_time}, we compare the regret performances under different values of $H_{\min}$ over various time horizon lengths. Specifically, we consider the cases when $H_{\min}\in\{0,\ 0.1T,\ T,\ T\log T\}$.
As shown in the figure, when the value of $H_{\min}$ is small (\textit{e.g.}, when $H_{\min}\leq 0.1 T$), even a slight increase in the offline historical information brings a noticeable improvement to the regret performance.
However, as the value of $H_{\min}$ increases, the degree of regret reduction becomes less significant.
For example, given $T=10^{5}$, the regret reduces by $5.71\%$ as the value of $H_{\min}$ increases from $0$ to $0.1T$, but only by $1.09\%$ from $0.1T$ to $T$.
All of the above results verify our theoretical analysis in Theorem \ref{theorem: regret bound} (see Section \ref{sec: analysis}).

%\vspace{-0.2em}
% Conclusion
\section{Conclusion}\label{sec: conclusion}
%\vspace{-0.2em}

In this paper, we considered the cache placement problem with unknown file popularities in caching-enabled Fog-assisted IoT systems. 
By formulating the problem as a constrained CMAB problem, we devised a novel proactive cache placement scheme called CPHBL with an effective integration of online control, online learning and offline historical information.
Results from our theoretical analysis and numerical simulations showed that our devised scheme achieves a near-optimal total cache hit reward under storage cost constraints with a sublinear time-averaged regret.
To the best of our knowledge, our work provided the first systematic study on the synergy of online control, online learning, and offline historical information.
Our results not only revealed novel insights to the designers of caching-enabled Fog-assisted IoT systems, but also verified the advantage of CPHBL over the deep reinforcement learning based approach.

\ifCLASSOPTIONcaptionsoff
  \newpage
\fi

\bibliographystyle{IEEEtran}
\bibliography{references.bib, IEEEabrv}

\begin{thebibliography}{10}
\providecommand{\url}[1]{#1}
\csname url@samestyle\endcsname
\providecommand{\newblock}{\relax}
\providecommand{\bibinfo}[2]{#2}
\providecommand{\BIBentrySTDinterwordspacing}{\spaceskip=0pt\relax}
\providecommand{\BIBentryALTinterwordstretchfactor}{4}
\providecommand{\BIBentryALTinterwordspacing}{\spaceskip=\fontdimen2\font plus
\BIBentryALTinterwordstretchfactor\fontdimen3\font minus
  \fontdimen4\font\relax}
\providecommand{\BIBforeignlanguage}[2]{{%
\expandafter\ifx\csname l@#1\endcsname\relax
\typeout{** WARNING: IEEEtran.bst: No hyphenation pattern has been}%
\typeout{** loaded for the language `#1'. Using the pattern for}%
\typeout{** the default language instead.}%
\else
\language=\csname l@#1\endcsname
\fi
#2}}
\providecommand{\BIBdecl}{\relax}
\BIBdecl

\bibitem{bastug2014living}
E.~Bastug, M.~Bennis, and M.~Debbah, ``Living on the edge: The role of
  proactive caching in 5g wireless networks,'' \emph{IEEE Communications
  Magazine}, vol.~52, no.~8, pp. 82--89, 2014.

\bibitem{zhao2017online}
S.~Zhao, Z.~Shao, H.~Qian, and Y.~Yang, ``Online user-ap association with
  predictive scheduling in wireless caching networks,'' in \emph{Proceedings of
  IEEE GLOBECOM}, 2017.

\bibitem{jiang2017novel}
Y.~Jiang, M.~Ma, M.~Bennis, F.~Zheng, and X.~You, ``A novel caching policy with
  content popularity prediction and user preference learning in fog-ran,'' in
  \emph{Proceedings of IEEE GLOBECOM Workshops}, 2017.

\bibitem{zhao2018femos}
S.~Zhao, Y.~Yang, Z.~Shao, X.~Yang, H.~Qian, and C.-X. Wang, ``Femos:
  Fog-enabled multitier operations scheduling in dynamic wireless networks,''
  \emph{IEEE Internet of Things Journal}, vol.~5, no.~2, pp. 1169--1183, 2018.

\bibitem{gao2020pora}
X.~Gao, X.~Huang, S.~Bian, Z.~Shao, and Y.~Yang, ``Pora: Predictive offloading
  and resource allocation in dynamic fog computing systems,'' \emph{IEEE
  Internet of Things Journal}, vol.~7, no.~1, pp. 72--87, 2020.

\bibitem{bharath2016learning}
B.~Bharath, K.~G. Nagananda, and H.~V. Poor, ``A learning-based approach to
  caching in heterogenous small cell networks,'' \emph{IEEE Transactions on
  Communications}, vol.~64, no.~4, pp. 1674--1686, 2016.

\bibitem{pang2016joint}
H.~Pang, L.~Gao, and L.~Sun, ``Joint optimization of data sponsoring and edge
  caching for mobile video delivery,'' in \emph{Proceedings of IEEE GLOBECOM},
  2016.

\bibitem{li2019combinatorial}
F.~Li, J.~Liu, and B.~Ji, ``Combinatorial sleeping bandits with fairness
  constraints,'' in \emph{Proceedings of IEEE INFOCOM}, 2019.

\bibitem{shivaswamy2012multi}
P.~Shivaswamy and T.~Joachims, ``Multi-armed bandit problems with history,'' in
  \emph{Proceedings of AISTATS}, 2012.

\bibitem{neely2010stochastic}
M.~J. Neely, ``Stochastic network optimization with application to
  communication and queueing systems,'' \emph{Synthesis Lectures on
  Communication Networks}, vol.~3, no.~1, pp. 1--211, 2010.

\bibitem{kwak2018hybrid}
J.~Kwak, Y.~Kim, L.~B. Le, and S.~Chong, ``Hybrid content caching in 5g
  wireless networks: Cloud versus edge caching,'' \emph{IEEE Transactions on
  Wireless Communications}, vol.~17, no.~5, pp. 3030--3045, 2018.

\bibitem{wang2018distributed}
Y.~Wang, W.~Wang, Y.~Cui, K.~G. Shin, and Z.~Zhang, ``Distributed packet
  forwarding and caching based on stochastic network utility maximization,''
  \emph{IEEE/ACM Transactions on Networking}, vol.~26, no.~3, pp. 1264--1277,
  2018.

\bibitem{xu2018joint}
J.~Xu, L.~Chen, and P.~Zhou, ``Joint service caching and task offloading for
  mobile edge computing in dense networks,'' in \emph{Proceedings of IEEE
  INFOCOM}, 2018.

\bibitem{blasco2014learning}
P.~Blasco and D.~G{\"u}nd{\"u}z, ``Learning-based optimization of cache content
  in a small cell base station,'' in \emph{Proceedings of IEEE ICC}, 2014.

\bibitem{blasco2014multi}
------, ``Multi-armed bandit optimization of cache content in wireless
  infostation networks,'' in \emph{Proceedings of IEEE ISIT}, 2014.

\bibitem{muller2016smart}
S.~M{\"u}ller, O.~Atan, M.~van~der Schaar, and A.~Klein, ``Smart caching in
  wireless small cell networks via contextual multi-armed bandits,'' in
  \emph{Proceedings of IEEE ICC}, 2016.

\bibitem{zhang2019learning}
X.~Zhang, G.~Zheng, S.~Lambotharan, M.~R. Nakhai, and K.-K. Wong, ``A learning
  approach to edge caching with dynamic content library in wireless networks,''
  in \emph{Proceedings of IEEE GLOBECOM}, 2019.

\bibitem{song2017learning}
J.~Song, M.~Sheng, T.~Q. Quek, C.~Xu, and X.~Wang, ``Learning-based content
  caching and sharing for wireless networks,'' \emph{IEEE Transactions on
  Communications}, vol.~65, no.~10, pp. 4309--4324, 2017.

\bibitem{xu2020collaborative}
X.~Xu, M.~Tao, and C.~Shen, ``Collaborative multi-agent multi-armed bandit
  learning for small-cell caching,'' \emph{IEEE Transactions on Wireless
  Communications}, vol.~19, no.~4, pp. 2570--2585, 2020.

\bibitem{ajmal2019survey}
S.~Ajmal, M.~B. Muzammil, A.~Jamil, S.~M. Abbas, U.~Iqbal, and P.~Touseef,
  ``Survey on cache schemes in heterogeneous networks using 5g internet of
  things,'' in \emph{Proceedings of ACM ICFNDS}, 2019.

\bibitem{li2018survey}
L.~Li, G.~Zhao, and R.~S. Blum, ``A survey of caching techniques in cellular
  networks: Research issues and challenges in content placement and delivery
  strategies,'' \emph{IEEE Communications Surveys \& Tutorials}, vol.~20,
  no.~3, pp. 1710--1732, 2018.

\bibitem{pang2018toward}
H.~Pang, J.~Liu, X.~Fan, and L.~Sun, ``Toward smart and cooperative edge
  caching for 5g networks: A deep learning based approach,'' in
  \emph{Proceedings of IEEE/ACM IWQoS}, 2018.

\bibitem{chen2019echo}
M.~Chen, W.~Saad, and C.~Yin, ``Echo-liquid state deep learning for
  $360^{\circ}$ content transmission and caching in wireless vr networks with
  cellular-connected uavs,'' \emph{IEEE Transactions on Communications},
  vol.~67, no.~9, pp. 6386--6400, 2019.

\bibitem{ndikumana2020deep}
A.~Ndikumana, N.~H. Tran, K.~T. Kim, C.~S. Hong \emph{et~al.}, ``Deep learning
  based caching for self-driving cars in multi-access edge computing,''
  \emph{IEEE Transactions on Intelligent Transportation Systems}, 2020.

\bibitem{liu2020distributed}
Z.~Liu, H.~Song, and D.~Pan, ``Distributed video content caching policy with
  deep learning approaches for d2d communication,'' \emph{IEEE Transactions on
  Vehicular Technology}, 2020.

\bibitem{bacstuug2015transfer}
E.~Ba{\c{s}}tu{\u{g}}, M.~Bennis, and M.~Debbah, ``A transfer learning approach
  for cache-enabled wireless networks,'' in \emph{Proceedings of IEEE WiOpt},
  2015.

\bibitem{sengupta2014learning}
A.~Sengupta, S.~Amuru, R.~Tandon, R.~M. Buehrer, and T.~C. Clancy, ``Learning
  distributed caching strategies in small cell networks.'' in \emph{Proceedings
  of ISWCS}, 2014.

\bibitem{sadeghi2019reinforcement}
A.~Sadeghi, F.~Sheikholeslami, A.~G. Marques, and G.~B. Giannakis,
  ``Reinforcement learning for adaptive caching with dynamic storage pricing,''
  \emph{IEEE Journal on Selected Areas in Communications}, vol.~37, no.~10, pp.
  2267--2281, 2019.

\bibitem{zhu2018learn}
Z.~Zhu, T.~Liu, S.~Jin, and X.~Luo, ``Learn and pick right nodes to offload,''
  in \emph{Proceedings of IEEE GLOBECOM}, 2018.

\bibitem{yao2019energy}
J.~Yao and N.~Ansari, ``Energy-aware task allocation for mobile iot by online
  reinforcement learning,'' in \emph{Proceedings of IEEE ICC}, 2019.

\bibitem{mukherjee2018resource}
A.~Mukherjee, S.~Misra, V.~S.~P. Chandra, and M.~S. Obaidat,
  ``Resource-optimized multiarmed bandit-based offload path selection in edge
  uav swarms,'' \emph{IEEE Internet of Things Journal}, vol.~6, no.~3, pp.
  4889--4896, 2018.

\bibitem{lopez2009ofdma}
D.~L{\'o}pez-P{\'e}rez, A.~Valcarce, G.~De~La~Roche, and J.~Zhang, ``Ofdma
  femtocells: A roadmap on interference avoidance,'' \emph{IEEE Communications
  Magazine}, vol.~47, no.~9, pp. 41--48, 2009.

\bibitem{chen2013combinatorial}
W.~Chen, Y.~Wang, and Y.~Yuan, ``Combinatorial multi-armed bandit: General
  framework and applications,'' in \emph{Proceedings of ICML}, 2013.

\bibitem{slivkins2019introduction}
A.~Slivkins \emph{et~al.}, ``Introduction to multi-armed bandits,''
  \emph{Foundations and Trends{\textregistered} in Machine Learning}, vol.~12,
  no. 1-2, pp. 1--286, 2019.

\bibitem{martello1999dynamic}
S.~Martello, D.~Pisinger, and P.~Toth, ``Dynamic programming and strong bounds
  for the 0-1 knapsack problem,'' \emph{Management Science}, vol.~45, no.~3,
  pp. 414--424, 1999.

\bibitem{martello2000new}
------, ``New trends in exact algorithms for the 0--1 knapsack problem,''
  \emph{European Journal of Operational Research}, vol. 123, no.~2, pp.
  325--332, 2000.

\bibitem{gao2020proactive}
X.~Gao, X.~Huang, Y.~Tang, Z.~Shao, and Y.~Yang, ``Proactive cache placement
  with bandit learning in fog-assisted iot system,'' in \emph{Proceedings of
  IEEE ICC}, 2020.

\bibitem{cormen2009introduction}
T.~H. Cormen, C.~E. Leiserson, R.~L. Rivest, and C.~Stein, \emph{Introduction
  to Algorithms}.\hskip 1em plus 0.5em minus 0.4em\relax MIT press, 2009.

\bibitem{bian2019neural}
S.~Bian, X.~Huang, Z.~Shao, and Y.~Yang, ``Neural task scheduling with
  reinforcement learning for fog computing systems,'' in \emph{Proceedings of
  IEEE GLOBECOM}, 2019.

\bibitem{pei2019optimal}
J.~Pei, P.~Hong, M.~Pan, J.~Liu, and J.~Zhou, ``Optimal vnf placement via deep
  reinforcement learning in sdn/nfv-enabled networks,'' \emph{IEEE Journal on
  Selected Areas in Communications}, vol.~38, no.~2, pp. 263--278, 2019.

\bibitem{irie1988capabilities}
B.~Irie and S.~Miyake, ``Capabilities of three-layered perceptrons.'' in
  \emph{Proceedings of ICNN}, 1988.

\bibitem{sutton2000policy}
R.~S. Sutton, D.~A. McAllester, S.~P. Singh, and Y.~Mansour, ``Policy gradient
  methods for reinforcement learning with function approximation,'' in
  \emph{Proceedings of NeurIPS}, 2000.

\bibitem{auer2002finite}
P.~Auer, N.~Cesa-Bianchi, and P.~Fischer, ``Finite-time analysis of the
  multiarmed bandit problem,'' \emph{Machine Learning}, vol.~47, no. 2-3, pp.
  235--256, 2002.

\bibitem{lee2001lrfu}
D.~Lee, J.~Choi, J.-H. Kim, S.~H. Noh, S.~L. Min, Y.~Cho, and C.~S. Kim,
  ``Lrfu: A spectrum of policies that subsumes the least recently used and
  least frequently used policies,'' \emph{IEEE Transactions on Computers},
  vol.~50, no.~12, pp. 1352--1361, 2001.

\bibitem{hinton2012overview}
G.~Hinton, N.~Srivastava, and K.~Swersky, ``Overview of mini-batch gradient
  descent,'' \emph{Neural Networks for Machine Learning}, vol. 575, no.~8,
  2012.

\bibitem{hoeffding1994probability}
W.~Hoeffding, ``Probability inequalities for sums of bounded random
  variables,'' in \emph{The Collected Works of Wassily Hoeffding}.\hskip 1em
  plus 0.5em minus 0.4em\relax Springer, 1994, pp. 409--426.

\end{thebibliography}

\appendices

\section{Algorithm Development}\label{appendix: alg develop}
We define a Lyapunov function as follows:
\begin{equation}\label{def: Lyapunov function}
	L\left(\boldsymbol{Q}\left(t\right)\right)\triangleq \frac{1}{2}\sum_{n\in\mathcal{N}}\left(Q_{n}\left(t\right)\right)^{2},
\end{equation}
in which $\boldsymbol{Q}(t)=(Q_{1}(t),Q_{2}(t),\cdots,Q_{N}(t))$ is the vector of all virtual queues.
Then we have 
\begin{equation}
\begin{split}
	&L\left(\boldsymbol{Q}\left(t+1\right)\right)-L\left(\boldsymbol{Q}\left(t\right)\right)\\
	&=\frac{1}{2}\sum_{n\in\mathcal{N}}\left[\left(Q_{n}\left(t+1\right)\right)^{2}-\left(Q_{n}\left(t\right)\right)^{2}\right]\\
	&\leq\frac{1}{2}\sum_{n\in\mathcal{N}}\left[b_{n}^{2}+\left(C_{n}\left(t\right)\right)^{2}+2Q_{n}\left(t\right)\left(C_{n}\left(t\right)-b_{n}\right)\right].
\end{split}
\end{equation}
Since $C_{n}\left(t\right)=\sum_{f\in\mathcal{F}}\alpha L_{f}X_{n,f}\left(t\right)\leq \alpha M_{n}$, it follows that 
\begin{multline}\label{eq: drift upper bound}
	L\left(\boldsymbol{Q}\left(t+1\right)\right)-L\left(\boldsymbol{Q}\left(t\right)\right)\\
	\leq B+\sum_{n\in\mathcal{N}}Q_{n}\left(t\right)\left(C_{n}\left(t\right)-b_{n}\right),
\end{multline}
where $B\triangleq\frac{1}{2}\sum_{n\in\mathcal{N}}\left(b_{n}^{2}+\alpha^{2}M_{n}^{2}\right)$. 

We consider an optimal cache placement scheme which makes \textit{i.i.d.} cache placement decisions $\boldsymbol{X}^{*}(t)$ in each time slot $t$, then the optimal time-averaged expected total reward of all EFSs is 
\begin{equation}\label{eq: optimal reward}
	R^{*}=\frac{1}{T}\sum_{t=0}^{T-1}\sum_{n\in\mathcal{N}}\mathbb{E}\left[\hat{R}_{n}\left(\boldsymbol{X}_{n}^{*}\left(t\right)\right)\right].
\end{equation}
According to (\ref{def: regret}), the regret of cache placement scheme $\left\{ \boldsymbol{X}\left(t\right)\right\} _{t}$ over $T$ time slots is
\begin{equation}
	Reg\left(T\right)=\frac{1}{T}\sum_{t=0}^{T-1}\sum_{n\in\mathcal{N}}\mathbb{E}\left[\hat{R}_{n}\left(\boldsymbol{X}_{n}^{*}\left(t\right)\right)-\hat{R}_{n}\left(\boldsymbol{X}_{n}\left(t\right)\right)\right].
\end{equation}
By the definition of reward $\hat{R}_{n}(\cdot)$ in (\ref{eq: EFS reward}), it follows that
\begin{multline}
	Reg\left(T\right)=\frac{1}{T}\sum_{t=0}^{T-1}\sum_{n\in\mathcal{N}}\sum_{f\in\mathcal{F}}L_{f}\big(\mathbb{E}\left[D_{n,f}\left(t\right)X_{n,f}^{*}\left(t\right)\right]\\
	-\mathbb{E}\left[D_{n,f}\left(t\right)X_{n,f}\left(t\right)\right]\big).
\end{multline}
Since the cache placement decision $X_{n,f}(t)$ is determined when $D_{n,f}(t)$ is unknown, $X_{n,f}(t)$ is independent of the $D_{n,f}(t)$. On the other hand, $X_{n,f}^{*}(t)$ is \textit{i.i.d.} over time slots and it is also independent of $D_{n,f}(t)$. Then by $\mathbb{E}[D_{n,f}(t)]=d_{n,f}$, we have
\begin{equation}
	Reg(T)\!=\!\frac{1}{T}\sum_{t=0}^{T-1}\sum_{n\in\mathcal{N}}\sum_{f\in\mathcal{F}}L_{f}d_{n,f}\mathbb{E}[X_{n,f}^{*}(t)\!-\!X_{n,f}(t)].
\end{equation}
We define the one-time-slot regret in each time slot $t$ as 
\begin{equation}\label{def: one-slot regret}
	\Delta_{Reg}\left(t\right)\triangleq\sum_{n\in\mathcal{N}}\sum_{f\in\mathcal{F}}L_{f}d_{n,f}\left(X_{n,f}^{*}\left(t\right)-X_{n,f}\left(t\right)\right).
\end{equation}
The regret $Reg(T)$ can be expressed as
\begin{equation}
	Reg\left(T\right)=\frac{1}{T}\sum_{t=0}^{T-1}\mathbb{E}\left[\Delta_{Reg}\left(t\right)\right].
\end{equation}
Then we define the Lyapunov drift-plus-regret as
\begin{multline}\label{def: drift-plus-regret}
	\Delta_{V}(\boldsymbol{Q}(t))\triangleq\mathbb{E}\left[L(\boldsymbol{Q}(t+1))-L(\boldsymbol{Q}(t))|\boldsymbol{Q}(t)\right]\\
	+V\mathbb{E}\left[\Delta_{Reg}\left(t\right)|\boldsymbol{Q}\left(t\right)\right].
\end{multline}
By (\ref{eq: drift upper bound}) and (\ref{def: one-slot regret}), it follows that
\begin{equation}\label{eq: drift-plus-regret bound 2}
\begin{split}
	&\Delta_{V}\left(\boldsymbol{Q}\left(t\right)\right)\\
	&\leq B+V\mathbb{E}\bigg[\sum_{n\in\mathcal{N}}\sum_{f\in\mathcal{F}}L_{f}d_{n,f}X_{n,f}^{*}\left(t\right)|\boldsymbol{Q}\left(t\right)\bigg]\\
	&~~~+\mathbb{E}\bigg[\sum_{n\in\mathcal{N}}Q_{n}\left(t\right)\left(C_{n}\left(t\right)-b_{n}\right)|\boldsymbol{Q}\left(t\right)\bigg]\\
	&~~~-V\mathbb{E}\bigg[\sum_{n\in\mathcal{N}}\sum_{f\in\mathcal{F}}L_{f}d_{n,f}X_{n,f}\left(t\right)|\boldsymbol{Q}\left(t\right)\bigg].
\end{split}
\end{equation}
Since $\tilde{d}_{n,f}\left(t\right)$ is the HUCB1 estimate of $d_{n,f}$ in time slot $t$ such that $\tilde{d}_{n,f}\left(t\right)\in [0,K_{n}]$, we have
\begin{equation}
\begin{split}
	&\sum_{n\in\mathcal{N}}\sum_{f\in\mathcal{F}}L_{f}d_{n,f}X_{n,f}\left(t\right)\\
	&=\sum_{n\in\mathcal{N}}\sum_{f\in\mathcal{F}}L_{f}\tilde{d}_{n,f}\left(t\right)X_{n,f}\left(t\right)\\
	&~~~+\sum_{n\in\mathcal{N}}\sum_{f\in\mathcal{F}}L_{f}\left(d_{n,f}-\tilde{d}_{n,f}\left(t\right)\right)X_{n,f}\left(t\right)\\
	&\stackrel{(a)}{\geq}\sum_{n\in\mathcal{N}}\sum_{f\in\mathcal{F}}L_{f}\tilde{d}_{n,f}\left(t\right)X_{n,f}\left(t\right)\\
	&~~~-\sum_{n\in\mathcal{N}}K_{n}\sum_{f\in\mathcal{F}}L_{f}X_{n,f}(t)\\
	&\stackrel{(b)}{\geq}\sum_{n\in\mathcal{N}}\sum_{f\in\mathcal{F}}L_{f}\tilde{d}_{n,f}\left(t\right)X_{n,f}\left(t\right)-\sum_{n\in\mathcal{N}}K_{n}M_{n},
\end{split}
\end{equation}
where inequality (a) holds because that $d_{n,f},\tilde{d}_{n,f}(t)\in [0,K_{n}]$ and inequality (b) is due to that $\sum_{f\in\mathcal{F}}L_{f}X_{n,f}(t)\leq M_{n}$.
Then it follows that
\begin{equation}\label{eq: drift-plus-regret bound}
\begin{split}
	&\Delta_{V}\left(\boldsymbol{Q}\left(t\right)\right)\leq B+\sum_{n\in\mathcal{N}}VK_{n}M_{n}\\
	&~~~+V\mathbb{E}\bigg[\sum_{n\in\mathcal{N}}\sum_{f\in\mathcal{F}}L_{f}d_{n,f}X_{n,f}^{*}\left(t\right)|\boldsymbol{Q}\left(t\right)\bigg]\\
	&~~~+\mathbb{E}\bigg[\sum_{n\in\mathcal{N}}Q_{n}\left(t\right)\left(C_{n}\left(t\right)-b_{n}\right)|\boldsymbol{Q}\left(t\right)\bigg]\\
	&~~~-V\mathbb{E}\bigg[\sum_{n\in\mathcal{N}}\sum_{f\in\mathcal{F}}L_{f}\tilde{d}_{n,f}\left(t\right)X_{n,f}\left(t\right)|\boldsymbol{Q}\left(t\right)\bigg].
\end{split}
\end{equation}
Substituting (\ref{eq: EFS cost}) and (\ref{eq: EFS reward}) into above inequality, we have
\begin{equation}\label{eq: drift-plus-regret bound 1}
\begin{split}
	&\Delta_{V}\left(\boldsymbol{Q}\left(t\right)\right)\leq B+\sum_{n\in\mathcal{N}}VK_{n}M_{n}-\sum_{n\in\mathcal{N}}Q_{n}\left(t\right)b_{n}\\
	&~~~+V\mathbb{E}\bigg[\sum_{n\in\mathcal{N}}\sum_{f\in\mathcal{F}}L_{f}d_{n,f}X_{n,f}^{*}\left(t\right)|\boldsymbol{Q}\left(t\right)\bigg]\\
	&~~~-\mathbb{E}\bigg[\sum_{n\in\mathcal{N}}\sum_{f\in\mathcal{F}}\tilde{w}_{n,f}\left(t\right) X_{n,f}(t)|\boldsymbol{Q}\left(t\right)\bigg],
\end{split}
\end{equation}
where $\tilde{w}_{n,f}\left(t\right)$ is defined as
\begin{equation}\label{def: estimate value func}
	\tilde{w}_{n,f}\left(t\right)\triangleq L_{f}\left(V\tilde{d}_{n,f}\left(t\right)-\alpha Q_{n}\left(t\right)\right).
\end{equation}

To minimize the upper bound of drift-plus-regret $\Delta_{V}(\boldsymbol{Q}(t))$ in (\ref{eq: drift-plus-regret bound 1}), we switch to solving the following problem in each time slot $t$:
\begin{equation}\label{p: one-slot}
\begin{array}{cl}
	\underset{\boldsymbol{X}(t)}{\text{maximize}}
	&\displaystyle \sum_{n\in\mathcal{N}}\sum_{f\in\mathcal{F}}\tilde{w}_{n,f}\left(t\right)X_{n,f}\left(t\right)\\
	\text{subject to}
	&\displaystyle \sum_{f\in\mathcal{F}}L_{f}X_{n,f}\left(t\right)\leq M_{n},\ \forall n\in\mathcal{N},\\
	&\displaystyle X_{n,f}(t)\in\{0,1\}, \forall n\in\mathcal{N}, f\in\mathcal{F}.
\end{array}
\end{equation}
In fact, problem (\ref{p: one-slot}) can be further decoupled into $N$ subproblems. For each EFS $n\in\mathcal{N}$, we solve the following subproblem for the cache placement vector $\boldsymbol{X}_{n}(t)$ in time slot $t$:
\begin{equation}
\begin{array}{cl}
	\underset{\boldsymbol{X}_{n}(t)}{\text{maximize}}
	&\displaystyle \sum_{f\in\mathcal{F}}\tilde{w}_{n,f}\left(t\right)X_{n,f}\left(t\right)\\
	\text{subject to}
	&\displaystyle \sum_{f\in\mathcal{F}}L_{f}X_{n,f}\left(t\right)\leq M_{n},\\
	&\displaystyle X_{n,f}(t)\in\{0,1\}, \forall f\in\mathcal{F}.
\end{array}
\end{equation}
\hfill \IEEEQED

% Appendix B: Theorem 1
\section{Proof of Theorem \ref{theorem: feasibility}}\label{proof: feasibility}

First, we have the following lemma.
\begin{lemma}\label{lemma}
If the budget vector $\boldsymbol{b}$ is an interior point of the maximal feasible region $\mathcal{B}$, then there exists a feasible scheme which makes \textit{i.i.d.} decision over time independent of the virtual queue backlog sizes.
\end{lemma}
The proof is omitted since it is quite standard as shown in the proof of Lemma 1 in \cite{li2019combinatorial}.

Then based on Lemma \ref{lemma}, we begin to prove Theorem \ref{theorem: feasibility}. 
By our assumption in Theorem \ref{theorem: feasibility} that $\boldsymbol{b}$ is an interior point of $\mathcal{B}$, there must exist some $\epsilon>0$ such that $\boldsymbol{b}-\epsilon\boldsymbol{1}$ is also an interior point of $\mathcal{B}$. Here $\boldsymbol{1}$ denotes the $N$-dimensional all-ones vector. 
Then by Lemma \ref{lemma}, since $\boldsymbol{b}-\epsilon\boldsymbol{1}$ lies in the interior of $\mathcal{B}$, there exists a feasible scheme which makes \textit{i.i.d.} decision over time independent of the virtual queue backlog sizes such that
\begin{equation}\label{eq: slackness}
	\mathbb{E}\left[\hat{C}_{n}\left(\boldsymbol{X}^{\epsilon}_{n}\left(t\right)\right)\right]\leq b_{n}-\epsilon,\ \forall n\in\mathcal{N}, t,
\end{equation}
where $\boldsymbol{X}^{\epsilon}(t)\triangleq (\boldsymbol{X}_{1}^{\epsilon}(t),\boldsymbol{X}_{2}^{\epsilon}(t),\cdots,\boldsymbol{X}_{N}^{\epsilon}(t))$ is the cache placement decision vector during time slot $t$ under the scheme.
We denote the cache placement decision vector during time slot $t$ under our scheme CPHBL by $\boldsymbol{X}^{c}(t)\triangleq (\boldsymbol{X}_{1}^{c}(t),\boldsymbol{X}_{2}^{c}(t),\cdots,\boldsymbol{X}_{N}^{c}(t))$, which is the optimal solution of problem (\ref{p: one-slot}). Then based on (\ref{eq: drift-plus-regret bound}), we have
\begin{equation}
\begin{split}
	&\Delta_{V}\left(\boldsymbol{Q}\left(t\right)\right)\leq B+\sum_{n\in\mathcal{N}}VK_{n}M_{n}\\
	&~~~+V\mathbb{E}\bigg[\sum_{n\in\mathcal{N}}\sum_{f\in\mathcal{F}}L_{f}d_{n,f}X_{n,f}^{*}\left(t\right)|\boldsymbol{Q}\left(t\right)\bigg]\\
	&~~~+\mathbb{E}\bigg[\sum_{n\in\mathcal{N}}Q_{n}\left(t\right)\left(\hat{C}_{n}\left(\boldsymbol{X}^{c}_{n}\left(t\right)\right)-b_{n}\right)|\boldsymbol{Q}\left(t\right)\bigg]\\
	&~~~-V\mathbb{E}\bigg[\sum_{n\in\mathcal{N}}\sum_{f\in\mathcal{F}}L_{f}\tilde{d}_{n,f}\left(t\right)X_{n,f}^{c}\left(t\right)|\boldsymbol{Q}\left(t\right)\bigg]\\
	&\leq B+\sum_{n\in\mathcal{N}}VK_{n}M_{n}\\
	&~~~+V\mathbb{E}\left[\sum_{n\in\mathcal{N}}\sum_{f\in\mathcal{F}}L_{f}d_{n,f}X_{n,f}^{*}\left(t\right)|\boldsymbol{Q}\left(t\right)\right]\\
	&~~~+\mathbb{E}\left[\sum_{n\in\mathcal{N}}Q_{n}\left(t\right)\left(\hat{C}_{n}\left(\boldsymbol{X}_{n}^{\epsilon}\left(t\right)\right)-b_{n}\right)|\boldsymbol{Q}\left(t\right)\right]\\
	&~~~-V\mathbb{E}\left[\sum_{n\in\mathcal{N}}\sum_{f\in\mathcal{F}}L_{f}\tilde{d}_{n,f}\left(t\right)X_{n,f}^{\epsilon}\left(t\right)|\boldsymbol{Q}\left(t\right)\right],
\end{split}
\end{equation}
where $B\triangleq\frac{1}{2}\sum_{n\in\mathcal{N}}\left(b_{n}^{2}+\alpha^{2}M_{n}^{2}\right)$. 
Since $\boldsymbol{X}^{\epsilon}(t)$ is independent of $\boldsymbol{Q}(t)$, we have
\begin{equation}
\begin{split}
	&\Delta_{V}\left(\boldsymbol{Q}\left(t\right)\right)\leq B+\sum_{n\in\mathcal{N}}VK_{n}M_{n}\\
	&+V\mathbb{E}\left[\sum_{n\in\mathcal{N}}\sum_{f\in\mathcal{F}}L_{f}d_{n,f}X_{n,f}^{*}\left(t\right)|\boldsymbol{Q}\left(t\right)\right]\\
	&+\sum_{n\in\mathcal{N}}Q_{n}\left(t\right)\mathbb{E}\left[\hat{C}_{n}\left(\boldsymbol{X}_{n}^{\epsilon}\left(t\right)\right)-b_{n}\right]\\
	&-V\mathbb{E}\left[\sum_{n\in\mathcal{N}}\sum_{f\in\mathcal{F}}L_{f}\tilde{d}_{n,f}\left(t\right)X_{n,f}^{\epsilon}\left(t\right)|\boldsymbol{Q}\left(t\right)\right].
\end{split}
\end{equation}
By (\ref{eq: slackness}), it follows that
\begin{equation}
\begin{split}
	&\Delta_{V}\left(\boldsymbol{Q}\left(t\right)\right)\leq B+\sum_{n\in\mathcal{N}}VK_{n}M_{n}-\epsilon\sum_{n\in\mathcal{N}}Q_{n}\left(t\right)\\
	&+V\mathbb{E}\left[\sum_{n\in\mathcal{N}}\sum_{f\in\mathcal{F}}L_{f}d_{n,f}X_{n,f}^{*}\left(t\right)|\boldsymbol{Q}\left(t\right)\right]\\
	&-V\mathbb{E}\left[\sum_{n\in\mathcal{N}}\sum_{f\in\mathcal{F}}L_{f}\tilde{d}_{n,f}\left(t\right)X_{n,f}^{\epsilon}\left(t\right)|\boldsymbol{Q}\left(t\right)\right].
\end{split}
\end{equation}
Since $\sum_{f\in\mathcal{F}}L_{f}d_{n,f}X_{n,f}^{*}\left(t\right)\leq K_{n}\sum_{f\in\mathcal{F}}L_{f}X_{n,f}^{*}\left(t\right)\leq K_{n}b_{n}$ and $\tilde{d}_{n,f}X_{n,f}^{\epsilon}\left(t\right)\geq0$, we have 
\begin{equation}
	\Delta_{V}\left(\boldsymbol{Q}\left(t\right)\right)\leq B+V\sum_{n\in\mathcal{N}}2K_{n}M_{n}-\epsilon\sum_{n\in\mathcal{N}}Q_{n}\left(t\right).
\end{equation}
Substituting (\ref{def: drift-plus-regret}) into above inequality, we have 
\begin{multline}
	\mathbb{E}\left[L\left(\boldsymbol{Q}\left(t+1\right)\right)-L\left(\boldsymbol{Q}\left(t\right)\right)|\boldsymbol{Q}\left(t\right)\right]+V\mathbb{E}\left[\Delta_{Reg}\left(t\right)|\boldsymbol{Q}\left(t\right)\right]\\
	\leq B+V\sum_{n\in\mathcal{N}}2K_{n}M_{n}-\epsilon\sum_{n\in\mathcal{N}}Q_{n}\left(t\right).
\end{multline}
Taking expectation at both sides of above inequality and summing it over time slots $\{0,1,\cdots,T'-1\}$, we have 
\begin{multline}
	\mathbb{E}\left[L\left(\boldsymbol{Q}\left(T'\right)\right)\right]-\mathbb{E}\left[L\left(\boldsymbol{Q}\left(0\right)\right)\right]+V\sum_{t=0}^{T'-1}\mathbb{E}\left[\Delta_{Reg}\left(t\right)\right]\\
	\leq T'B+T'V\sum_{n\in\mathcal{N}}2K_{n}M_{n}-\epsilon\sum_{t=0}^{T'-1}\sum_{n\in\mathcal{N}}\mathbb{E}\left[Q_{n}\left(t\right)\right].
\end{multline}
Dividing at both sides by $T'\epsilon$ and rearrange the terms, we have
\begin{equation}
\begin{split}
	&\frac{1}{T'}\sum_{t=0}^{T'-1}\sum_{n\in\mathcal{N}}\mathbb{E}\left[Q_{n}\left(t\right)\right]\leq\frac{1}{\epsilon}\left(B+V\sum_{n\in\mathcal{N}}2K_{n}M_{n}\right)\\
	&+\frac{\mathbb{E}[L(\boldsymbol{Q}(0))]}{T'\epsilon}\!-\!\frac{\mathbb{E}[L(\boldsymbol{Q}(T'))]}{T'\epsilon}\!-\!\frac{V}{T'\epsilon}\sum_{t=0}^{T'-1}\mathbb{E}[\Delta_{Reg}(t)].
\end{split}	
\end{equation}
It follows by the fact $L(\boldsymbol{Q}(0))=0$, $L(\boldsymbol{Q}(T'))\geq 0$, and $\frac{1}{T'}\sum_{t=0}^{T'-1}\mathbb{E}\left[\Delta_{Reg}\left(t\right)\right]=Reg\left(T'\right)\geq 0$ that
\begin{equation}
	\frac{1}{T'}\sum_{t=0}^{T'-1}\sum_{n\in\mathcal{N}}\mathbb{E}\left[Q_{n}(t)\right]\leq\frac{B+V\sum_{n\in\mathcal{N}}2K_{n}M_{n}}{\epsilon}.
\end{equation}
By taking the limsup of the left-hand-side term in above inequality as $T'\rightarrow\infty$, we obtain
\begin{equation}
	\limsup_{T'\rightarrow\infty}\frac{1}{T'}\!\sum_{t=0}^{T'-1}\!\sum_{n\in\mathcal{N}}\mathbb{E}[Q_{n}(t)]\!\leq\!\frac{B\!+\!V\!\sum_{n\in\mathcal{N}}2K_{n}M_{n}}{\epsilon}.
\end{equation}
This implies that $\limsup_{T'\rightarrow\infty}\frac{1}{T'}\sum_{t=0}^{T'-1}\mathbb{E}[Q_{n}(t)]<\infty$ and the virtual queueing process $\{Q_{n}\}_{t}$ defined in (\ref{eq: queue update}) is strongly stable for each EFS $n\in\mathcal{N}$. Hence, the time-averaged storage cost constraints in (\ref{constraint: cost}) are satisfied. 
\hfill \IEEEQED

\section{Proof of Theorem \ref{theorem: regret bound}}\label{proof: regret bound}

By Lemma \ref{lemma}, since $\boldsymbol{b}$ lies in the interior of $\mathcal{B}$, there exists an optimal scheme which makes \textit{i.i.d.} decision over time independent of the virtual queue backlog sizes such that
\begin{equation}\label{eq: opt slackness}
	\mathbb{E}\left[\hat{C}_{n}\left(\boldsymbol{X}^{*}_{n}\left(t\right)\right)\right]\leq b_{n},\ \forall n\in\mathcal{N}, t,
\end{equation}
where $\boldsymbol{X}^{*}(t)\triangleq (\boldsymbol{X}_{1}^{*}(t),\boldsymbol{X}_{2}^{*}(t),\cdots,\boldsymbol{X}_{N}^{*}(t))$ is the cache placement decision vector in time slot $t$ under the optimal scheme.
By the inequality in (\ref{eq: drift upper bound}) and the definition (\ref{def: one-slot regret}), under CPDBL, we have
\begin{equation}
\begin{split}
	&L\left(\boldsymbol{Q}\left(t+1\right)\right)-L\left(\boldsymbol{Q}\left(t\right)\right)+V\Delta_{Reg}\left(t\right)\\
	&\leq B+\sum_{n\in\mathcal{N}}Q_{n}\left(t\right)\left(\hat{C}_{n}\left(\boldsymbol{X}_{n}^{c}\left(t\right)\right)-b_{n}\right)\\
	&~~~-V\sum_{n\in\mathcal{N}}\sum_{f\in\mathcal{F}}L_{f}d_{n,f}\left(X_{n,f}^{*}\left(t\right)-X_{n,f}^{c}\left(t\right)\right).
\end{split}	
\end{equation}
The inequality above can be equivalently written as
\begin{equation}
\begin{split}
	&L\left(\boldsymbol{Q}\left(t+1\right)\right)-L\left(\boldsymbol{Q}\left(t\right)\right)+V\Delta_{Reg}\left(t\right)\\
	&\leq B+\sum_{n\in\mathcal{N}}Q_{n}\left(t\right)\left(\hat{C}_{n}\left(\boldsymbol{X}_{n}^{*}\left(t\right)\right)-b_{n}\right)\\
	&+\!\sum_{n\in\mathcal{N}}\!\bigg(\!\sum_{f\in\mathcal{F}}VL_{f}d_{n,f}X_{n,f}^{*}(t)\!-\!Q_{n}(t)\hat{C}_{n}(\boldsymbol{X}_{n}^{*}(t))\!\bigg)\\
	&-\!\sum_{n\in\mathcal{N}}\!\bigg(\!\sum_{f\in\mathcal{F}}VL_{f}d_{n,f}X_{n,f}^{c}(t)\!-\!Q_{n}(t)\hat{C}_{n}(\boldsymbol{X}_{n}^{c}(t))\!\bigg).
\end{split}
\end{equation}
Substituting (\ref{eq: EFS cost}) into the above inequality, we have
\begin{equation}\label{eq: drift + one-slot regret bound}
\begin{split}
	&L\left(\boldsymbol{Q}\left(t+1\right)\right)-L\left(\boldsymbol{Q}\left(t\right)\right)+V\Delta_{Reg}\left(t\right)\\
	&\leq B+\sum_{n\in\mathcal{N}}Q_{n}\left(t\right)\left(\hat{C}_{n}\left(\boldsymbol{X}_{n}^{*}\left(t\right)\right)-b_{n}\right)\\
	&+\!\sum_{n\in\mathcal{N}}\sum_{f\in\mathcal{F}}L_{f}\left(Vd_{n,f}-\alpha Q_{n}(t)\right)X_{n,f}^{*}(t)\\
	&-\!\sum_{n\in\mathcal{N}}\sum_{f\in\mathcal{F}}L_{f}\left(Vd_{n,f}-\alpha Q_{n}(t)\right)X_{n,f}^{c}(t).
\end{split}
\end{equation}
For each EFS $n\in\mathcal{N}$ and each file $f\in\mathcal{F}$, we define
\begin{equation}\label{def: accurate value func}
w_{n,f}\left(t\right)\triangleq L_{f}\left(Vd_{n,f}-\alpha Q_{n}\left(t\right)\right).
\end{equation}
Then inequality (\ref{eq: drift + one-slot regret bound}) can be simplified as:
\begin{equation}
\begin{split}
	&L\left(\boldsymbol{Q}\left(t+1\right)\right)-L\left(\boldsymbol{Q}\left(t\right)\right)+V\Delta_{Reg}\left(t\right)\\
	&\leq B+\sum_{n\in\mathcal{N}}Q_{n}\left(t\right)\left(\hat{C}_{n}\left(\boldsymbol{X}_{n}^{*}\left(t\right)\right)-b_{n}\right)\\
	&~~~ +\sum_{n\in\mathcal{N}}\sum_{f\in\mathcal{F}}w_{n,f}\left(t\right)\left(X_{n,f}^{*}\left(t\right)-X_{n,f}^{c}(t)\right).
\end{split}
\end{equation}
For simplicity of expression, we define 
\begin{equation}\label{def: Phi1}
\Phi_{1}\left(t\right)\triangleq \sum_{n\in\mathcal{N}}\sum_{f\in\mathcal{F}}w_{n,f}\left(t\right)\left(X_{n,f}^{*}(t)-X_{n,f}^{c}\left(t\right)\right). 
\end{equation}
It follows that
\begin{multline}
	L\left(\boldsymbol{Q}\left(t+1\right)\right)-L\left(\boldsymbol{Q}\left(t\right)\right)+V\Delta_{Reg}(t)\\
	\leq B+\Phi_{1}(t)\!+\!\sum_{n\in\mathcal{N}}Q_{n}(t)\big(\hat{C}_{n}(\boldsymbol{X}_{n}^{*}(t))-b_{n}\big).
\end{multline}
Taking conditional expectation at both sides of above inequality, we have
\begin{align}
	&\mathbb{E}\left[L(\boldsymbol{Q}(t+1))-L(\boldsymbol{Q}(t))|\boldsymbol{Q}(t)\right]+V\mathbb{E}\left[\Delta_{Reg}(t)|\boldsymbol{Q}(t)\right]\nonumber\\
	&\leq B+\mathbb{E}\left[\Phi_{1}(t)|\boldsymbol{Q}(t)\right]\\
	&~~~+\mathbb{E}\bigg[\sum_{n\in\mathcal{N}}Q_{n}(t)\big(\hat{C}_{n}(\boldsymbol{X}_{n}^{*}(t))-b_{n}\big)|\boldsymbol{Q}(t)\bigg]\nonumber\\
	&=B+\mathbb{E}\left[\Phi_{1}\left(t\right)|\boldsymbol{Q}\left(t\right)\right]\nonumber\\
	&~~~+\sum_{n\in\mathcal{N}}Q_{n}(t)\left(\mathbb{E}\big[\hat{C}_{n}\big(\boldsymbol{X}_{n}^{*}(t)\big)\big]-b_{n}\right).
\end{align}
The last equality holds because that $\hat{C}_{n}\left(\boldsymbol{X}_{n}^{*}(t)\right)$ is independent of $\boldsymbol{Q}(t)$.
By the inequalities in (\ref{eq: opt slackness}), it follows that
\begin{multline}
	\mathbb{E}\left[L\left(\boldsymbol{Q}\left(t+1\right)\right)-L\left(\boldsymbol{Q}\left(t\right)\right)|\boldsymbol{Q}\left(t\right)\right]\\
	+V\mathbb{E}\left[\Delta_{Reg}\left(t\right)|\boldsymbol{Q}\left(t\right)\right]
	\leq B+\mathbb{E}\left[\Phi_{1}\left(t\right)|\boldsymbol{Q}\left(t\right)\right].
\end{multline}
Taking expectation at both sides of above inequality, we have
\begin{multline}
	\mathbb{E}\left[L\left(\boldsymbol{Q}\left(t+1\right)\right)-L\left(\boldsymbol{Q}\left(t\right)\right)\right]+V\mathbb{E}\left[\Delta_{Reg}\left(t\right)\right]\\
	\leq B+\mathbb{E}\left[\Phi_{1}\left(t\right)\right].
\end{multline}
Summing above inequality over time slots $\{0,1,\cdots,T-1\}$ and dividing it at both sides by $TV$, we have
\begin{multline}
	\frac{\mathbb{E}\left[L\left(\boldsymbol{Q}\left(T\right)\right)\right]}{TV}-\frac{\mathbb{E}\left[L\left(\boldsymbol{Q}\left(0\right)\right)\right]}{TV}+\frac{1}{T}\sum_{t=0}^{T-1}\mathbb{E}\left[\Delta_{Reg}\left(t\right)\right]\\
	\leq\frac{B}{V}+\frac{1}{TV}\sum_{t=0}^{T-1}\mathbb{E}\left[\Phi_{1}\left(t\right)\right].
\end{multline}
Since $L\left(\boldsymbol{Q}(0)\right)$ and $L\left(\boldsymbol{Q}(T)\right)$ are both non-negative, it follows that
\begin{equation}\label{eq: regret bound 1}
\begin{split}
	Reg(T)&=\frac{1}{T}\sum_{t=0}^{T-1}\mathbb{E}\left[\Delta_{Reg}\left(t\right)\right]\\
	&\leq\frac{B}{V}+\frac{1}{TV}\sum_{t=0}^{T-1}\mathbb{E}\left[\Phi_{1}\left(t\right)\right].
\end{split}
\end{equation}
Next, we bound $\Phi_{1}(t)$ to obtain the upper bound of the regret $Reg(T)$.

\subsection{Bounding $\Phi_{1}(t)$}

To find the upper bound of $\Phi_{1}(t)$.
Consider a cache placement scheme which makes a placement decision during each time slot $t$, denoted by vector $\boldsymbol{X}'(t)\triangleq (\boldsymbol{X}_{1}'(t),\boldsymbol{X}_{2}'(t),\cdots,\boldsymbol{X}_{N}'(t))$ with each entry $\boldsymbol{X}_{n}'(t)$ as the optimal solution of the following problem:
\begin{equation}\label{p: subproblem without uncertainty}
\begin{array}{cl}
	\underset{\boldsymbol{X}_{n}(t)}{\text{maximize}}
	&\displaystyle \sum_{f\in\mathcal{F}}w_{n,f}\left(t\right)X_{n,f}\left(t\right)\\
	\text{subject to}
	&\displaystyle \sum_{f\in\mathcal{F}}L_{f}X_{n,f}\left(t\right)\leq M_{n},\\
	&\displaystyle X_{n,f}(t)\in\{0,1\}, \forall f\in\mathcal{F}.
\end{array}
\end{equation}
Since $\boldsymbol{X}_{n}^{*}(t)$ is a feasible solution of problem (\ref{p: subproblem without uncertainty}), we have
\begin{equation}
	\sum_{f\in\mathcal{F}}w_{n,f}\left(t\right)X_{n,f}'\left(t\right)\geq\sum_{f\in\mathcal{F}}w_{n,f}\left(t\right)X_{n,f}^{*}\left(t\right).
\end{equation}
It follows that
\begin{equation}\label{eq: Phi1 bound 1}
\begin{split}
	\Phi_{1}\left(t\right)=&\sum_{n\in\mathcal{N}}\sum_{f\in\mathcal{F}}w_{n,f}\left(t\right)\left(X_{n,f}^{*}\left(t\right)-X_{n,f}^{c}\left(t\right)\right)\\
	\leq &\sum_{n\in\mathcal{N}}\sum_{f\in\mathcal{F}}w_{n,f}\left(t\right)\left(X_{n,f}'\left(t\right)-X_{n,f}^{c}\left(t\right)\right)\\
	\leq &\sum_{n\in\mathcal{N}}\sum_{f\in\mathcal{F}}w_{n,f}\left(t\right)\left(X_{n,f}'\left(t\right)-X_{n,f}^{c}\left(t\right)\right)\\
	&+\sum_{n\in\mathcal{N}}\sum_{f\in\mathcal{F}}\tilde{w}_{n,f}\left(t\right)\left(X_{n,f}^{c}\left(t\right)-X_{n,f}'\left(t\right)\right).
\end{split}
\end{equation}
The last inequality holds since $\boldsymbol{X}^{c}_{n}(t)$ is the optimal solution of problem (\ref{p: one-slot subproblem}) but $\boldsymbol{X}_{n}'(t)$ is only a feasible solution.
Rearranging the right-hand side of (\ref{eq: Phi1 bound 1}), we obtain
\begin{multline}\label{eq: Phi1 bound 2}
	\Phi_{1}\left(t\right)\leq\sum_{n\in\mathcal{N}}\sum_{f\in\mathcal{F}}\left(\tilde{w}_{n,f}\left(t\right)-w_{n,f}\right)X_{n,f}^{c}\left(t\right)\\
	+\sum_{n\in\mathcal{N}}\sum_{f\in\mathcal{F}}\left(w_{n,f}-\tilde{w}_{n,f}\left(t\right)\right)X_{n,f}'\left(t\right).
\end{multline}
By (\ref{def: estimate value func}) and (\ref{def: accurate value func}), we have
\begin{equation}\label{eq: weight sub mean weight}
\begin{split}
	&\tilde{w}_{n,f}\left(t\right)-w_{n,f}\\
	&=L_{f}\left(V\tilde{d}_{n,f}(t)-\alpha Q_{n}(t)\right)-L_{f}\left(Vd_{n,f}-\alpha Q_{n}(t)\right)\\
	&=VL_{f}\left(\tilde{d}_{n,f}\left(t\right)-d_{n,f}\right).
\end{split}
\end{equation}
Substituting (\ref{eq: weight sub mean weight}) into (\ref{eq: Phi1 bound 2}), we obtain
\begin{equation}\label{eq: Phi1 bound 3}
\begin{split}
	\Phi_{1}\left(t\right)\leq &\sum_{n\in\mathcal{N}}\sum_{f\in\mathcal{F}}VL_{f}\left(\tilde{d}_{n,f}\left(t\right)-d_{n,f}\right)X_{n,f}^{c}\left(t\right)\\
	&+\sum_{n\in\mathcal{N}}\sum_{f\in\mathcal{F}}VL_{f}\left(d_{n,f}-\tilde{d}_{n,f}\left(t\right)\right)X_{n,f}'\left(t\right).
\end{split}
\end{equation}
Next, we define
\begin{equation}\label{def: phi2}
	\Phi_{2}\left(t\right)\triangleq \sum_{n\in\mathcal{N}}\sum_{f\in\mathcal{F}}L_{f}\left(\tilde{d}_{n,f}\left(t\right)-d_{n,f}\right)X_{n,f}^{c}\left(t\right)
\end{equation}
and 
\begin{equation}\label{def: Phi3}
	\Phi_{3}\left(t\right)\triangleq\sum_{n\in\mathcal{N}}\sum_{f\in\mathcal{F}}L_{f}\left(d_{n,f}-\tilde{d}_{n,f}\left(t\right)\right)X_{n,f}'\left(t\right).
\end{equation}
Then the upper bound of $\Phi_{1}(t)$ in (\ref{eq: Phi1 bound 3}) can be rewritten as 
\begin{equation}\label{eq: Phi1 bound 4}
	\Phi_{1}\left(t\right)\leq V\left(\Phi_{2}\left(t\right)+\Phi_{3}\left(t\right)\right).
\end{equation}
In the following subsections, we obtain the upper bounds of $\Phi_{2}(t)$ and $\Phi_{3}(t)$ respectively to bound $\Phi_{1}(t)$.

\subsection{Bounding $\Phi_{2}(t)$}

To derive the upper bound of $\Phi_{2}(t)$, we define the event $G_{n,f}(t)\triangleq\{ \tilde{d}_{n,f}(t)\geq d_{n,f}\} $ for each $n\in\mathcal{N}$ and $f\in\mathcal{F}$. Then we have
\begin{equation}\label{eq: Phi2 bound 1}
\begin{split}
	&\Phi_{2}\left(t\right)=\sum_{n\in\mathcal{N}}\sum_{f\in\mathcal{F}}L_{f}\left(\tilde{d}_{n,f}\left(t\right)-d_{n,f}\right)X_{n,f}^{c}\left(t\right)\\
	&~~~~~~~~~~~~~~~~~~~~~~\cdot\left(\mathds{1}\{ G_{n,f}(t)\} +\mathds{1}\{ G_{n,f}^{c}(t)\} \right)\\
	&=\sum_{n\in\mathcal{N}}\sum_{f\in\mathcal{F}}L_{f}\left(\tilde{d}_{n,f}(t)-d_{n,f}\right)X_{n,f}^{c}(t)\mathds{1} \{ G_{n,f}(t)\} \\
	&~~~+\sum_{n\in\mathcal{N}}\sum_{f\in\mathcal{F}}L_{f}\big(\tilde{d}_{n,f}(t)-d_{n,f}\big)X_{n,f}^{c}(t)\mathds{1}\{ G_{n,f}^{c}(t)\} \\
	&\leq \sum_{n\in\mathcal{N}}\sum_{f\in\mathcal{F}}L_{f}\big(\tilde{d}_{n,f}(t)-d_{n,f}\big)X_{n,f}^{c}(t)\mathds{1}\left\{ G_{n,f}(t)\right\} .
\end{split}
\end{equation}
The last inequality holds since when event $G_{n,f}^{c}(h)$ occurs, we have $\tilde{d}_{n,f}(t)< d_{n,f}$ and $(\tilde{d}_{n,f}(t)-d_{n,f})\mathds{1}\{ G_{n,f}^{c}(t)\}< 0$.
Next, we define 
\begin{equation}\label{def: phi2}
	\phi_{2,n,f}\left(t\right)\triangleq\left(\tilde{d}_{n,f}\left(t\right)-d_{n,f}\right)X_{n,f}^{c}\left(t\right)\mathds{1}\left\{ G_{n,f}\left(t\right)\right\}.
\end{equation}
Then we rewrite the upper bound of $\Phi_{2}(t)$ in (\ref{eq: Phi2 bound 1}) as
\begin{equation}\label{eq: Phi2 bound 2}
	\Phi_{2}\left(t\right)\leq\sum_{n\in\mathcal{N}}\sum_{f\in\mathcal{F}}L_{f}\phi_{2,n,f}\left(t\right).
\end{equation}

Let $t_{n,f}^{(1)}$ be the index of the first time slot in which file $f$ is cached on EFS $n$.
We define event $U_{n,f}(t)\triangleq\Big\{\bar{d}_{n,f}(t)- d_{n,f}> K_{n}\sqrt{\frac{3\log t}{2(h_{n,f}(t)+H_{n,f})}}\Big\} $ for each $n\in\mathcal{N}$ and $f\in\mathcal{F}$.
Summing $\phi_{2,n,f}(t)$ over time slots $\{0,1,\cdots,T-1\}$, it turns out that
\begin{equation}\label{eq: phi2 bound 1}
\begin{split}
	&~~~\sum_{t=0}^{T-1}\phi_{2,n,f}\left(t\right)\\
	&=\sum_{t=0}^{T-1}\left(\tilde{d}_{n,f}\left(t\right)-d_{n,f}\right)X_{n,f}^{c}\left(t\right)\mathds{1}\left\{ G_{n,f}\left(t\right)\right\} \\
	&\leq K_{n}X_{n,f}^{c}(t)\\
	&~~~+\sum_{t=t_{n,f}^{(1)}+1}^{T-1}\left(\tilde{d}_{n,f}\left(t\right)-d_{n,f}\right)X_{n,f}^{c}\left(t\right)\mathds{1}\left\{ G_{n,f}\left(t\right)\right\} \\
	&=K_{n}X_{n,f}^{c}(t)\\
	&~~~+\sum_{t=t_{n,f}^{(1)}+1}^{T-1}\left(\tilde{d}_{n,f}\left(t\right)-d_{n,f}\right)X_{n,f}^{c}\left(t\right)\mathds{1}\left\{ G_{n,f}\left(t\right)\right\} \\
	&~~~\cdot\left(\mathds{1}\left\{ U_{n,f}\left(t\right)\right\} +\mathds{1}\left\{ U_{n,f}^{c}\left(t\right)\right\} \right)	\\
	&=K_{n}X_{n,f}^{c}(t)\\
	&~~~+\sum_{t=t_{n,f}^{(1)}+1}^{T-1}\left(\tilde{d}_{n,f}\left(t\right)-d_{n,f}\right)X_{n,f}^{c}\left(t\right)\\
	&~~~\cdot\mathds{1}\left\{ G_{n,f}\left(t\right)\cap U_{n,f}\left(t\right)\right\} \\
	&~~~+\sum_{t=t_{n,f}^{(1)}+1}^{T-1}\left(\tilde{d}_{n,f}\left(t\right)-d_{n,f}\right)X_{n,f}^{c}\left(t\right)\\
	&~~~\cdot\mathds{1}\left\{ G_{n,f}\left(t\right)\cap U_{n,f}^{c}\left(t\right)\right\}.
\end{split}
\end{equation}
Next, we define
\begin{multline}\label{def: phi2_1}
	\phi_{2,n,f}^{\left(1\right)}\left(t\right)\triangleq\left(\tilde{d}_{n,f}\left(t\right)-d_{n,f}\right)X_{n,f}^{c}\left(t\right)\\
	\cdot\mathds{1}\left\{ G_{n,f}\left(t\right)\cap U_{n,f}\left(t\right)\right\}
\end{multline}
and 
\begin{multline}\label{def: phi2_2}
	\phi_{2,n,f}^{\left(2\right)}\left(t\right)\triangleq\left(\tilde{d}_{n,f}\left(t\right)-d_{n,f}\right)X_{n,f}^{c}\left(t\right)\\
	\cdot\mathds{1}\left\{ G_{n,f}\left(t\right)\cap U_{n,f}^{c}\left(t\right)\right\}.
\end{multline}
Then we rewrite inequality (\ref{eq: phi2 bound 1}) as the following equivalent form:
\begin{multline}\label{eq: phi2 bound 2}
	\sum_{t=0}^{T-1}\phi_{2,n,f}\left(t\right)\leq K_{n}X_{n,f}^{c}(t)\\
	+\sum_{t=t_{n,f}^{(1)}+1}^{T-1}\phi_{2,n,f}^{\left(1\right)}\left(t\right)+\sum_{t=t_{n,f}^{(1)}+1}^{T-1}\phi_{2,n,f}^{\left(2\right)}\left(t\right).
\end{multline}
By (\ref{eq: phi2 bound 2}), to bound $\phi_{2,n,f}(t)$, we switch to bounding $\phi_{2,n,f}^{(1)}(t)$ and $\phi_{2,n,f}^{(2)}(t)$. In the following, we derive upper bounds for such two terms, respectively.

First, we bound $\sum_{t=t_{n,f}^{(1)}+1}^{T-1}\phi_{2,n,f}^{\left(1\right)}\left(t\right)$. 
According to (\ref{def: phi2_1}), we have
\begin{multline}\label{eq: phi2_1 bound}
	\sum_{t=t_{n,f}^{(1)}+1}^{T-1}\phi_{2,n,f}^{\left(1\right)}\left(t\right)=\sum_{t=t_{n,f}^{(1)}+1}^{T-1}\left(\tilde{d}_{n,f}\left(t\right)-d_{n,f}\right)X_{n,f}^{c}\left(t\right)\\
	\cdot\mathds{1}\left\{ G_{n,f}\left(t\right)\cap U_{n,f}\left(t\right)\right\} .
\end{multline}
When event $U_{n,f}(t)\!=\!\Big\{\bar{d}_{n,f}(t)\!-\! d_{n,f}\!>\! K_{n}\sqrt{\frac{\!3\log t}{2(h_{n,f}(t)+H_{n,f})}}\Big\}$ occurs, we consider the following two cases:
\begin{enumerate}[(i)]
	\item If $\tilde{d}_{n,f}(t)=\min\Big\{\bar{d}_{n,f}(t)+K_{n}\sqrt{\frac{3\log t}{2(h_{n,f}(t)+H_{n,f})}},$ $K_{n}\Big\} =K_{n}$, then $\tilde{d}_{n,f}\left(t\right)\geq d_{n,f}$, \textit{i.e.}, event $G_{n,f}(t)$ occurs.
	\item If $\tilde{d}_{n,f}(t)=\min\left\{ \bar{d}_{n,f}\left(t\right)+K_{n}\sqrt{\frac{3\log t}{2(h_{n,f}(t)+H_{n,f})}},K_{n}\right\}$ $=\bar{d}_{n,f}\left(t\right)+K_{n}\sqrt{\frac{3\log t}{2(h_{n,f}(t)+H_{n,f})}}$, then event $G_{n,f}(t)$ still occurs, \textit{i.e.}, $\tilde{d}_{n,f}(t)>d_{n,f}+2K_{n}\sqrt{\frac{3\log t}{2(h_{n,f}(t)+H_{n,f})}}$.
\end{enumerate}
Therefore, we have  $U_{n,f}(t)\subset G_{n,f}(t)$, or equivalently, $\mathds{1}\left\{ G_{n,f}\left(t\right)\cap U_{n,f}\left(t\right)\right\}=\mathds{1}\left\{U_{n,f}\left(t\right)\right\}$. It follows that
\begin{multline}
	\sum_{t=t_{n,f}^{(1)}+1}^{T-1}\phi_{2,n,f}^{\left(1\right)}\left(t\right)\\
	=\sum_{t=t_{n,f}^{(1)}+1}^{T-1}\left(\tilde{d}_{n,f}\left(t\right)-d_{n,f}\right)X_{n,f}^{c}\left(t\right)\mathds{1}\left\{ U_{n,f}\left(t\right)\right\}.
\end{multline}
Since $\tilde{d}_{n,f}(t),d_{n,f}\in [0,K_{n}]$, we have $\tilde{d}_{n,f}(t)-d_{n,f}\leq K_{n}$. Then we have
\begin{equation}\label{eq: phi2_1 bound 2}
	\sum_{t=t_{n,f}^{(1)}+1}^{T-1}\phi_{2,n,f}^{(1)}(t)\leq \sum_{t=t_{n,f}^{(1)}+1}^{T-1}K_{n}X_{n,f}^{c}(t)\mathds{1}\left\{ U_{n,f}(t)\right\}.
\end{equation}
Taking expectation of (\ref{eq: phi2_1 bound 2}) at both sides, we have
\begin{equation}
\begin{split}
	&\sum_{t=t_{n,f}+1}^{T-1}\mathbb{E}\left[\phi_{2,n,f}^{\left(1\right)}\left(t\right)\right]\\
	&\leq\sum_{t=t_{n,f}^{(1)}+1}^{T-1}K_{n}\mathbb{E}[X_{n,f}^{c}(t)]\Pr\left\{ U_{n,f}\left(t\right)\right\}\\
	&=\sum_{t=t_{n,f}^{(1)}+1}^{T-1}K_{n}\mathbb{E}[X_{n,f}^{c}(t)]\\
	&~~~\cdot\Pr\left\{ \bar{d}_{n,f}(t)-d_{n,f}>K_{n}\sqrt{\frac{3\log t}{2(h_{n,f}(t)+H_{n,f})}}\right\} .
\end{split}
\end{equation}
Using the Chernoff-Hoeffding bound\cite{hoeffding1994probability}, we have
\begin{equation}
\begin{split}
	&\Pr\left\{ \bar{d}_{n,f}(t)-d_{n,f}>K_{n}\sqrt{\frac{3\log t}{2(h_{n,f}(t)+H_{n,f})}}\right\}\\
	&\leq \exp\left(-\frac{2\left(h_{n,f}(t)+H_{n,f}\right)^{2}}{(h_{n,f}(t)+H_{n,f})K_{n}^{2}}\cdot K_{n}^{2}\frac{3\log t}{2(h_{n,f}(t)+H_{n,f})}\right)\\
	&=\exp(-3\log t)=t^{-3}.
\end{split}
\end{equation}
Then it follows that
\begin{equation}\label{eq: phi2_1 bound 1}
\begin{split}
	&\sum_{n\in\mathcal{N}}\sum_{f\in\mathcal{F}}\sum_{t=t_{n.f}^{\left(1\right)}+1}^{T-1}L_{f}\mathbb{E}\left[\phi_{2,n,f}^{\left(1\right)}\left(t\right)\right]\\
	&\leq\sum_{t=1}^{\infty}\sum_{n\in\mathcal{N}}K_{n}\mathbb{E}\bigg[\sum_{f\in\mathcal{F}}L_{f}X_{n,f}^{c}(t)\bigg]t^{-3}\\
	&\leq\sum_{t=1}^{\infty}\sum_{n\in\mathcal{N}}K_{n}M_{n}t^{-3}\\
	&=\sum_{n\in\mathcal{N}}K_{n}M_{n}\left(1+\sum_{t=2}^{\infty}t^{-3}\right)\\
	&\leq\sum_{n\in\mathcal{N}}K_{n}M_{n}\left(1+\int_{1}^{\infty}t^{-3}dt\right)=\frac{3}{2}\sum_{n\in\mathcal{N}}K_{n}M_{n}.
\end{split}
\end{equation}

Next, we consider the upper bound of $\sum_{t=t_{n,f}^{(1)}+1}^{T-1}\phi_{2,n,f}^{\left(2\right)}\left(t\right)$. According to (\ref{def: phi2_2}), we have
\begin{multline}
	\sum_{t=t_{n,f}^{(1)}+1}^{T-1}\phi_{2,n,f}^{\left(2\right)}\left(t\right)=\sum_{t=t_{n,f}^{(1)}+1}^{T-1}\left(\tilde{d}_{n,f}\left(t\right)-d_{n,f}\right)X_{n,f}^{c}\left(t\right)\\
	\cdot\mathds{1}\left\{ G_{n,f}\left(t\right)\cap U_{n,f}^{c}\left(t\right)\right\} .
\end{multline}
When event $U_{n,f}^{c}(t)$ occurs, we have 
\begin{multline}
	\tilde{d}_{n,f}(t)=\min\left\{ \bar{d}_{n,f}(t)+K_{n}\sqrt{\frac{3\log t}{2(h_{n,f}(t)+H_{n,f})}},K_{n}\right\} \\
	\leq\bar{d}_{n,f}(t)+K_{n}\sqrt{\frac{3\log t}{2(h_{n,f}(t)+H_{n,f})}},
\end{multline}
thus
\begin{multline}\label{eq: estimate error bound}
	\tilde{d}_{n,f}\left(t\right)-d_{n,f}=\left(\tilde{d}_{n,f}(t)-\bar{d}_{n,f}(t)\right)\\
	+\left(\bar{d}_{n,f}\left(t\right)-d_{n,f}\right)\leq2K_{n}\sqrt{\frac{3\log t}{2(h_{n,f}(t)+H_{n,f})}}.
\end{multline}
Then by (\ref{eq: estimate error bound}) and $X_{n,f}(t)\leq 1$, we have
\begin{equation}\label{eq: phi2_2 bound 1}
\begin{split}
	&\sum_{t=t_{n,f}^{(1)}+1}^{T-1}\phi_{2,n,f}^{(2)}(t)\\
	&=\sum_{t=t_{n,f}^{(1)}+1}^{T-1}2K_{n}X_{n,f}^{c}(t)\sqrt{\frac{3\log t}{2(h_{n,f}(t)+H_{n,f})}}\\
	&~~~\cdot\mathds{1}\left\{ G_{n,f}\left(t\right)\cap U_{n,f}^{c}\left(t\right)\right\}\\
	&\leq\sum_{t=t_{n,f}^{(1)}+1}^{T-1}2K_{n}X_{n,f}^{c}\left(t\right)\sqrt{\frac{3\log t}{2(h_{n,f}(t)+H_{n,f})}}\\
	&\leq \sum_{t=t_{n,f}^{(1)}+1}^{T-1}K_{n}\sqrt{6\log T}\frac{X_{n,f}^{c}(t)}{\sqrt{h_{n,f}(t)+H_{n,f}}}.
\end{split}
\end{equation}
Since $h_{n,f}(t)\leq T$, we have
\begin{equation}
\begin{split}
	\frac{1}{\sqrt{h_{n,f}\left(t\right)+H_{n,f}}}&=\sqrt{\frac{h_{n,f}\left(t\right)}{h_{n,f}\left(t\right)+H_{n,f}}}\cdot\frac{1}{\sqrt{h_{n,f}\left(t\right)}}\\
	&\leq\sqrt{\frac{T}{T+H_{n,f}}}\cdot\frac{1}{\sqrt{h_{n,f}\left(t\right)}}.
\end{split}
\end{equation}
Then it follows that
\begin{equation}
	\sum_{t=t_{n,f}^{(1)}+1}^{T-1}\!\phi_{2,n,f}^{(2)}(t)\leq \!\!\sum_{t=t_{n,f}^{(1)}+1}^{T-1}\!K_{n}\sqrt{\frac{6T\log T}{T+H_{n,f}}}\frac{1}{\sqrt{h_{n,f}(t)}}.
\end{equation}
Let $t_{n,f}^{(i)}$ be the $i$-th time slot in which file $f$ is cached on EFS $n$. Then $t_{n,f}^{(h_{n,f}(T))}$ is the time slot in which file $f$ is lastly cached before time slot $T$. Accordingly, we have
\begin{equation}
\begin{split}
	&\sum_{t=t_{n,f}^{(1)}+1}^{T-1}\frac{1}{\sqrt{h_{n,f}\left(t\right)}} =\sum_{i=2}^{h_{n,f}\left(T\right)}\frac{1}{\sqrt{h_{n,f}(t_{n,f}^{(i)})}}\\
	&=\sum_{i=2}^{h_{n,f}\left(T\right)}\frac{1}{\sqrt{i-1}}=\sum_{i=1}^{h_{n,f}\left(T\right)-1}\frac{1}{\sqrt{i}}\\
	&\leq\int_{1}^{h_{n,f}\left(T\right)}\frac{1}{\sqrt{i}}di=2\left(\sqrt{h_{n,f}\left(T\right)}-1\right)\\
	&\leq 2\sqrt{h_{n,f}\left(T\right)}.
\end{split}
\end{equation}
It follows that
\begin{equation}\label{eq: phi2_2 bound 2}
	\sum_{t=t_{n,f}^{\left(1\right)}+1}^{T-1}\phi_{2,n,f}^{\left(2\right)}\left(t\right)\leq
	2K_{n}\sqrt{\frac{6T\log T}{T+H_{n,f}}}\sqrt{h_{n,f}\left(T\right)}.
\end{equation}
Combining (\ref{eq: Phi2 bound 2}), (\ref{eq: phi2 bound 2}), (\ref{eq: phi2_1 bound 1}) and (\ref{eq: phi2_2 bound 2}), we have
\begin{equation}
\begin{split}
	&\sum_{t=0}^{T-1}\mathbb{E}\left[\Phi_{2}\left(t\right)\right]
	\leq \sum_{n\in\mathcal{N}}\sum_{f\in\mathcal{F}}L_{f}\sum_{t=0}^{T-1}\mathbb{E}\left[\phi_{2,n,f}\left(t\right)\right]\\
	&\leq \frac{5}{2}\sum_{n\in\mathcal{N}}K_{n}M_{n}\\
	&~~~+2\sum_{n\in\mathcal{N}}\sum_{f\in\mathcal{F}}L_{f}K_{n}\sqrt{\frac{6T\log T}{T+H_{n,f}}}\sqrt{h_{n,f}\left(T\right)}\\
	&\leq \frac{5}{2}\sum_{n\in\mathcal{N}}K_{n}M_{n}\\
	&~~~+2\sqrt{\frac{6T\log T}{T+H_{\min}}}\sum_{n\in\mathcal{N}}K_{n}\sum_{f\in\mathcal{F}}L_{f}\sqrt{h_{n,f}\left(T\right)},
\end{split}
\end{equation}
where we define a non-negative integer $H_{\min}\triangleq\min_{n,f}H_{n,f}$. The last inequality holds because that $\sum_{f\in\mathcal{F}}L_{f}X_{n,f}^{c}\left(t\right)\leq M_{n}$ for each $n\in\mathcal{N}$.
On the other hand, by Jensen's inequality, we have 
\begin{equation}
\begin{split}
\sum_{f\in\mathcal{F}}\frac{L_{f}}{\sum_{f\in\mathcal{F}}L_{f}}\sqrt{h_{n,f}\left(T\right)}&\leq \sqrt{\frac{\sum_{f\in\mathcal{F}}L_{f}h_{n,f}\left(T\right)}{\sum_{f\in\mathcal{F}}L_{f}}}\\
&\leq\sqrt{\frac{M_{n}T}{\sum_{f\in\mathcal{F}}L_{f}}}.
\end{split}
\end{equation}
Then it follows that
\begin{multline}\label{eq: Phi2 bound 3}
	\sum_{t=0}^{T-1}\mathbb{E}\left[\Phi_{2}\left(t\right)\right]
	\leq \frac{5}{2}\sum_{n\in\mathcal{N}}K_{n}M_{n}\\
	+2\left(\sum_{n\in\mathcal{N}}K_{n}\sqrt{M_{n}\sum_{f\in\mathcal{F}}L_{f}}\right)\sqrt{\frac{6T^{2}\log T}{T+H_{\min}}}.
\end{multline}

\subsection{Bounding $\Phi_{3}(t)$}

Recall by (\ref{def: Phi3}) and $G_{n,f}(t)\triangleq\{ \tilde{d}_{n,f}(t)\geq d_{n,f}\}$ that 
\begin{equation}\label{eq: Phi3 bound 1}
\begin{split}
	&\Phi_{3}\left(t\right)
	=\sum_{n\in\mathcal{N}}\sum_{f\in\mathcal{F}}L_{f}\left(d_{n,f}-\tilde{d}_{n,f}\left(t\right)\right)X_{n,f}'\left(t\right)\\
	&=\sum_{n\in\mathcal{N}}\sum_{f\in\mathcal{F}}L_{f}\left(d_{n,f}-\tilde{d}_{n,f}\left(t\right)\right)X_{n,f}'\left(t\right)\\
	&~~~~~~~~~~~~~~~\cdot\left(\mathds{1}\left\{ G_{n,f}\left(t\right)\right\} +\mathds{1}\left\{ G_{n,f}^{c}\left(t\right)\right\} \right)\\
	&\leq\sum_{n\in\mathcal{N}}\sum_{f\in\mathcal{F}}L_{f}\left(d_{n,f}-\tilde{d}_{n,f}(t)\right)X_{n,f}'(t)\mathds{1}\left\{ G_{n,f}^{c}(t)\right\}.
\end{split}
\end{equation}
Then we define
\begin{equation}\label{def: phi3}
	\phi_{3,n,f}\left(t\right)\triangleq\left(d_{n,f}-\tilde{d}_{n,f}\left(t\right)\right)X_{n,f}'\left(t\right)\mathds{1}\left\{ G_{n,f}^{c}\left(t\right)\right\} ,
\end{equation}
whereby the upper bound of $\Phi_{3}(t)$ in (\ref{eq: Phi3 bound 1}) can be written as 
\begin{equation}\label{eq: Phi3 bound 2}
	\Phi_{3}\left(t\right)\leq\sum_{n\in\mathcal{N}}\sum_{f\in\mathcal{F}}L_{f}\phi_{3,n,f}\left(t\right).
\end{equation}
Next, we consider the case where $t\leq t_{n,f}^{(1)}$ and the case where $t\geq t_{n,f}^{(1)}+1$, respectively. 
When $t\leq t_{n,f}^{(1)}$, we have $\tilde{d}_{n,f}(t)=K_{n}$. Then the event $G_{n,f}^{c}(t)=\{ \tilde{d}_{n,f}(t)<d_{n,f}\}$ would not occur since $d_{n,f}\leq K_{n}$. Therefore, $\phi_{3,n,f}(t)=0$ when $t\leq t_{n}$.

When $t\geq t_{n,f}^{(1)}+1$, suppose that event $G_{n,f}^{c}(t)$ occurs. Then we have $\tilde{d}_{n,f}(t)<d_{n,f}\leq K_{n}$, which implies that $\tilde{d}_{n,f}(t)=\bar{d}_{n,f}(t)+K_{n}\sqrt{\frac{3\log t}{2(h_{n,f}(t)+H_{n,f})}}$. It follows that $d_{n,f}>\bar{d}_{n,f}(t)+K_{n}\sqrt{\frac{3\log t}{2(h_{n,f}(t)+H_{n,f})}}$. Hence, we bound $\mathbb{E}[\phi_{3,n,f}(t)]$ as follows:
\begin{equation}
\begin{split}
	&\mathbb{E}\left[\phi_{3,n,f}\left(t\right)\right]\\
	&=\mathbb{E}\left[\left(d_{n,f}-\tilde{d}_{n,f}\left(t\right)\right)X_{n,f}'\left(t\right)\mathds{1}\left\{ G_{n,f}^{c}\left(t\right)\right\} \right]\\
	&\leq\mathbb{E}\left[K_{n}X_{n,f}'\left(t\right)\mathds{1}\left\{ G_{n,f}^{c}\left(t\right)\right\} \right]\\
	&=K_{n}X_{n,f}'\left(t\right)\mathbb{E}\left[\mathds{1}\left\{ G_{n,f}^{c}\left(t\right)\right\} \right]\\
	&=K_{n}X_{n,f}'\left(t\right)\Pr\left\{ G_{n,f}^{c}\left(t\right)\right\} \\
	&\leq K_{n}X_{n,f}'(t)\\
	&~~~\cdot\Pr\left\{ \!d_{n,f}\!>\!\bar{d}_{n,f}(t)+K_{n}\sqrt{\frac{3\log t}{2(h_{n,f}(t)+H_{n,f})}}\right\} .
\end{split}
\end{equation}
By Chernoff-Hoeffding bound, we have
\begin{equation}
\begin{split}
	&\Pr\left\{ d_{n,f}>\bar{d}_{n,f}\left(t\right)+K_{n}\sqrt{\frac{3\log t}{2(h_{n,f}(t)+H_{n,f})}}\right\} \\
	&=\Pr\left\{ \bar{d}_{n,f}\left(t\right)< d_{n,f}-K_{n}\sqrt{\frac{3\log t}{2(h_{n,f}(t)+H_{n,f})}}\right\} \\
	&\leq\exp\left(-3\log t\right)=t^{-3}.
\end{split}
\end{equation}
Hence, we have
\begin{equation}
	\mathbb{E}\left[\phi_{3,n,f}\left(t\right)\right]\leq K_{n}X_{n,f}'\left(t\right)t^{-3}.
\end{equation}
Based on the above inequality, we have
\begin{equation}\label{eq: phi3 bound}
\begin{split}
	&\sum_{t=0}^{T-1}\sum_{f\in\mathcal{F}}L_{f}\mathbb{E}\left[\phi_{3,n,f}\left(t\right)\right]\\
	&\leq K_{n}\sum_{t=t_{n,f}^{\left(1\right)}+1}^{T-1}\sum_{f\in\mathcal{F}}L_{f}X_{n,f}'\left(t\right)t^{-3}\\
	&\leq K_{n}\sum_{t=t_{n,f}^{\left(1\right)}+1}^{T-1}M_{n}t^{-3},
\end{split}
\end{equation}
where the last inequality holds because $\sum_{f\in\mathcal{F}}L_{f}X_{n,f}'\left(t\right)\leq M_{n}$.
Then by (\ref{eq: phi3 bound}), it follows that
\begin{equation}\label{eq: phi3 bound 1}
\begin{split}
	&\sum_{t=0}^{T-1}\sum_{f\in\mathcal{F}}L_{f}\mathbb{E}\left[\phi_{3,n,f}\left(t\right)\right]\\
	&\leq K_{n}M_{n}\sum_{t=t_{n,f}^{\left(1\right)}+1}^{T-1}t^{-3}\leq K_{n}M_{n}\sum_{t=1}^{\infty}t^{-3}\\
	&\leq K_{n}M_{n}\left(1+\sum_{t=2}^{\infty}t^{-3}\right)\\
	&\leq K_{n}M_{n}\left(1+\int_{1}^{\infty}t^{-3}dt\right)=\frac{3}{2}K_{n}M_{n}.
\end{split}
\end{equation}
By (\ref{eq: Phi3 bound 2}) and (\ref{eq: phi3 bound 1}), we have
\begin{equation}\label{eq: Phi3 bound 3}
\begin{split}
	\sum_{t=0}^{T-1}\mathbb{E}\left[\Phi_{3}\left(t\right)\right]&\leq\sum_{n\in\mathcal{N}}\sum_{t=0}^{T-1}\sum_{f\in\mathcal{F}}L_{f}\mathbb{E}\left[\phi_{3,n,f}\left(t\right)\right]\\
	&\leq\sum_{n\in\mathcal{N}}\frac{3}{2}K_{n}M_{n}.
\end{split}
\end{equation}
Combining (\ref{eq: Phi1 bound 4}), (\ref{eq: Phi2 bound 3}) and (\ref{eq: Phi3 bound 3}), we obtain
\begin{equation}\label{eq: Phi1 bound 5}
\begin{split}
	&\sum_{t=0}^{T-1}\mathbb{E}\left[\Phi_{1}\left(t\right)\right]\\
	&\leq V\left(\sum_{t=0}^{T-1}\mathbb{E}\left[\Phi_{2}\left(t\right)\right]+\sum_{t=0}^{T-1}\mathbb{E}\left[\Phi_{3}\left(t\right)\right]\right)\\
	&\leq 4V\sum_{n\in\mathcal{N}}K_{n}M_{n}\\
	&~~~+2V\left(\sum_{n\in\mathcal{N}}K_{n}\sqrt{M_{n}\sum_{f\in\mathcal{F}}L_{f}}\right)\sqrt{\frac{6T^{2}\log T}{T+H_{\min}}}.
\end{split}
\end{equation}
Substituting (\ref{eq: Phi1 bound 5}) into (\ref{eq: regret bound 1}), we obtain a regret bound as follows:
\begin{multline}
	Reg\left(T\right)\leq \frac{B}{V}+\frac{4\sum_{n\in\mathcal{N}}K_{n}M_{n}}{T}\\
	+2\left(\sum_{n\in\mathcal{N}}K_{n}\sqrt{M_{n}\sum_{f\in\mathcal{F}}L_{f}}\right)\sqrt{\frac{6\log T}{T+H_{\min}}},
\end{multline}
where $B=\frac{1}{2}\sum_{n\in\mathcal{N}}(b_{n}^{2}+\alpha^{2}M_{n}^{2})$ and $H_{\min}=\min_{n,f}H_{n,f}$.

\hfill \IEEEQED
\ifCLASSOPTIONcaptionsoff
  \newpage
\fi

\end{document}